%% file: eurosp-2022-template.tex
\documentclass[compsoc,conference,a4paper,10pt,times]{IEEEtran}
\IEEEoverridecommandlockouts
\usepackage{cite}
\usepackage{amsmath,amssymb,amsfonts}
\usepackage{graphicx}
\usepackage{textcomp}
\usepackage{bmpsize}
\usepackage{xcolor}
\usepackage{lipsum}
\usepackage{comment}
\usepackage{enumitem}
\usepackage[colorlinks=true,urlcolor=black]{hyperref}
\def\BibTeX{{\rm B\kern-.05em{\sc i\kern-.025em b}\kern-.08em
    T\kern-.1667em\lower.7ex\hbox{E}\kern-.125emX}}
    
\UseRawInputEncoding
\usepackage{algorithmicx,algpseudocode}

\newcommand{\gm}[1]{\mynote{GMG}{#1}}

\usepackage[numbers]{natbib}
\usepackage{amsthm}
\makeatletter
\def\els@aparagraph[#1]#2{\elsparagraph[#1]{#2\@addpunct{.}}}
\def\els@bparagraph#1{\elsparagraph*{#1\@addpunct{.}}}
\makeatother

\usepackage{amsthm}
\makeatletter
\def\els@aparagraph[#1]#2{\elsparagraph[#1]{#2\@addpunct{.}}}
\def\els@bparagraph#1{\elsparagraph*{#1\@addpunct{.}}}
\makeatother

\usepackage{caption} 
\usepackage{float}
\floatstyle{plaintop}
\restylefloat{table}

\usepackage{xcolor}

\newcommand\norm[1]{\left\lVert#1\right\rVert}
\usepackage{array}
\newcolumntype{M}[1]{>{\centering\arraybackslash}m{#1}}

\usepackage{comment}

\usepackage{booktabs}

\def\Plus{\texttt{+}}

\usepackage{pdfpages}

\usepackage{algorithm}

\usepackage{scalerel,stackengine}
\stackMath
\newcommand\reallywidehat[1]{%
\savestack{\tmpbox}{\stretchto{%
  \scaleto{%
    \scalerel*[\widthof{\ensuremath{#1}}]{\kern-.6pt\bigwedge\kern-.6pt}%
    {\rule[-\textheight/2]{1ex}{\textheight}}
  }{\textheight}%
}{0.5ex}}%
\stackon[1pt]{#1}{\tmpbox}%
}

\usepackage{color}
\newcommand{\mynote}[2]
    {{\color{red} \fbox{\bfseries\sffamily\scriptsize#1}
    {\small$\blacktriangleright$\textsf{\emph{#2}}$\blacktriangleleft$}}~}

\begin{document}

\title{Do I Get the Privacy I Need? Benchmarking Utility in Differential Privacy Libraries\\ {\footnotesize
} }


\author{
\IEEEauthorblockN{Gonzalo Munilla Garrido*\thanks{*Corresponding author. Email: gonzalo.munilla-garrido@tum.de}}
\IEEEauthorblockA{\textit{TUM}
}
\\
\IEEEauthorblockN{Warren He}
\IEEEauthorblockA{Oasis Labs}
\and
\IEEEauthorblockN{Joseph P. Near}
\IEEEauthorblockA{\textit{University of Vermont}}
\\
\IEEEauthorblockN{Roman Matzutt}
\IEEEauthorblockA{RWTH Aachen}
\and
\IEEEauthorblockN{Aitsam Muhammad}
\IEEEauthorblockA{RWTH Aachen}
\\
\IEEEauthorblockN{Florian Matthes}
\IEEEauthorblockA{TUM}
}

\maketitle
\thispagestyle{plain}
\pagestyle{plain}

\begin{abstract}
An increasing number of open-source libraries promise to bring differential privacy to practice, even for non-experts.
This paper studies five libraries that offer differentially private analytics: Google DP, SmartNoise, diffprivlib, diffpriv, and Chorus.
We compare these libraries qualitatively (capabilities, features, and maturity) and quantitatively (utility and scalability) across four analytics queries (count, sum, mean, and variance) executed on synthetic and real-world datasets.
We conclude that these libraries provide similar utility (except in some notable scenarios).
However, there are significant differences in the features provided, and we find that no single library excels in all areas.
Based on our results, we provide guidance for practitioners to help in choosing a suitable library, guidance for library designers to enhance their software, and guidance for researchers on open challenges in differential privacy tools for non-experts.
\end{abstract}


\newcommand{\chk}[1]{{\textcolor{blue}{#1}}}
\newcommand{\fix}[1]{{\textcolor{red}{#1}}}
\newcommand{\todo}[1]{{\textcolor{red}{TODO: #1}}}
\newcommand{\RM}[1]{{\textcolor{magenta}{Roman: #1}}}

\renewcommand{\paragraph}[1]{\vspace*{1mm}\noindent \textbf{#1}\hspace*{1mm}}

\input{01_Introduction_new}

\input{02_relatedWork}
\input{03_DP}

\input{04_Methodology}

\input{05_Datasets_Overview}
\input{06_Qualitative_comparison}
\input{07_Experiments}
\input{08_Discussion}

\input{09_Conclusion}

\bibliographystyle{IEEEtranN}
\bibliography{references}

\pagebreak
\appendix

\subsection{Experiments workflow}

\begin{algorithm} 
\caption{Logic of the experiments to measure utility from the outputs of the analytics queries.}
\begin{flushleft}
\hspace*{\algorithmicindent} 
\textbf{Input}: $\mathcal{P}$, set of libraries to benchmark; $\mathcal{M}_\mathcal{P}$, set of randomized analytics queries to benchmark within a library; $\mathcal{D}$, set of datasets; $\mathcal{E}$, set of $\varepsilon$; $\mathcal{N}$, number of experiments for a single $\varepsilon$; $\mathcal{W}_{\mathcal{M}_\mathcal{P}}$, set of baseline truthful analytics queries mapped to a given $\mathcal{M}_\mathcal{P}$. \\

\hspace*{\algorithmicindent} \textbf{Output}: $\hat{y}$, noisy query result; $y$, truthful query result; $E$, set of size $\mathcal{N}$ of $L_1$ errors; $RE$, set of size $\mathcal{N}$ of $L_1$ relative errors; $\reallywidehat{RE}$, set of cardinality $|\mathcal{E}|$ of sample means of each $RE$; $S$, set of cardinality $|\mathcal{E}|$ of sample std of each scaled $E$.
\end{flushleft}

\begin{algorithmic}[1] 
\State 
    \textbf{for each:} {$p \in \mathcal{P} $} \par 
    \State \hskip\algorithmicindent \textbf{for each:} {$\mathcal{M}_p \in \mathcal{M}_\mathcal{P} $}\par 
    \State \hskip\algorithmicindent \hskip\algorithmicindent \textbf{for each:} {$\mathcal{I} \in \mathcal{D} $}\par
    \State \hskip\algorithmicindent \hskip\algorithmicindent \hskip\algorithmicindent $\reallywidehat{RE}$ = \{\}; $S$ = \{\};
    \State \hskip\algorithmicindent \hskip\algorithmicindent \hskip\algorithmicindent \textbf{for each:} {$\varepsilon \in \mathcal{E} $}\par
    \State \hskip\algorithmicindent \hskip\algorithmicindent \hskip\algorithmicindent \hskip\algorithmicindent $RE$ = \{\};  $E$ = \{\}
    \State \hskip\algorithmicindent \hskip\algorithmicindent \hskip\algorithmicindent \hskip\algorithmicindent $y = \mathcal{W}_{\mathcal{M}_p}(\mathcal{I},\varepsilon)$
    \State \hskip\algorithmicindent \hskip\algorithmicindent \hskip\algorithmicindent \hskip\algorithmicindent \textbf{for i $=$ 1 to $\mathcal{N}$ do}\par
    \State \hskip\algorithmicindent \hskip\algorithmicindent \hskip\algorithmicindent
    \hskip\algorithmicindent  \hskip\algorithmicindent $\hat{y}$ = $\mathcal{M}_p(\mathcal{I},\varepsilon)$ \par
    \State \hskip\algorithmicindent \hskip\algorithmicindent \hskip\algorithmicindent
    \hskip\algorithmicindent  \hskip\algorithmicindent $E$ = $E$ $\cup$ $\norm{\hat{y} - y}_1$ \par
    \State \hskip\algorithmicindent
    \hskip\algorithmicindent \hskip\algorithmicindent \hskip\algorithmicindent \hskip\algorithmicindent $RE$ = $RE$ $\cup$ $\dfrac{\norm{\hat{y} - y}_1}{\norm{y}_1}$ \par
    \State \hskip\algorithmicindent \hskip\algorithmicindent \hskip\algorithmicindent \hskip\algorithmicindent $\reallywidehat{RE}$ = $\reallywidehat{RE}$ $\cup$ $mean(RE)$
    \State \hskip\algorithmicindent \hskip\algorithmicindent \hskip\algorithmicindent \hskip\algorithmicindent $S$ = $S$ $\cup$ $SampleStd\bigg(\dfrac{E}{|\mathcal{I}|}\bigg)$
    \State \hskip\algorithmicindent \hskip\algorithmicindent \hskip\algorithmicindent plot($\reallywidehat{RE}$, $\mathcal{E}$)\par
    \State \hskip\algorithmicindent \hskip\algorithmicindent \hskip\algorithmicindent plot($S$, $\mathcal{E}$)\par
\end{algorithmic}
\label{basic_query_experiment}
\end{algorithm}

\subsection{Differential privacy mechanisms}

\input{DP_Mechs_Table.tex}

\subsection{Experiment Results}


\input{plots_table}

\end{document}

%% file: 01_introduction_new.tex
\section{Introduction}
\label{introduction}

In recent years, a confluence of trends drives academics and industry practitioners to research and invest in more powerful privacy-enhancing measures to protect people's privacy while leveraging their data.
One of the drivers is the increase in costs for the institutions due to more frequent data breaches~\cite{security_2020_nodate}, e.g. loss of brand equity, customer turnover, or legal expenditure.
Furthermore, white-hat academics have performed demonstration attacks on ``de-identified'' public data in different domains, effectively deprecating legacy privacy-enhancing methodologies.
Notable examples of re-identification have taken place
in genome sequencing~\cite{sweeney_identifying_2013}, in the automotive, telecommunications, and entertainment industry~\cite{gao_elastic_2014, kondor_towards_2020, narayanan_robust_2008}, and in e-commerce~\cite{archie_-anonymization_nodate}.
Moreover, the promise of privacy-enhancing products and services can also bring benefits, such as avoiding price discrimination, allow the utilization of data across organizations, new business models, 
and develop less pervasive forms of social media that can prevent malicious social engineering~\cite{social_engineering}, among others. 
In this context, differential privacy (DP) has received increasing attention because of its mathematical guarantees of privacy unique among anonymization techniques.
Moreover, DP's inherent implementation complexity~\cite{dwork_exposed_2017} has driven organizations and researchers to create libraries for practitioners to include DP in their stack.

In this paper, we examine a set of mainstream libraries through the lens of a benchmark.
We consider open-source libraries that provide support for analytics queries, come from prominent institutions or researchers, and offer DP functionality that eases integration and usage.
Thus, we consider the following libraries: Microsoft's SmartNoise~\cite{SmartNoise_repo},  
IBM's diffprivlib~\cite{IBM_repo}, 
Google-DP~\cite{Google_repo}, 
Chorus~\cite{Chorus_repo, uber_dp}, and diffpriv~\cite{diffpriv_repo, sensitivity_sampler_R}.
There exist other libraries that comply with the targeted qualities, e.g, the pioneering PINQ~\cite{mcsherry_privacy_nodate} and GUPT~\cite{gupt}; however, they are not maintained anymore.

At the core, these libraries aim to abstract DP to a level where \emph{non-experts} can implement DP applications.
As interest in differential privacy grows, research, governmental, and private institutions will gravitate towards these open-source libraries.
Our work aims to help guide practitioners, library designers, and researchers in navigating the coming adoption of tools for DP.

\paragraph{Our Contributions.}
Specifically, this paper makes the following contributions:

\begin{itemize}[topsep=1mm,leftmargin=5mm]
\itemsep1.0mm
\item We conduct a comprehensive comparison and evaluation of five mainstream open-source DP libraries
\item We provide \textit{guidance for practitioners} to aid in selecting a specific library (\S~\ref{sec:recommendations_for_practitioners})
\item We provide \textit{guidance for library designers} on how to make their software more useful (\S~\ref{sec:recommendations_designers})
\item We provide \textit{guidance for researchers} on important open research challenges remaining in differential privacy tools for non-experts (\S~\ref{key_findings})
\end{itemize}

\noindent We provide the \href{https://github.com/gonzalo-munillag/Benchmarking_Differential_Privacy_Analytics_Libraries}{source code}~\cite{benchmark_repo} to reproduce our study and for practitioners to quickly implement further benchmarks. 
  
\paragraph{Methodology.}
Our guidance is based on the answers to three research questions (RQ), described in Section~\ref{methodology}.
To answer these RQs, we compare these libraries both qualitatively (in terms of capabilities, features, and maturity) and quantitatively (in terms of scalability and utility through a set of principled experiments). 
Our benchmarks include both synthetic data (to explore the differences between libraries systematically) and real-world data (to confirm these results in a realistic setting). 
We compare the libraries across four query types: count, sum, mean, and var. 

\paragraph{Results \& Key Findings.}
Our qualitative comparison (\S~\ref{qualitative_comparison_results}) demonstrates clear feature differences between libraries (see Table~\ref{tab:Comparison_Table}).
First, libraries offer \textbf{different capabilities}: \emph{e.g.} some target only analytics queries, while others provide machine learning capabilities as well.
Second, libraries differ in \textbf{analyst support}: \emph{e.g.} some calculate sensitivity automatically, while others require the analyst to provide it.
Third, some libraries are designed for \textbf{direct use} by analysts, while others provide frameworks for \textbf{building applications}.
In addition, libraries differ in their protections against \textbf{side channels} (such as floating-point vulnerabilities~\cite{mironov_significance_2012}) and support for \textbf{privacy budgeting} over multiple queries.

All of the libraries in our study provided \textbf{similar utility} for corresponding queries in our benchmarks, with some important exceptions (\S~\ref{experiment}). We find that all five libraries offer similar performance characteristics, but that \textbf{none of the libraries} scales to truly massive datasets (\S~\ref{performance_results}).

\paragraph{Guidance for Practitioners.}
Our results suggest that the largest differences between libraries come in their support for analysts and protections against side-channels.
We therefore advise practitioners to prioritize these factors when choosing a library.
Libraries like diffprivlib provide the best support for data scientists, while Google-DP is designed for building new applications and provides strong protection against side-channels.
Our complete guidance appears in Section~\ref{sec:recommendations_for_practitioners}.

\paragraph{Guidance for Library Designers.}
The results of our study indicate that library designers have generally done a good job implementing basic mechanisms and providing sufficient performance for small-scale analytics.
We suggest that library designers prioritize support for the analyst, protections against side-channels, and the addition of ``simple'' mechanisms that can provide good real-world performance (like the Geometric mechanism).
We also note the danger of implementation bugs in these libraries, and support the use of tools like Google-DP's stochastic tester to ensure correctness. Our complete guidance appears in Section~\ref{sec:recommendations_designers}.

\paragraph{Guidance for Researchers.}
Research in differential privacy has historically focused on developing mechanisms that improve utility.
Our study suggests that the increasing adoption of differential privacy opens important new avenues for researchers in this area.
In particular, practitioners need better tools for understanding \textbf{how much utility to expect} and \textbf{how to improve utility}.
They also need help understanding the \textbf{non-privacy implications} of each mechanism, such as output consistency.
Finally, as differential privacy sees increasing adoption, tools for \textbf{privacy budgeting} become even more important. Our complete guidance appears in Section~\ref{key_findings}.

The rest of this paper is organized as follows. We discuss related work in Section~\ref{related_work} and give background on differential privacy in Section~\ref{background}. We describe our research questions and methodology in Section~\ref{methodology} and the datasets used in our study in Section~\ref{Datasets_overview}. The results of our study appear in Section~\ref{qualitative_comparison_results}--\ref{performance_results} and our guidance appears in Sections~\ref{sec:recommendations_for_practitioners}--\ref{key_findings}. 
Finally, we discuss our conclusion in Section~\ref{conclusion}.

%% file: 02_relatedWork.tex
\section{Related work}
\label{related_work}
DP is a powerful concept that causes ever-increasing interest among privacy experts and is currently an active field of research in academia and industry.
Ever since the introduction of DP by Dwork~\cite{DP_dwork_original}, there has been abundant research conducted on the theoretical aspects of DP, addressing questions such as choosing the correct amount of noise and the appropriate values for DP parameters~\cite{hardt2010geometry, lee2011much, de2012lower}, as well as on the practical application of DP, for instance, in data mining \cite{friedman_data_2010}, data publication and analysis \cite{yang2012datapub}, and deep learning \cite{abadi2016deep}.

Consequently, the expansion of the research and application of DP in recent years prompted researchers to perform systematic reviews and comparative studies of the work conducted in the field.
Xiong~et~al.~\cite{xiong2020comprehensive} and Yang~et~al.~\cite{yang2020local} present comprehensive surveys on local DP algorithms and their applications, providing a source of reference for different privacy-related scenarios, as well as identifying research gaps and possible directions for future research.
Hassan~et~al.~\cite{hassan2019differential} conducted a detailed survey on the implementation of various DP techniques in cyber-physical systems, such as energy, transportation, healthcare, and industrial Internet of things. 
The authors covered all dimensions and aspects of implementing DP in these domains and discussed related issues and challenges.
Furthermore, motivated by the argument that even aggregated data such as histograms can result in privacy leakage, Nelson and Reuben \cite{nelson2020sok} conducted a systematic literature review, a qualitative analysis of $27$ papers that address the application of DP for histogram and synthetic data. 
The authors identified trends in the field and explained a crucial issue in adopting DP to tackle the privacy-utility trade-off. 

Aside from these qualitative surveys, there exist quantitative comparisons in the use of DP in range queries \cite{range_queries_DP}, geo-spatial data \cite{DP_grids}, spatial decomposition \cite{DP_hierarchy}, data mining \cite{friedman_data_2010}, and a test framework of DP \cite{zhang_testing_2020}.
However, the work most closely related to ours was conducted by Hay~et~al.~\cite{DP_Bench} in 2016. 
In their paper, the authors propose a principled framework called DPBench to evaluate $1$- and $2$-dimensional range queries.
However, none of the extant literature benchmarks the mainstream open-source DP libraries towards which non-experts gravitate; with this study, we fill this research gap.


%% file: 03_DP.tex
\section{Differential Privacy}
\label{background}

Traditional privacy protection methods are vulnerable to auxiliary information attacks against sensitive data analysis public releases~\cite{sweeney_identifying_2013, gao_elastic_2014, kondor_towards_2020, narayanan_robust_2008, archie_-anonymization_nodate}.
On the other hand, DP, introduced in 2006 by Dwork et al.~\cite{DP_dwork_original}, addresses the traditional techniques' limitations by mathematically formalizing a differential guarantee of privacy agnostic to auxiliary information.
DP maintains this promise by ensuring that an informed adversary analyzing a query output cannot determine whether an individual's data was used to compute such output. 

DP ensures that outcomes are similarly likely, \emph{with} or \emph{without} the data contributed by a particular individual.
The similarity of outcomes is parameterized by the parameter $\varepsilon$, which tunes the strength of the privacy guarantee (a lower $\varepsilon$ leads to better privacy).
We define a state of privacy as the prevention of the re-identification of an individual~\cite{Wu2012}, whose protection is adjusted by $\varepsilon$.
DP is formally described in Definition 1, which is based on~\cite{goos_lecture_nodate_2}:

\textbf{Definition 1} (\textit{($\varepsilon$,$\delta$)-Differential Privacy}). A randomized algorithm $\mathcal{M}$ is ($\varepsilon$,$\delta$)-differentially private if for any two datasets $\mathcal{D}$ and $\mathcal{D}^{\prime}$ differing on at most one element, and any set of possible outputs $\mathcal{S}$ $\in$ $Range(\mathcal{M})$:
\begin{center}
	Pr[$\mathcal{M}$($\mathcal{D}$)  $\in$  $\mathcal{S}$] $\leq$ $e^\varepsilon$ $\times$ Pr[$\mathcal{M}$($\mathcal{D}^{\prime}$)  $\in$  $\mathcal{S}$] + $\delta$.
\end{center}

For $\delta = 0$, Definition 1 is considered \emph{pure DP}.
The weaker guarantee when $\delta > 0$ is called \emph{approximate DP}.
This relaxed form of DP lowers the privacy of the individuals in exchange for utility~\cite{dwork_algorithmic_2013}.
%
Additionally, one may choose how $\mathcal{D}$ and $\mathcal{D}^{\prime}$ differ in one individual, leading to two possibilities: \emph{bounded} or \emph{unbounded} DP~\cite{kifer2011no}.

Mechanisms ensure DP by adding carefully-chosen random noise, typically to the output of a deterministic function.
The functions used in this study are based on the analytics queries: count, sum, mean, and var.
The scale of the noise used is based on the deterministic function's \emph{$\ell_1$-sensitivity} (\textit{global sensitivity})~\cite{dwork_algorithmic_2013}:

\textbf{Definition 2} ($\ell_1$-sensitivity). The $\ell_1$-sensitivity of an algorithm $\mathcal{W}: \mathbb{R}^{m} \rightarrow \mathbb{R}^{n}$, executed over datasets $\mathcal{D}$, $\mathcal{D}^{\prime}$ $\in$ $\mathbb{R}^k$ at a Hamming distance of $d_h\left(\mathcal{D}, \mathcal{D}^{\prime}\right) = 1$, is:
\begin{center}
    $
    \Delta f = \max_{\substack{
    {\mathcal{D}, \mathcal{D}^{\prime} \in \mathbb{R}^{k}} \\
    d_h\left(\mathcal{D}, \mathcal{D}^{\prime}\right) = 1  \\
    }} \|\mathcal{W}\left(\mathcal{D}\right)-\mathcal{W}\left(\mathcal{D}^{\prime}\right)\|_{1}.$
\end{center}

There are multiple implementations of mechanisms complying with Definition 1 and 2.
The most commonly known are the Laplace mechanism~\cite{agrawal_differential_2008} and the Gaussian mechanism~\cite{dwork_algorithmic_2013} for numerical data, and the Exponential mechanism~\cite{mcsherry_mechanism_nodate} for categorical and numerical data.
The Laplace and exponential mechanisms ensure pure DP, while the Gaussian mechanism requires approximate DP.
In this paper, we benchmark algorithms derived from the Laplace mechanism~\cite{agrawal_differential_2008}:

\textbf{Definition 3} (Laplace mechanism). For an algorithm $\mathcal{W}$ executed over a dataset $\mathcal{D}$, its differentially private version $\mathcal{M}$ adds Laplace noise: 
$\mathcal{M}$($\mathcal{D}$) = $\mathcal{W}$($\mathcal{D}$) + $Lap(x|\mu, b)$, with center $\mu = 0$ and scale $b = \frac{\Delta f}{\varepsilon}$.

One may observe that the lower the $\varepsilon$, the larger the standard deviation (std) of the noise, improving privacy (but reducing utility).
The noise magnitude is independent of the number of records in a dataset (dataset size), so analyzing more data yields better relative utility.

DP algorithms obey \textit{sequential composition} \cite{dwork_algorithmic_2013}, i.e. if a randomized query $\mathcal{M}$ is executed $n$ times over $\mathcal{D}$ with $\varepsilon_i$, the total $\varepsilon$ employed is equal to $\sum \varepsilon_i$, which is the consumed \emph{privacy budget}. 
Because an adversary can use the $n$ outputs to average out the noise because it is centered around $0$, some libraries implement privacy budget trackers to help practitioners preventing these attacks.
These trackers monitor the consumed budget and block further queries once the budget is consumed.

A function's sensitivity ($\Delta f$) depends on the function itself, and is sometimes difficult to calculate.
Counting queries are easy: they have a sensitivity of 1.
Other queries are more difficult: for example, the sensitivity of a summation query depends on the minimum and maximum possible values being summed (which themselves need to be kept private!).
Thus, practitioners either set estimates without looking at the data, or employ libraries offering \emph{private sensitivity estimation}, which consumes privacy budget.

Another practical consideration of DP algorithms is their \emph{output consistency}, e.g., a var query should not output values $\leq 0$, counts should not be decimal values, or the outputs of a DP count and sum from a library should yield similar values to the library's DP mean.
Lastly, it is important to consider \emph{side-channels} that may leak more information than the mathematical definition of the mechanism (like the \emph{floating-point vulnerabilities}~\cite{mironov_significance_2012}).

Additional challenges of deploying DP include choosing $\varepsilon$ and tracking budgets across different systems~\cite{DP_census_problems}.
Despite its flaws, DP holds a set of advantages.
Primarily, DP allows to mathematically limit the privacy loss of an analysis on sensitive data~\cite{dwork_algorithmic_2013}, and practitioners can adapt DP to different use cases, e.g., deep learning~\cite{DP_SGD}.
Moreover, the community of researchers continuously tackles the flaws of DP, such as solving its floating-point vulnerability with the Snapping mechanism~\cite{mironov_significance_2012} and improving this mechanism's utility (SmartNoise, Google DP). 
Overall, because of its unique features, DP may become the de facto standard of privacy in the near future.

%% file: 04_Methodology.tex
\section{Methodology}
\label{methodology}

\subsection{Research questions}
\label{qualitative_comparison}

To provide practitioners, library designers, and researchers with a comprehensive picture of the qualities and performance of the five selected open-source libraries, we aim to answer the following research questions (RQ):

{\textbf{RQ1}} \textit{How do available open-source libraries compare in terms of functionality?} To answer this RQ, we conduct a qualitative comparison in Section~\ref{qualitative_comparison_results}.
\newline 
\indent \textbf{RQ2} \textit{How do available open-source libraries compare regarding utility in the count, sum, mean, and var queries?} 
In answer to this RQ, we perform a utility benchmark of the analytics queries in Section~\ref{experiment}.
\newline 
\indent \textbf{RQ3} \textit{Which libraries perform best regarding execution time and memory consumption as the dataset size increases?} 
Section~\ref{performance_results} tackles this RQ by executing a scalability benchmark.

After answering these RQs, we extracted a set of recommendations for practitioners (\S~\ref{sec:recommendations_for_practitioners}) and library designers (\S~\ref{sec:recommendations_designers}), and guidance for researchers (\S~\ref{key_findings}).

\input{Principles.tex}

\subsection{Principles for evaluation}
\label{benchmarking_basic_queries}

Our evaluation focuses on four commonly-used analytics queries: count, sum, mean, and var. 
We base our evaluation principles on DPBench, developed by Hay~et~al.~\cite{DP_Bench}; most of the other literature focuses on comparing algorithms without formalizing the process~\cite{range_queries_DP, DP_grids, DP_hierarchy, friedman_data_2010}. 
We formulate our comparison Principles in Table~\ref{tab:Principles_tab}.


Principles from I to V are necessary to lay the basis for a holistic comparison of the libraries.
Specifically, we employed real datasets and created synthetic datasets (I) to make our comparisons comprehensive and reproducible.
We employed the skew-normal distribution to add diversity to our synthetic data and to provide a reasonable model of realistic data distributions; consequently, we varied the number of records (II), how these records are distributed over the domain by adjusting the skew parameter (III), and the spread of the datasets by changing the scale parameter (IV).
Our synthetic datasets contain continuous data, since the benchmarked algorithms work equally well for continuous and discrete data (and are not impacted by domain size).
Consecutively, to ensure $\varepsilon$ diversity (V), we ran experiments for $73$ $\varepsilon$ values from $0.01$ to $100$ at logarithmic steps, except for $\varepsilon$ values between $0.01$ to $0.5$, for which we employed steps of $0.01$.
We chose this fine granularity to reveal the behavior at small $\varepsilon$ values because practitioners mainly adhere to this range~\cite{dwork_exposed_2017}, as privacy is conserved most.

Principle VI deals with the fact that the outputs of the algorithms are r.v.'s.
Thus, to measure the results' utility, we perform $500$ experiments for each $\varepsilon$ ($500$ x $73 = 36500$ experiments) to have a large enough sample for each library and input dataset.
Moreover, we set the outputs of our benchmark to be the relative error and the absolute scaled error w.r.t the dataset scale. 
Consequently, the values we compare between libraries are the sample std of the absolute scaled error (SASE), and the sample mean of the relative error (MRE), providing two summary values per $\varepsilon$.
The baselines for the error calculations are the deterministic outputs of the analytics queries.
We choose the MRE because it is familiar to practitioners and widely used in utility comparisons in the extant literature~\cite{DP_Bench}, and allows to compare datasets of different scales; it is a measure of the accuracy of the random query.
In contrast, the SASE serves to understand the variability of the outputs across datasets of different scales, i.e., it measures the precision of the random query.
Without scaling the error, the SASE would remain the same across scales; however, in practicality, an error spread of equal magnitude imbues more uncertainty in small datasets than in large ones. Scaling makes the comparison across dataset scales fair.

The MRE is informative for "risk-neutral" practitioners, while the SASE is for the "risk-averse" because the metric indicates the presence of outliers in the outputs~\cite{DP_Bench}. 
For each $\varepsilon$, we consider the libraries with smaller SASE and MRE to perform better. Furthermore, measuring the MRE for large values of $\varepsilon$ (up to $100$) will reveal the presence of an algorithm's bias.
Because the MRE's underlying value is the relative error, we enable practitioners to use our results as a baseline for comparing other DP algorithms (VII).
Finally, we did not employ outlier values for the generation of the synthetic datasets to avoid comparisons that may not be equivalent to reality (VIII).

For the scalability benchmark, we measure execution time ($t$) and memory consumption ($m$).
We perform this benchmark by obtaining $t$ and $m$ from running an analytics query at varying dataset scales from $10$ to $10$ million data points with a fixed skewness and scale.
We limit the measurements to the query execution itself, not considering: pre-processing steps, dispatching, and receiving the query result.
This experimental setting is preferred because these three steps may vary between practitioners' system architectures.

%% file: Principles.tex
\begin{table*}
\centering
\caption{Principles of comparison adopted from or inspired by DPBench~\cite{DP_Bench}.}
\label{tab:Principles_tab}
\begin{tabular}{
>{\arraybackslash}p{0.45cm}
>{\raggedright\arraybackslash}p{2.15cm} 
>{\raggedright\arraybackslash}p{5.5cm} 
>{\raggedright\arraybackslash}p{7.5cm} 
}
\toprule
& \textbf{Principle} & \textbf{Description} & \textbf{Implementation} \\
\hline 

(I) & Datasets of synthetic and real nature 
& \textit{The algorithms’ inputs should comprise synthetic and real data.} 
& We employ synthetic data to surface differences between algorithms and real data to confirm these results. \\ 
\hline 

(II) & Dataset size diversity 
& \textit{Execute algorithms on datasets of varying numbers of records.} 
& The synthetic datasets contain $1000$, $10000$, or $100000$ records. \\  
\hline 

(III) & Shape diversity
& \textit{Execute algorithms on datasets of varying record distribution over the domain.}
& To make synthetic datasets more diverse, we set the \emph{skew} parameter of Scipy's skew-normal noise generator~\cite{scipy_guide, Azzalini_1999} to the values $0$, $5$, $50$ (The location parameter was set to $0$). \\ 
\hline 

(IV) & Spread diversity & 
\textit{Execute algorithms on datasets of varying spread.}
& Likewise, aiming to produce more diverse datasets, we tuned the synthetic datasets’ spread with the \emph{scale} parameter of Scipy's skew-normal noise generator~\cite{scipy_guide, Azzalini_1999} with the values $50$, $250$, and $500$. \\ 
\hline 

(V) & $\varepsilon$ diversity 
& \textit{Execute algorithms under different $\varepsilon$ values.} 
& We perform experiments for $73$ $\varepsilon$ values from $0.01$ to $100$. \\ 
\hline 

(VI) & Measurement of expectation, variability, and bias 
& \textit{The benchmark’s results should register accuracy, precision, and bias to measure utility.} 
& The output of our benchmark measures the outputs' sample mean of the relative error (accuracy) and sample std of the absolute scaled error (precision).
Measuring accuracy at values of $\varepsilon$ up to $100$ can reveal an algorithm's bias.  \\ 
\hline 

(VII) & Comparable results 
& \textit{The results should be comparable to other algorithms beyond the ones compared in the benchmark.} 
& We measure the relative error so that other practitioners can compare their algorithms. \\ \hline 

(VIII) & Avoid extreme input settings 
& \textit{The algorithms’ inputs should not lead to edge-case comparisons.} 
& We do not input extreme values such as scales of a handful of records. \\

\bottomrule
 
\end{tabular}
\end{table*}














%% file: 05_Datasets_Overview.tex
\section{Datasets overview}
\label{Datasets_overview}




\subsection{Synthetic datasets}
\label{normally_dist_datasets}

\begin{figure}[!htbp]
    \centering
    \includegraphics[scale=0.52]{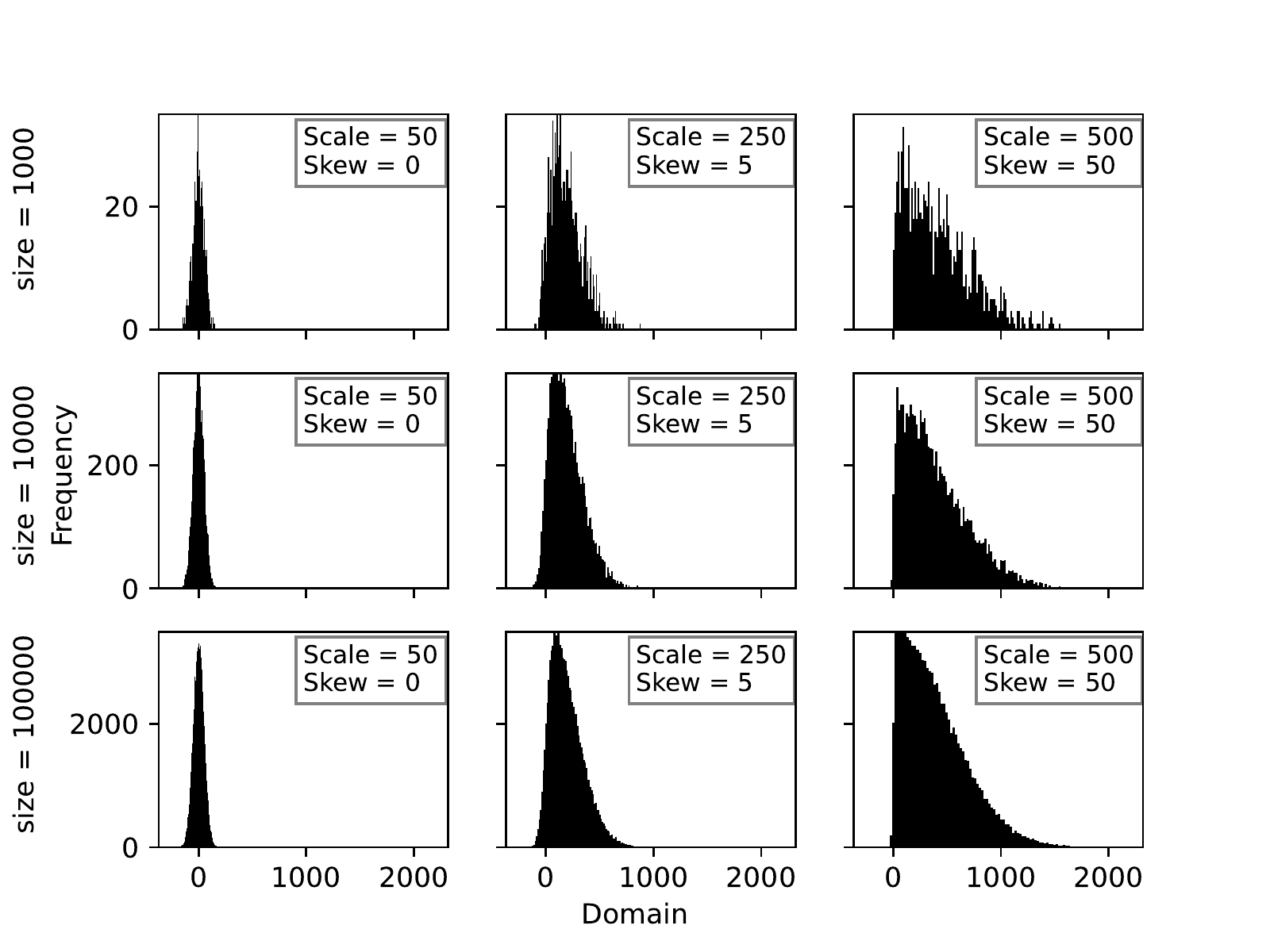}
    \caption{ 9 of the 27 normally distributed synthetic dataset histograms.}
    \label{fig:normally_distributed_datasets}
\end{figure}

We chose three values for each of the characteristics proposed by principles II (\textit{dataset size diversity}), III (\textit{shape diversity}), and IV (\textit{spread diversity}) following our implementation guidelines of Table~\ref{tab:Principles_tab}, obtaining $27$ datasets\footnote{\label{Scipy} The $27$ datasets were generated using skew-normal noise with the Python package Scipy~\cite{scipy_guide, Azzalini_1999}.} (see Fig.~\ref{fig:normally_distributed_datasets}).

Some queries require knowledge about the range of possible values in the dataset. Sometimes, domain knowledge can be used to define this range (e.g. human ages are typically between 0 and 100), but in other cases (like our synthetic data), this process is more challenging. 
Google DP provides a DP mechanism for estimating the range of a dataset, but the other libraries require the analyst to specify the range (including Google DP's Python wrapper). 
To ensure consistency in our synthetic data experiments, we use the \emph{actual} maximum and minimum values of the dataset when such a range is required. 

More formally, we represent each $D_i$ as a data vector $\textbf{x}$, which contains a list of \emph{float} values (within the set of real numbers), that varies in size, shape, and scale (based on Principles II to IV) across different datasets $D_i$, for $i \in [1, 27]$.
For example, dataset $D_1$ was generated with a skewness of $0$, a scale of $50$, and $1000$ data points.
The \textit{workload}, $\textbf{W}$, is a set of four one-dimensional functions (analytics queries, $\mathcal{W} \in \textbf{W}$) executed over $\textbf{x}$. 
Applying the $\mathcal{W}$ to $\textbf{x}$ yields the vector of deterministic results $\textbf{y} = \mathcal{W}\left(\textbf{x}\right)$.
The libraries transform an analytics query $\mathcal{W}$ from the workload $\textbf{W}$ into a DP algorithm $\mathcal{M}$ by adding noise to the result of $\mathcal{W}\left(\cdot\right)$, i.e., the noise generates randomized analytics queries.
The noisy result is $\hat{\textbf{y}} = \mathcal{M}\left(\textbf{x}\right) = \mathcal{W}\left(\textbf{x}\right) +  Noise$, and we use the $\textit{L}_1$ norm as the error of $\mathcal{M}$, $\norm{\mathcal{W}\left(\textbf{x}\right) - \hat{\textbf{y}}}_1$, scaled by the deterministic value $\mathcal{W}\left(\textbf{x}\right)$ for the relative error, and by the cardinality of the dataset $|\mathcal{D}_i|$ for the absolute scaled error.
The libraries' goal is to report the approximate results of the queries in $\textbf{W}$ on the private datasets in $\mathcal{D}$ while incurring minimal error.

Lastly, to test memory consumption and execution time, we carried out experiments on datasets of fixed \textit{shape} and \textit{spread} (generated with a skewness of $5$ and a scale of $250$) but of varying \textit{dataset size} from $10$ to $10$ million data points.

\subsection{Real-World datasets}
\label{real_datasets}

\begin{figure}[!t]
    \centering
    \includegraphics[scale=0.25]{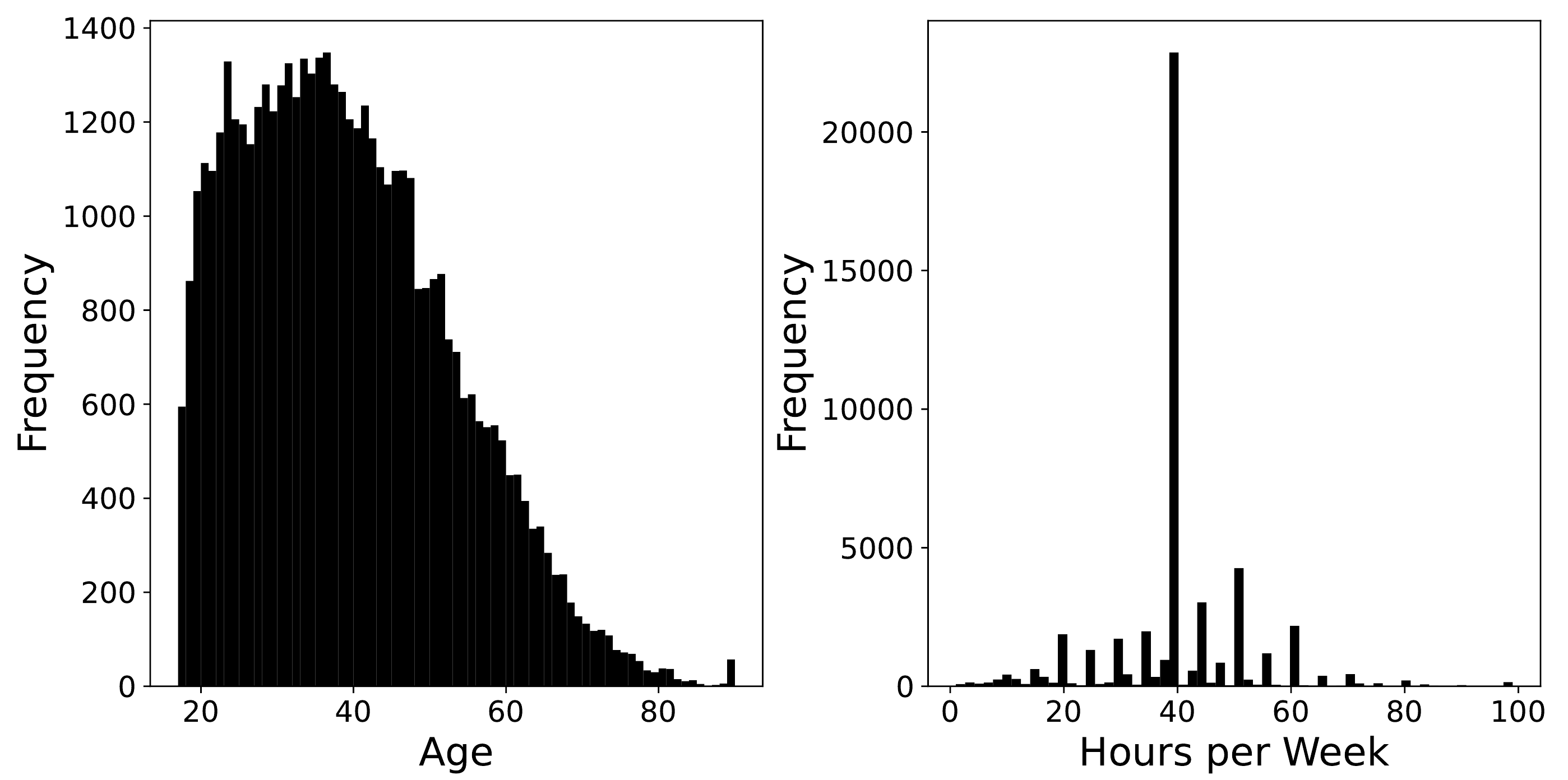}
    \caption{ 1984 USA Census dataset attributes of \textit{age} and worked \textit{hours per week} containing 48842 individuals.}
    \label{fig:real_distributed_datasets_census}
\end{figure}

\begin{figure}[!t]
    \centering
    \includegraphics[scale=0.25]{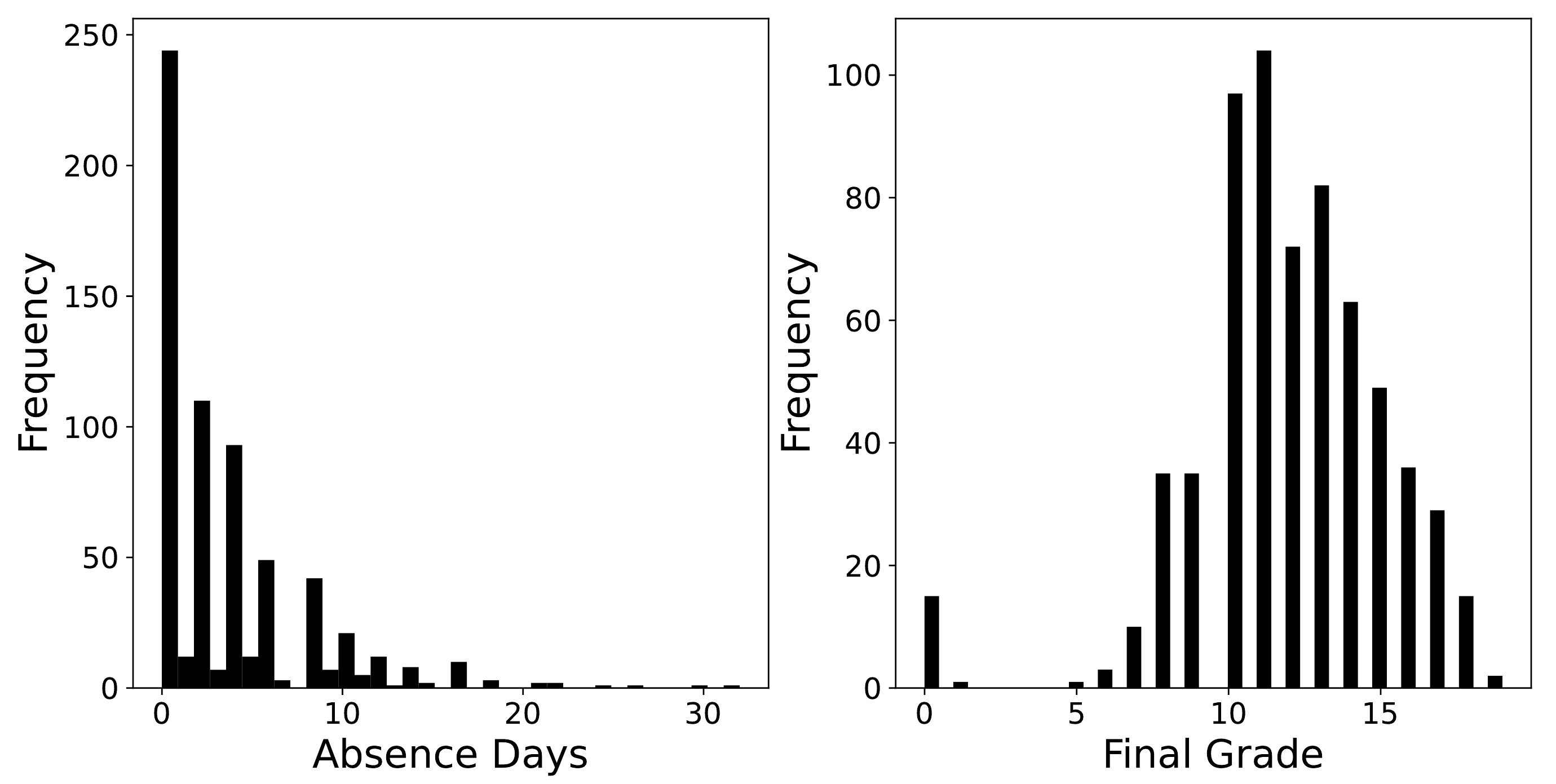}
    \caption{ Portuguese education dataset attributes of \textit{absence days} and \textit{final grade} containing 649 students.}
    \label{fig:real_distributed_datasets_education}
\end{figure}

Benchmarking analytics queries solely on synthetic data is not enough, as practitioners will ultimately execute the queries over real-world datasets.
Therefore, we must also select publicly available datasets for the benchmark.
Consequently, we have chosen two publicly available demographic datasets from different sources and contexts: the 1994 USA census~\cite{census_repo, census_dataset} with $48842$ individuals, and a Portuguese education dataset~\cite{student_repo, student_dataset} with $649$ students.
To increase the diversity of inputs of the benchmark, we select two sensitive numeric attributes from each of the two datasets: \textit{age} and\textit{ hours worked per week} from the census dataset, and \textit{absence days} and \textit{final exam grade} from the education dataset; histograms of these attributes may be observed in Fig.~\ref{fig:real_distributed_datasets_census} and Fig.~\ref{fig:real_distributed_datasets_education}. 
Unlike for the synthetic datasets, to set the range bounds for the sensitivity calculation, we selected values based on the domain knowledge of the real-world datasets' attributes, e.g., for the \textit{age} attribute, we selected a lower bound and an upper bound of $0$ and a $100$ years, respectively.
Lastly, we formally define the real-world datasets and the workload equally to the synthetic datasets.


%% file: 06_Qualitative_comparison.tex
\section{Qualitative comparison (RQ1)}
\label{qualitative_comparison_results}

Table~\ref{tab:Comparison_Table} provides an overview of the main characteristics of the benchmarked libraries. We detail these characteristics below. 

\input{Comparison_Table.tex}

\textbf{diffprivlib}. Developed in Python under MIT license, IBM's diffprivlib~\cite{IBM_repo} is a general-purpose library for data scientists.
diffprivlib differentiates itself by including a plethora of DP mechanisms that go beyond the Laplacian, the Gaussian, and the Exponential, by additionally including the Geometric~\cite{geometric_mech_IBM}, the Vector~\cite{vector_mech_IBM}, the Staircase~\cite{staircase_mech}, among others (see Table~\ref{tab:DP_Mechs_Table}). 
Additionally, diffprivlib offers some of these mechanisms in various forms, namely their truncated and the bounded form~\cite{bounded_truncated_IBM}.
However, not having floating-point safety for a differentially private mechanism is a vulnerability.
Altogether, the unique value proposition of diffprivlib is the comprehensive catalog of mechanisms, analytics queries, and DP machine learning algorithms, which run like Sklearn classifiers with the option to track the privacy budget.

\textbf{SmartNoise}. Implemented under MIT license in Rust (runtime) and bound to a Python wrapper, Microsoft in collaboration with the OpenDP initiative led by Harvard IQSS and SEAS 
provide SmartNoise~\cite{SmartNoise_repo} as an open-source library of DP mechanisms for the release of statistics, APIs for defining DP analysis, a validator to perform privacy budget accounting, and metadata concerning the utility of the outputs. 
Furthermore, SmartNoise offers floating-point safety, allows the user to choose from bounded and unbounded DP definitions, and offers the most extensive set of analytics queries. 
However, SmartNoise lacks DP machine learning implementations, and it is not easy to work with SmartNoise when there are multiple datasets involved, as one must follow a specific design pattern instead of a more familiar one using, e.g., Sklearn.
Despite its data science limitations, SmartNoise compensates by offering an architecture designed towards large-scale systems.


\textbf{Google DP}. Google's library~\cite{Google_repo} under Apache-2.0 license offers a suite of DP analytics queries.
The Python wrapper (PyDP) around C\Plus\Plus \space built by OpenMined makes Google DP accessible to data scientists. 
Furthermore, this library provides a layer for non-experts based on Apache Beam, while an expert may use the DP building blocks directly. 
Google DP can automatically approximate bounds for the analytical sensitivity formulas if none were input, enables privacy budget tracking, and checks whether any DP guarantee has been broken in the analysis through a stochastic tester. 
Moreover, developers may deploy operations-ready applications with the underlying C\Plus\Plus \space codebase.
However, Google DP does not contain any machine learning algorithms.


\textbf{diffpriv}. diffpriv~\cite{diffpriv_repo, sensitivity_sampler_R} is a package developed in R, under MIT license, that enables data scientists to execute user-defined functions in a DP manner, such as analytics queries or model fit. 
Furthermore, diffpriv's sampler can empirically calculate the sensitivity of a user-defined function under the bounded definition of DP~\cite{sensitivity_sampler_R, diffpriv_overview}, so that non-experts do not need to calculate sensitivity analytically, which may be arduous for machine learning algorithms.
Rubinstein~et~al.'s sensitivity sampler ensures that DP holds with high probability~\cite{sensitivity_sampler_R}, i.e., one assures \emph{random} DP, but not pure DP. 
We, however, decided to use the sensitivity sampler for the analytics queries because we target non-experts.
Additionally, we also ran another set of the experiments providing the same analytical sensitivity formulas for bounded DP that SmartNoise (Microsoft) has used from Harvard's Privacy Tools Project~\cite{sensitivity_repo} (coincidentally the same as in diffprivlib). 
However, there are also shortcomings:
the empirical calculation of sensitivity is computationally expensive even for simple queries, and diffpriv does not offer floating-point safety or a privacy budget accountant.

\input{Default_mechanisms.tex}

\textbf{Chorus}. Chorus~\cite{Chorus_repo, uber_dp} distributed as a Scala library is a research system developed specifically to explore the use of DP at scale---for example, in production datasets at tech companies, which often contain billions or trillions of records and do not fit in memory. To query data at this scale, organizations often build and deploy extensive infrastructure.
Chorus aims to work in \emph{cooperation} with the existing infrastructure by using an existing SQL database to perform the ``heavy lifting'' of executing queries on large-scale datasets.
Chorus provides three components: a query analysis framework (e.g., determining the sensitivity of an analyst-specified query), a query rewriting framework (e.g., modifying a query to perform clipping before executing it), and a privacy budget accountant. 
The resulting DP mechanisms aim to maintain roughly the same scalability properties as the underlying database infrastructure. 
Note that Chorus is a research framework and does not provide an out-of-the-box system ready for deployment; moreover, it has fewer built-in mechanisms and is less ready for production use. 
However, Chorus is unique in its ability to scale to large datasets by cooperating with an existing high-performance database.


%% file: Comparison_Table.tex
\begin{table*} [t!]

\small
\begin{tabular}{ |M{2.5cm}||M{2.5cm}|M{2.5cm}|M{2.5cm}|M{2.5cm}|M{2cm}| } 

\hline
\textbf{Features}  & \textbf{diffprivlib} & \textbf{SmartNoise} & \textbf{Google-DP \break (PyDP)} & \textbf{diffpriv} & \textbf{Chorus}\\
\hline

Contributor & IBM & Microsoft & Google \break (OpenMined) & B. Rubinstein et al. & J. P. Near et al. \\
\hline
Programming Language & Python & Python wrapper over Rust runtime & Google-DP: C\Plus\Plus, Java, Go \break (PyDP: Python wrapper over C\Plus\Plus) & R & Scala \\
\hline
Primary use & Data science facing operations (notebooks) & Data science facing operations (notebooks), and large-scale systems & Google-DP: Production-ready applications \break (PyDP: \break Data science) & Data science  & Large-scale systems \\\hline
Unique value proposition  & Numerous machine learning algorithms, and DP mechanisms for experimentation & Blend of data science and operations &  Google-DP: Deployment of applications, e.g., in mobile phones \break (PyDP: Data science) & Flexibility for data scientists: User-defined functions and empirical calculation of sensitivity & Scalability via cooperation with existing databases; extensibility \\
\hline
License & MIT & MIT & Apache-2.0 & MIT & MIT \\
\hline
Benchmarked version & 0.4.0 & 0.2.2 & 1.0.1 & 0.4.2 & 0.1.3 \\
\hline \hline

\multicolumn{6}{|l|}{\textit{\textbf{Functional features}}}\\
\cline{1-6}
Mechanisms & Laplace, Gaussian, Exponential, Geometric, Staircase, Binary, Bingham, Vector, and Uniform  & Laplace, Gaussian, Exponential, Geometric, and Snapping & Google-DP: Laplace, Gaussian, Exponential, and Snapping \newline (PyDP: Laplace) & Laplace, Gaussian, and Exponential  & Laplace, Gaussian, Noisy Max, FLEX, SVT, Sample \& Aggregate \\
\hline
Analytics queries & Count, Sum, Mean, Var, Std, and Histogram & Count, Sum, Mean, Var, Covar, Histogram, Quantile, Maximum, Minimum, Median, and Raw Moment & Count, Sum, Mean, Var, Std, Maximum, Minimum, and Median & Any, provided the sensitivity sampler. However, none are built-in &  Count, Sum, Mean, Histogram \\
\hline
DP definition for analytics queries & Bounded & User defined & Unbounded & User defined &  Unbounded (Most mechanisms); bounded (FLEX mechanism) \\
\hline
Privacy budget accounting & Available  & Available & Available & N/A &  Available \\
\hline
Sensitivity calculation (private sensitivity calculation) & Available \break (N/A)  & Available \break (N/A) & Available \break(Available) & Only with the sampler \break (N/A) &  Available \break (N/A) \\
\hline
Floating-point vulnerability  protection  & N/A & Snapping mechanism for the Laplacian distribution & Snapping mechanism for the Laplacian and the Gaussian distributions  &  N/A & N/A \\
\hline
Differentially private machine learning algorithms & K-means, Linear and Logistic Regression, Naive Bayes, and PCA (tools: Standard scaler) & Linear Regression & N/A & Bernstein (built-in). \break Provided the sensitivity sampler, user-defined: SVMs, Bayesian inference, feature selection, among others \cite{sensitivity_sampler_R} &  N/A \\
\hline

\end{tabular}

\caption{ Qualitative overview of the selected five open-source libraries.} 
\label{tab:Comparison_Table}
\vspace{1cm}
\end{table*}

%% file: Default_mechanisms.tex
\begin{table*} [t!]
\centering
\small 
\begin{tabular}{ |M{2.5cm}||M{2.5cm}|M{2.5cm}|M{2.5cm}|M{2.5cm}|M{2cm}| } 
\hline

\textbf{Library}  & \textbf{Count} & \textbf{Sum }& \textbf{Mean} & \textbf{Var} & \textbf{Floating-point safety} \\
\hline \hline

\textbf{diffprivlib \break (IBM)} & Uses the sum query to add the count (\textit{ints)} of non-zero values: Geometric Truncated & Laplace Truncated for \textit{floats} and Geometric Truncated for \textit{ints} & Laplace Truncated &  Laplace Bounded Domain & N/A \\ \cline{2-5} 
&  Unbounded DP & Bounded DP &  Bounded DP  & Bounded DP & \\ 
\hline

\textbf{SmartNoise \break (Microsoft)}  & Pure Geometric & Snapping Laplace for \textit{floats}, and Pure Geometric for \textit{ints} & Uses the sum and count query mechanisms: Snapping Laplace, and Pure Geometric, respectively & Uses the mean query to compute: \break $Var(X) = \mathbb{E}[X^2] - \mathbb{E}[X]^2$. Therefore, in turn, uses Snapping Laplace for the sum and Pure Geometric for the count & Default \\ \cline{2-5} 
&  Unbounded DP & Bounded DP &  Bounded DP  & Bounded DP & \\ 
\hline

\textbf{Google-DP} & Snapping Laplace & Snapping Laplace & Noisy average with normalization \cite{DP_in_practice}: Uses the sum and the count query mechanisms, i.e. Snapping Laplace & Uses the mean query to compute: \break $Var(X) = \mathbb{E}[X^2] - \mathbb{E}[X]^2$. Means computed according to \cite{DP_in_practice}, therefore, in turn, var uses the count and sum queries: Snapping Laplace & Only option \\ \cline{2-5} 
&  Unbounded DP & Unounded DP &  Unbounded DP  & Unbounded DP & \\ 
\hline

\textbf{diffpriv \break (Rubinstein et al.)}  & Pure Laplace & Pure Laplace & Pure Laplace & Pure Laplace &  N/A \\ \cline{2-5} 
&  Unbounded DP & Bounded DP &  Bounded DP  & Bounded DP & \\ 
\hline 

\textbf{Chorus \break (Johnson et al.) }  & Pure Laplace & Pure Laplace & Uses the sum and count query mechanisms, i.e., Pure Laplace &  N/A & N/A \\ \cline{2-5} 
&  Unbounded DP & Unbounded DP  & Unbounded DP & Unbounded DP &  \\ 

\hline 

\end{tabular}

\caption{ Default mechanisms and DP definitions (Bounded or unbounded) of the libraries used for each query in our benchmark. 
For more details, see Appendix Table \ref{tab:DP_Mechs_Table}.}

\label{tab:Default_mechanisms}

\end{table*}

%% file: 07_Experiments.tex
\section{Utility benchmark (RQ2)}
\label{experiment}

\subsection{Setup}
\label{setup}

We conducted the experiments running Ubuntu 20.04.1 LTS~\cite{Intel_processor} on one server (Intel Xeon E5-2650 v2 16-core CPU, 32 GiB of memory). 
We used Python 3.8.5, R 4.0.3, and Scala in version 2.10, to run the latest stable versions of each considered library at the time of the benchmark.
Using the available cores in parallel, we independently conducted each experiment $500$ times per $\varepsilon$ for the utility experiments, and $5$ times per dataset size in the scalability benchmark of Section~\ref{performance_results}.
To improve plot readability, in the body of this paper we depict the plots for $\varepsilon \leq 10$ because utility for $\varepsilon > 10$ was predominantly equal among libraries (see Appendix).
Algorithm~\ref{basic_query_experiment} of the Appendix depicts the workflow of the experiments concerning analytics queries executed on the $27$ synthetic datasets and the two real-world datasets, jointly represented by $\mathcal{D}$.

\subsection{Benchmarked algorithms}
\label{setup}

We benchmarked the default mechanisms because we target practitioners without in-depth knowledge of DP.
All the default mechanisms from Table~\ref{tab:Default_mechanisms} are derived from the Laplace mechanism, including the Snapping mechanism~\cite{mironov_significance_2012} and the Geometric mechanism~\cite{geometric_mech_IBM}.
The Snapping mechanism protects against the floating-point vulnerability by executing a succession of steps such as sampling from a uniform distribution, clamping, and rounding to the closest multiple of a power of $2$~\cite{mironov_significance_2012}.
Furthermore, the Snapping implementations from SmartNoise and Google DP differ, and add less noise than the original work of Mironov~\cite{mironov_significance_2012}. 
On the other hand, the Geometric mechanism is a discrete variant of the Laplace mechanism that satisfies DP with equality and, therefore, produces tighter guarantees for integer-value outputs, resulting in higher accuracy.
Note that the Geometric mechanism is inherently invulnerable to a floating-point attack because its distribution is supported on the integers. 
Aside from these variants, library developers (namely diffprivlib's) have truncated or bounded the domain of the Laplace distribution to preserve output consistency while holding under DP, e.g., preventing counts $< 0$ or vars $\leq 0$.

Aside from the mechanisms, there are different ways to implement a query (see Table~\ref{tab:Default_mechanisms}). 
diffprivlib implements specific DP algorithms for each query except for the count, which reuses the sum query.
Google, Microsoft, and Chorus use the count and the sum queries as building blocks for the mean and var queries (with $Var(X) = \mathbb{E}[X^2] - \mathbb{E}[X]^2$); however, Chorus does not implement the var query.
Lastly, diffpriv enables all queries with the Laplace mechanism.




\subsection{Experiments on synthetic datasets}
\label{synthetic_results}

In this Section, we introduce the experiments' results, namely the behaviour of the dependent variables presented in Section~\ref{methodology}, i.e., the sample mean of the relative error (MRE---accuracy) and the sample std of the absolute scaled error (SASE---precision), w.r.t. the independent variables (dataset size, skewness, and scale), from which we derive the recommendations of Sections~\ref{sec:recommendations_for_practitioners} and~\ref{sec:recommendations_designers}.
\medskip

\begin{figure}
    \centering
\includegraphics[scale=0.27]{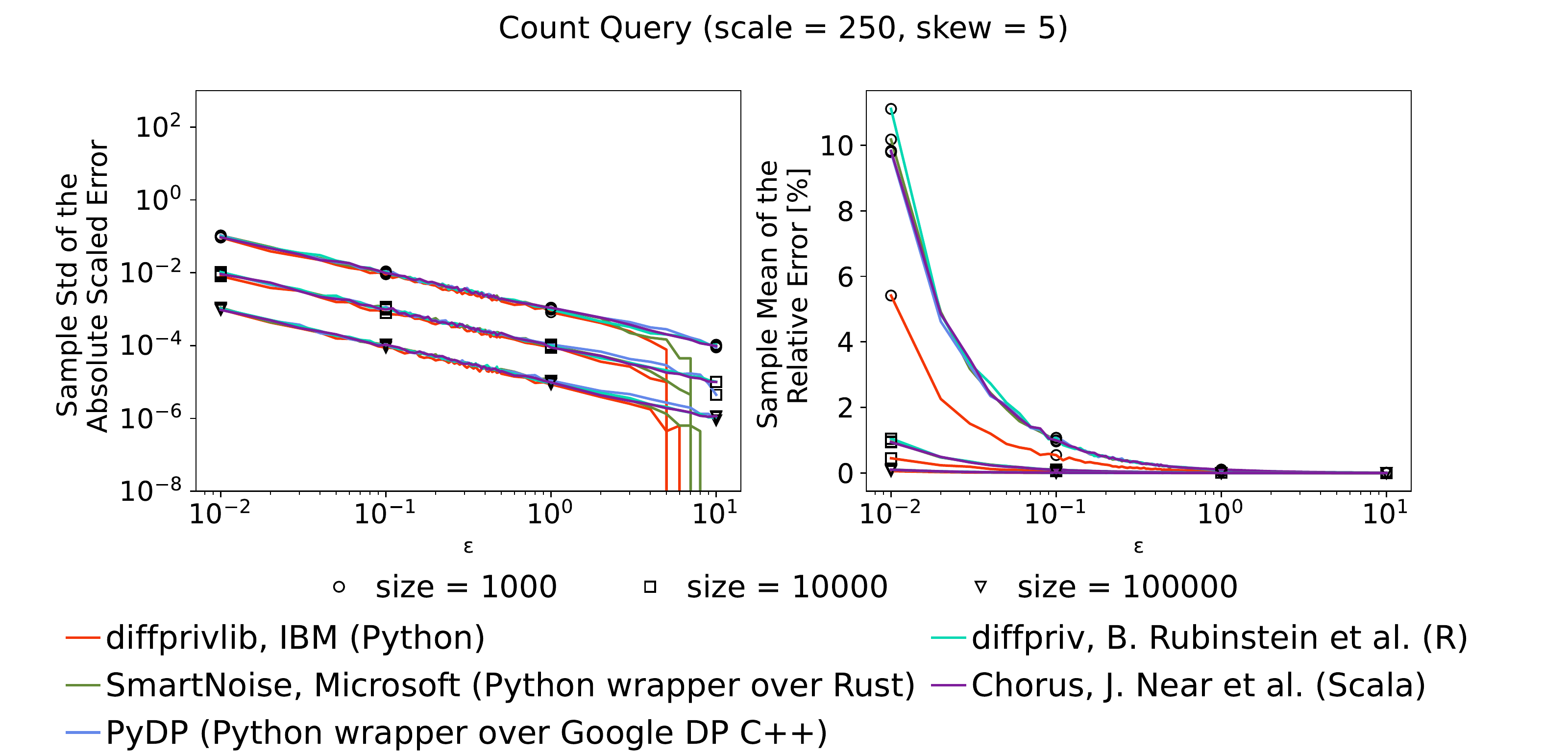} 
\caption{ SASE and MRE of the count query for experiments with  synthetic datasets.}
\label{fig:Synthetic_count}
\end{figure}

\textbf{Count}. The count of records in a dataset is only affected by the dataset size; thus, we dismiss the other independent variables, i.e., skewness and scale.
The most characteristic behavior in the left plot of Fig.~\ref{fig:Synthetic_count}, which measures the SASE, is the drops at around $\varepsilon = 10$ due to rounding.
The noise added at such values of $\varepsilon$ by diffprivlib, SmartNoise, and Google DP is small enough for the output rounding to the nearest integer to produce the same value consistently.
However, diffpriv and Chorus do not round the output, which is why the curves continue in the same manner beyond $\varepsilon =10$ (See Appendix).
Overall, because the global sensitivity of the count query remains constant at the value of $1$ for any dataset size, i.e., the DP noise distribution remains unchanged, the larger the dataset size, the lower the impact of the noise relative to the deterministic result (diminishing MRE), and the lower the impact of the noise spread on the absolute scaled error (diminishing SASE).

\begin{figure} 
    \centering
\includegraphics[scale=0.27]{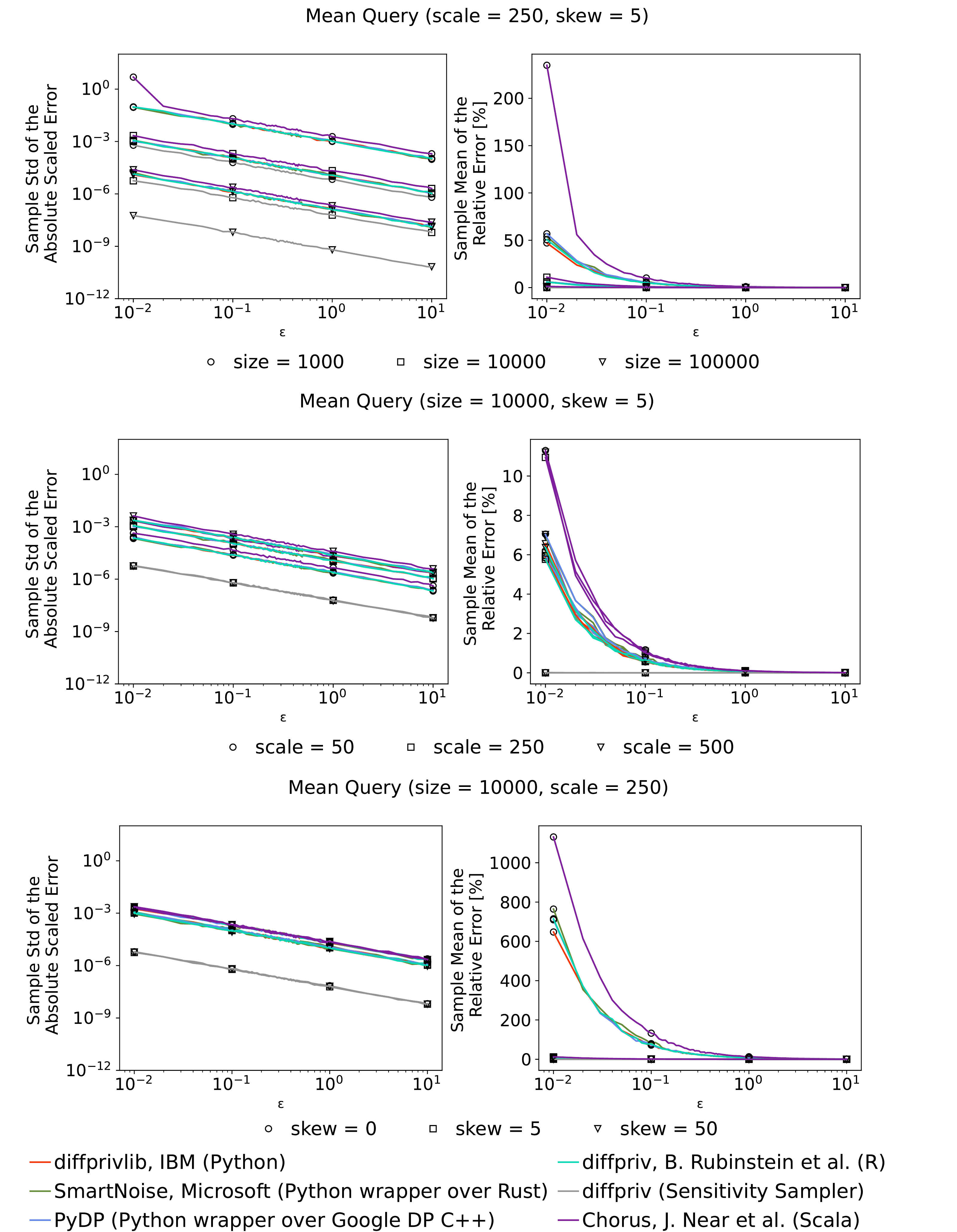}
\caption{ SASE and MRE of the mean query for experiments with synthetic datasets.}
\label{fig:Synthetic_mean}
\end{figure}

\textbf{Mean}. We selected the mean query to show the libraries' behavior across the three independent variables: dataset size, skewness, and scale\footnote{Given the similarity in behaviour across queries w.r.t. the independent variables, to describe the dependencies between the variables in the most concise manner, we consider the sensitivity of the mean and not of the combination of the count and sum.}. 
Fig.~\ref{fig:Synthetic_mean} shows that the independent variables with the highest impact on the SASE are dataset size, followed by the scale, and lastly by the skewness.
On the other hand, one can observe significant MRE values on the outputs for datasets generated with low skewness and small dataset size for $\varepsilon \leq 1$.

(i) Regarding dataset size, see Fig.~\ref{fig:Synthetic_mean} top plot. 
As dataset size decreases, the global sensitivity for the mean increases, which, in turn, increases the spread of the DP noise, resulting in higher SASE and MRE.
(ii) Concerning the datasets' scale, see Fig.~\ref{fig:Synthetic_mean} middle plot. Larger scales lead to higher dataset spreads and global sensitivity, and therefore also higher SASE and MRE.
(iii) Regarding skewness, see Fig.~\ref{fig:Synthetic_mean} bottom plot.
While not as much as the scale parameter, the skewness parameter of the skew-normal also affects the spread of the distribution, from which we sampled the datasets.
Thus, we observe diminishing MRE as skew increases with a fixed scale, and the underlying reason is the decrease of the datasets' spread. 
However, the impact in the SASE is not as notable.

\begin{figure}
    \centering
\includegraphics[scale=0.27]{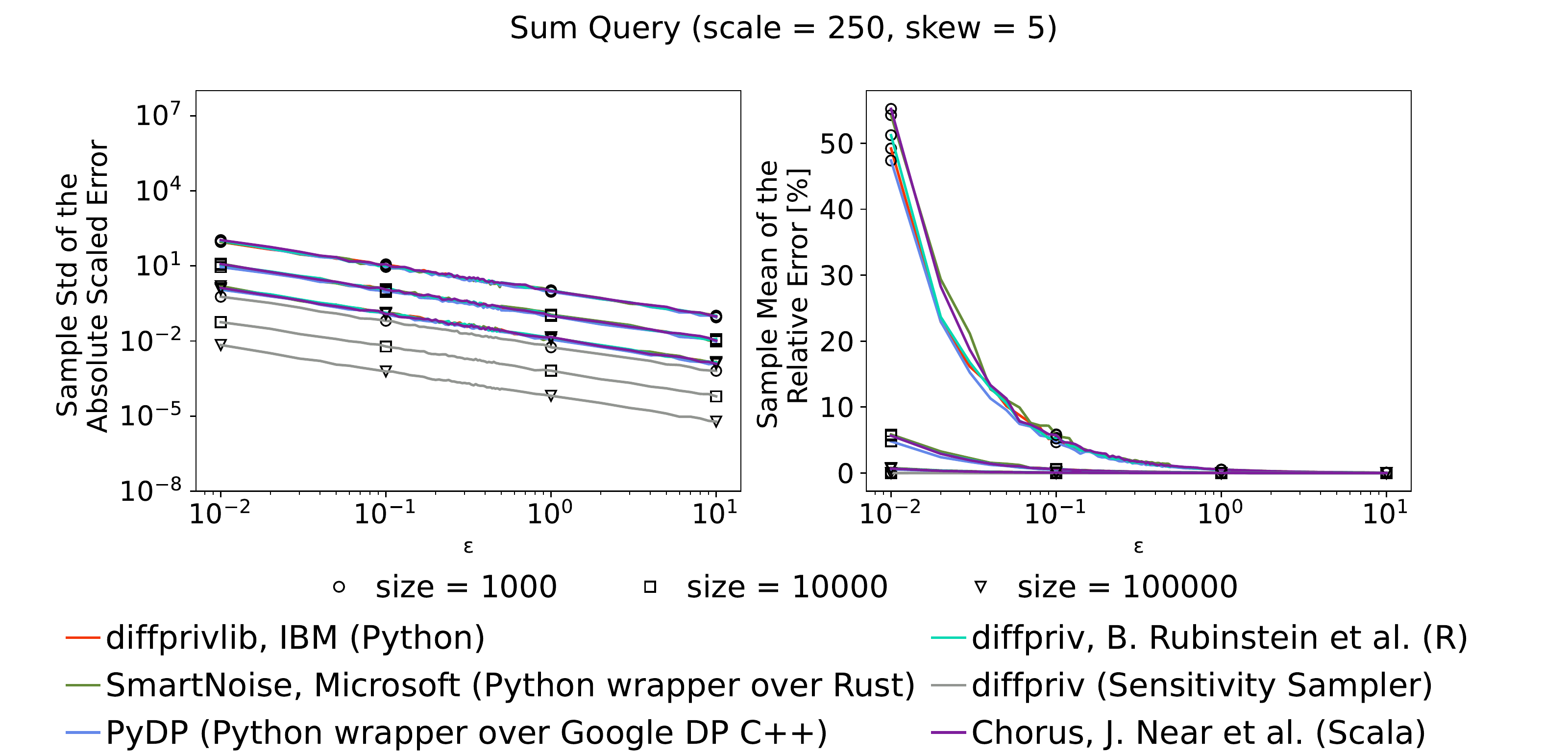}
\caption{SASE and MRE of the sum query for experiments with synthetic datasets.}
\label{fig:Synthetic_sum}
\end{figure}

\textbf{Sum}. Overall, the behavior induced by variations of the three independent variables in the sum query is roughly equivalent to the behavior noted in the mean query (see Fig.~\ref{fig:Synthetic_sum}).
This similarity is the result of some libraries using the sum and count queries as building blocks for the mean query algorithm (Google DP, Smartnoise, and Chorus) or the same underlying mechanism (diffprivlib, diffpriv) (see Table~\ref{tab:Default_mechanisms}).
Aside from the behavior covered discussing the mean, these commonalities make the sum and mean queries display a corner-case behavior when the scale is large while the skewness is low.
In these scenarios, the effect of the dataset's scale on the sensitivity could be significant enough to counter the effect of a larger dataset size, which would lower the relative error.
Consequently, in such a context, the SASE results from a larger dataset are not necessarily better than with a smaller dataset; this is because there is room for more outliers in the larger dataset that increase the difference between the range clipped bounds, which, in turn, increases the global sensitivity (see the top plots in Table~\ref{tab:all_sum_exps} of the Appendix).
In these cases, the tool that provides better utility for the sum query for $\varepsilon$ values less than $1$ is Google DP's Snapping Laplace mechanism; this makes Google DP's sum query relatively better in terms of privacy protection and utility. 
Outside of these conditions, however, the libraries perform similarly.   

\begin{figure}
    \centering
\includegraphics[scale=0.27]{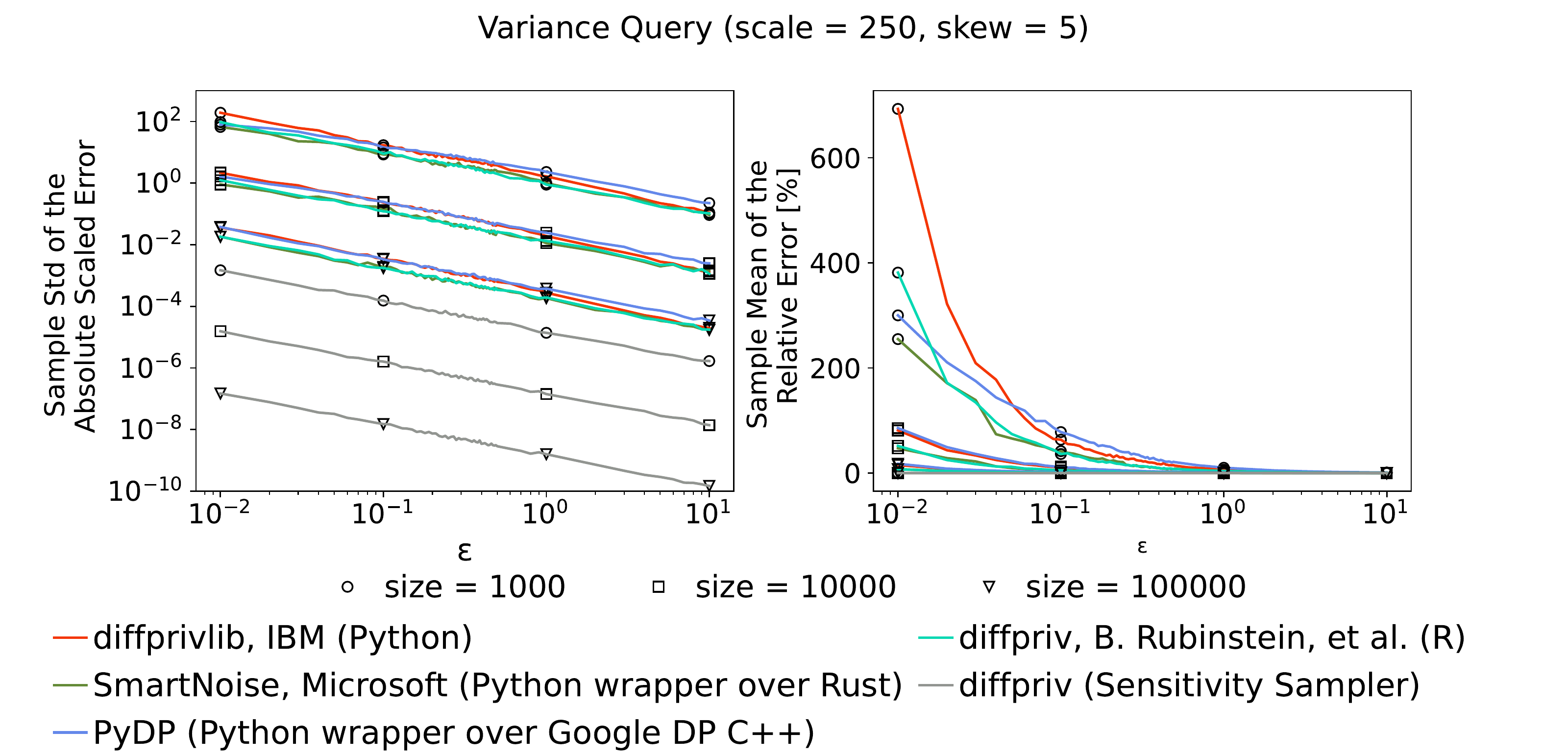}
\caption{ SASE and MRE of the var query for experiments with synthetic datasets.}
\label{fig:Synthetic_var}
\end{figure}

\textbf{Var}.
Equivalently to the mean, most of these libraries utilize the sum and the count as building blocks for the variance algorithm or the same mechanism (see Table~\ref{tab:Default_mechanisms}).
Thus, the overall behavior is also similar to the sum across the three independent variables (see Fig.~\ref{fig:Synthetic_var}), except in the case of diffprivlib, which uses a distinct algorithm for the var.
Furthermore, note that some libraries offer the std query as a built-in function (see Table~\ref{tab:Comparison_Table}); nonetheless, their algorithms calculate the std by the square root of the var query output, leveraging the post-processing properties of DP.
Prominently, the var is the query where diffpriv's sensitivity sampler has significantly outperformed the rest of the libraries.
Lastly, note that Chorus has no var query functionality.

\subsection{Experiments on real-world datasets}
\label{synthetic_results}

\begin{figure} 
    \centering
\includegraphics[ scale=0.27]{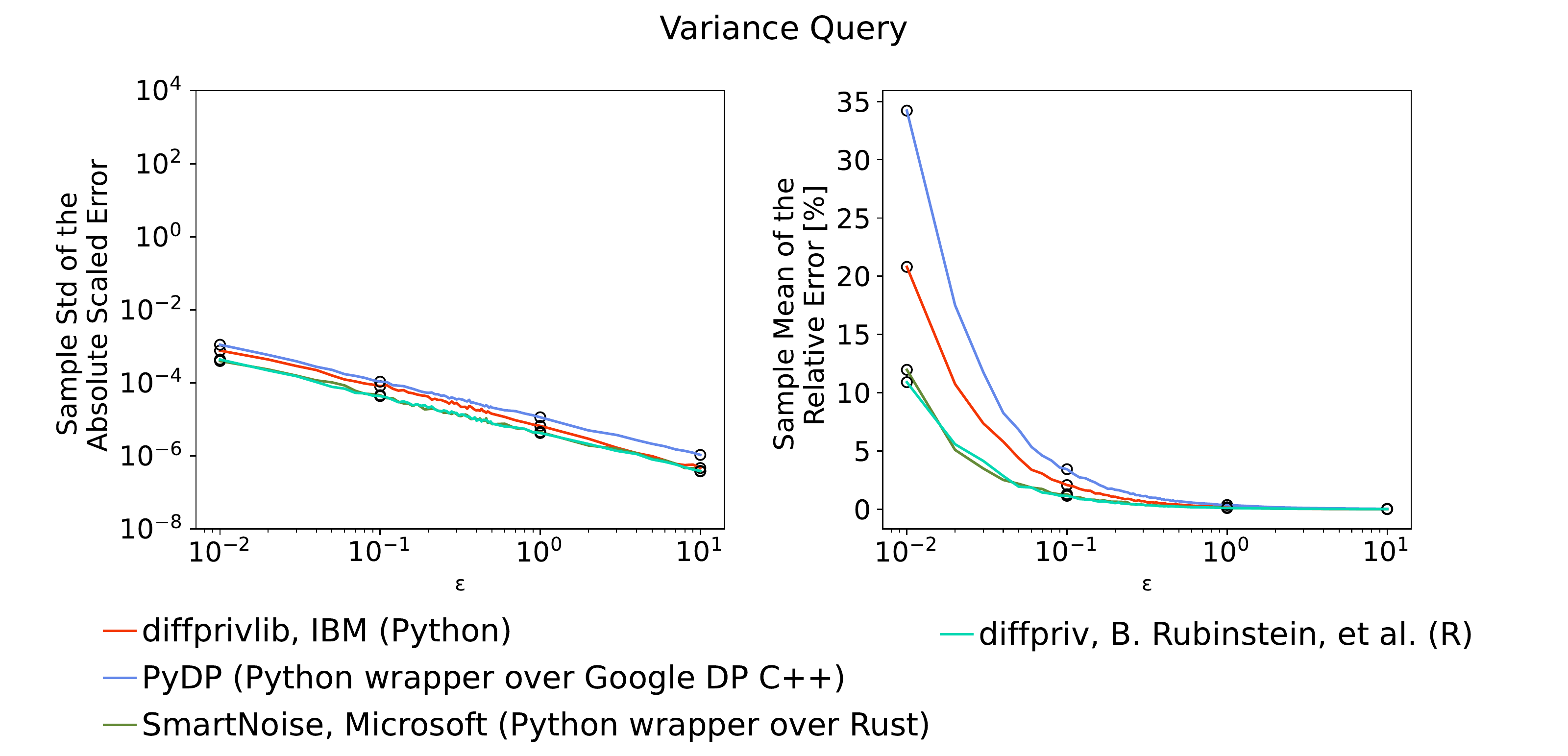}
\caption{ SASE and MRE of the mean of Age in the USA census dataset with 48842 data points.}
\label{fig:real_var}
\end{figure}

The utility results obtained with the real-world datasets show the same patterns across queries, e.g., in the var query of Fig.~\ref{fig:real_var}.
Notably, from the results, we observe that for the USA census dataset with $48842$ data points, the libraries provide similar values of SASE and MRE for the mean and the sum, except for Chorus.
Moreover, the utility results of the Portuguese Education dataset with $649$ data points suffer due to the sparsity over the domain of values in both attributes and the small size of the dataset, especially for the var query (see Tables~\ref{tab:real_absences} and~\ref{tab:real_grades} in the Appendix).
Moreover, Fig.~\ref{fig:real_count} depicts significant MRE values produced by diffprivlib in the count query.

\begin{figure} 
    \centering
\includegraphics[ scale=0.27]{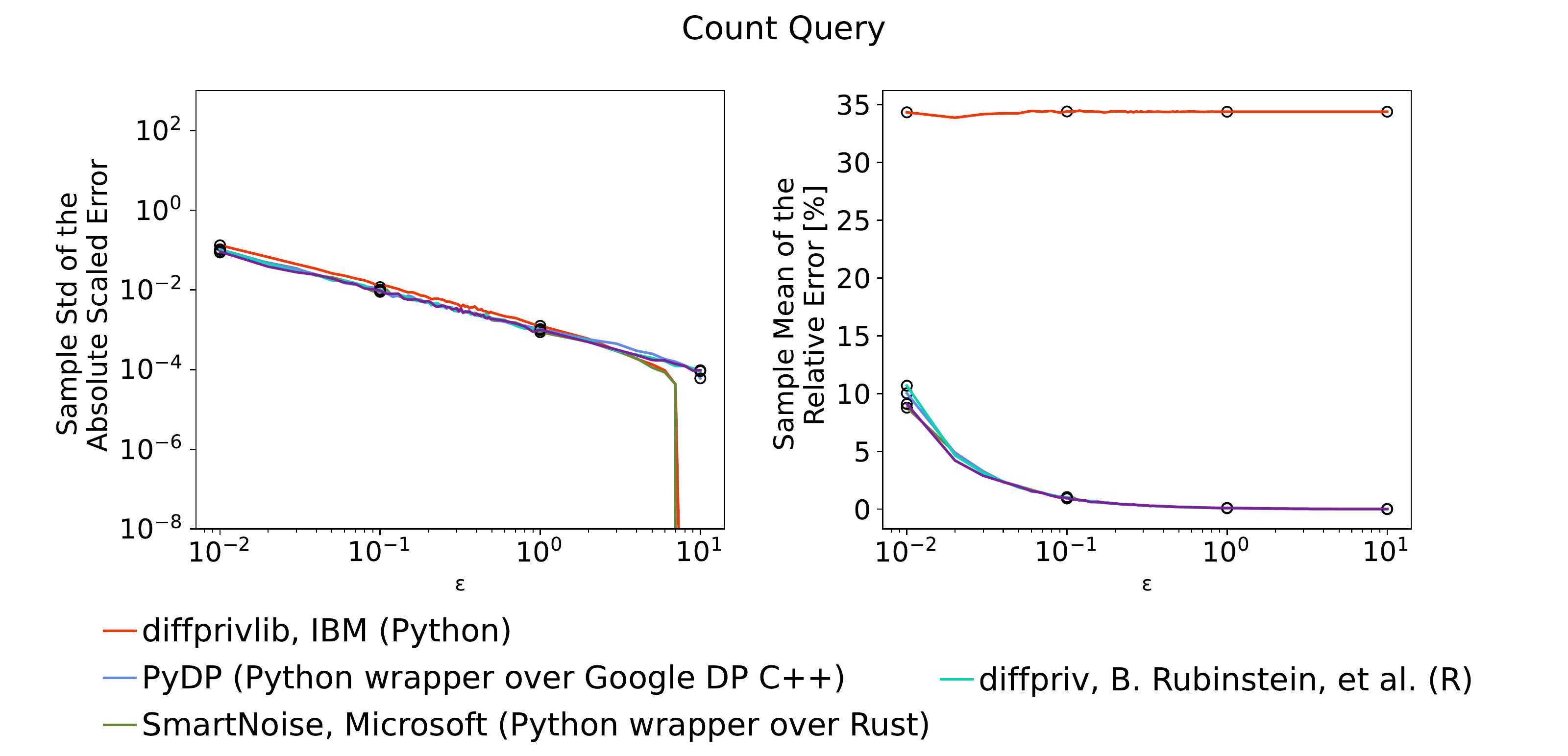}
\caption{ SASE and MRE of the count of Absences in the Portuguese Education Dataset with 649 data points.}
\label{fig:real_count}
\end{figure}

\section{Scalability Benchmark (RQ3)}
\label{performance_results}

We selected the mean query to discuss scalability; however, the behaviour remains uniform across queries (see Tables~\ref{tab:all_execution_time} and~\ref{tab:all_memory_comsumption} of the Appendix).
Note that we used different $\varepsilon$ values as sanity check for the implementation, as the scalability should be independent to the sampled noise.

\begin{figure}
    \centering
\includegraphics[ scale=0.29]{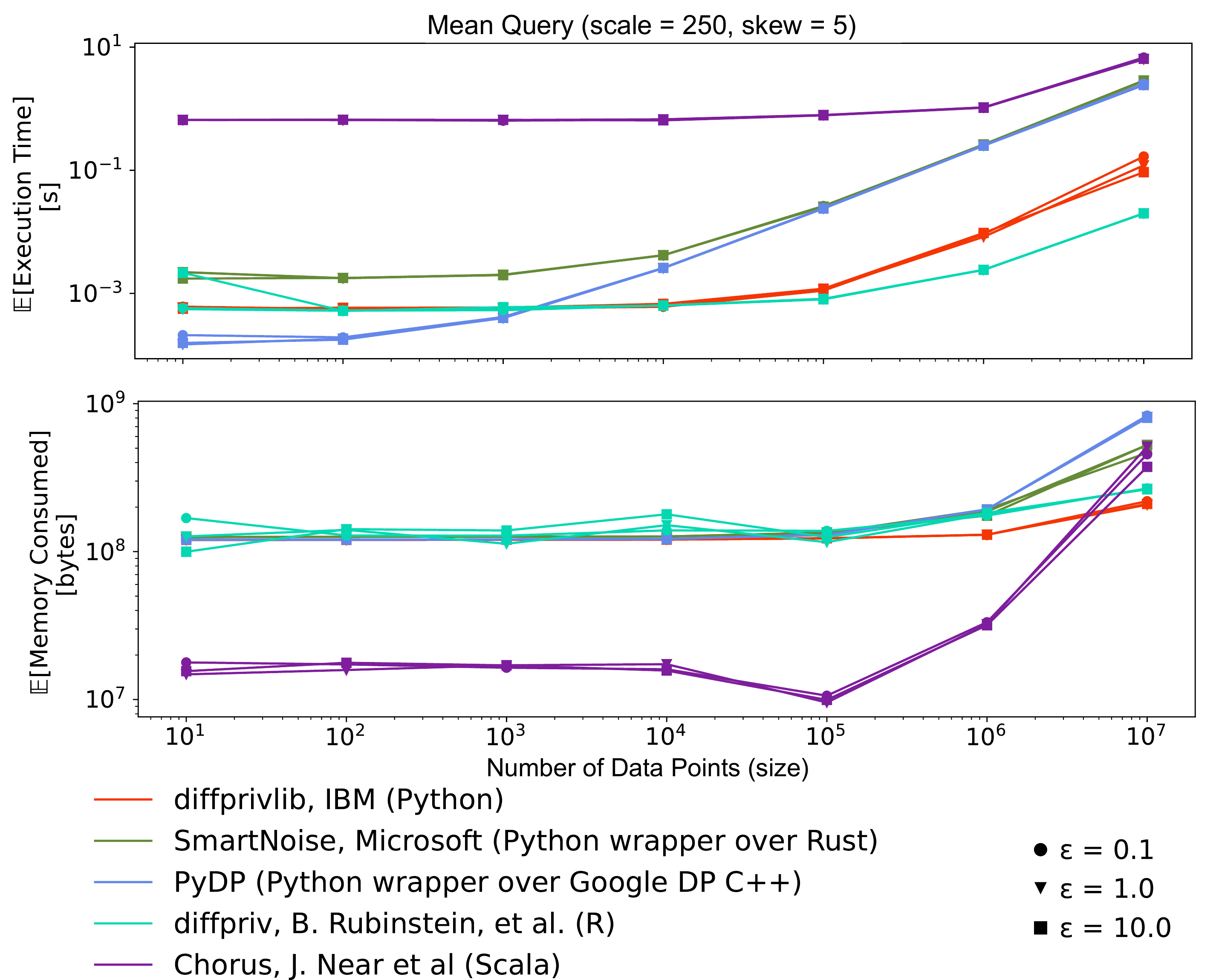}
\caption{ Libraries' execution time and memory consumption for experiments with synthetic datasets of varying dataset size.} 
\label{fig:perfomance_body}
\end{figure}

Based on the execution time analysis in Fig.~\ref{fig:perfomance_body}, we conclude that Chorus scales several orders of magnitude worse than the rest of the libraries across all dataset sizes. 
Furthermore, while the rest of the libraries perform similar up to $10000$ data points, all libraries tend to perform slower on larger datasets, specially the ones with a Python wrapper (Google DP and SmartNoise). 
Lastly, diffpriv scales best at the largest dataset size, followed by diffprivlib. 
Nonetheless, SmartNoise and Google DP might be faster without the wrappers; however, we have not conducted experiments in this direction.
On the other hand, analyzing the memory consumption of Fig.~\ref{fig:perfomance_body}, we conclude that Chorus outperforms the rest of the libraries for datasets of less than $10$ million data points. 
Moreover, despite the low-level implementations in C\Plus\Plus\, of Google DP and Rust of SmartNoise under their Python wrappers, their overall memory consumption is equal to or worse than the rest of libraries.


%% file: 08_Discussion.tex
\section{Guidance for Practitioners}
\label{sec:recommendations_for_practitioners}

\begin{figure*} [!t]
    \centering
\includegraphics[scale=0.15]{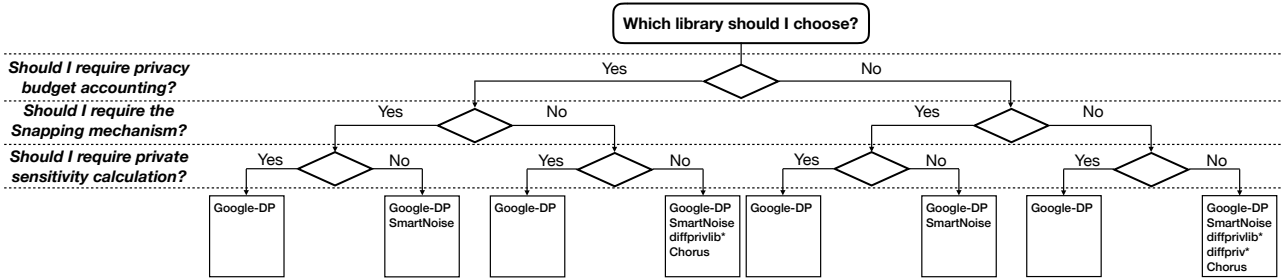}
\caption{Decision tree for choosing DP libraries based on side channels.*Provide DP machine learning functionality}
\label{fig:Decision_tree}
\end{figure*}

\paragraph{Do I get the privacy I need?}
\emph{Yes}---our results suggest that the libraries we studied contain robust implementations of well-studied mechanisms, and they can generally be used to provide differential privacy.
We see different noise levels across different libraries in our experiments, but not to an extent to suggest that some libraries would fail the DP criteria.
Depending on the threat model of a particular deployment, however, it may be vital to consider side channels like floating-point protection (discussed in detail later in this section).

\paragraph{Which library should I choose?}
Based on our results, the libraries share enough similarities for practitioners to feel comfortable choosing any of the libraries.
However, if the practitioner is an analyst or data scientist, we recommend diffprivlib primarily due to its output consistency, ease of use, and a wide variety of DP mechanisms and machine learning models.
On the other hand, if the practitioner is a developer whose applications will expose data to a broader audience, we recommend using Google DP chiefly because it tackles more side-channel attacks than the rest of libraries.
For practitioners concerned with side channels, we have compiled a decision tree presented in Fig.~\ref{fig:Decision_tree}. 
For practitioners primarily concerned with utility and scalability, we recommend referencing Table~\ref{tab:quantitative_comparison}. We detail additional considerations below.

\paragraph{RQ1.} From our qualitative comparison, we extracted the following recommendations:

\paragraph{Integration.}
(i) For endeavors concerning exposing data to the public or in a world-facing application, we recommend the libraries that offer privacy budget tracking and protection against the floating-point vulnerability with the Snapping mechanism (Google DP and SmartNoise).
(ii) Practitioners needing to integrate with existing large-scale query infrastructure can employ Chorus, Google DP, or SmartNoise; however, Chorus needs added code for deployment. 
(iii) On the other hand, in development, research, or data science, we recommend the mechanisms with the best utility for the desired level of privacy while being mindful about side channels that can influence privacy in ways not captured by $\varepsilon$.

\paragraph{Ease of use and available mechanisms.}
Notably, diffprivlib offers easy-to-use syntax, many mechanisms, and some machine learning features. 
Moreover, diffprivlib and diffpriv are designed for data scientists: the features of these libraries are integrated into existing Python and R syntax, e.g., diffprivlib can run DP classifiers with SKlearn, unlike the C\Plus\Plus\, (and its Python wrapper PyDP) or Scala implementations of Google DP and Chorus, respectively. 
On the other hand, Google DP, Chorus, and SmartNoise are more fitting for developing new applications due to their architectural components.

\paragraph{Side channels.}
(i) To address floating-point vulnerabilities, practitioners can employ the Snapping mechanism from SmartNoise or GoogleDP, and the Geometric mechanism from diffprivlib for counts or sums of integers. 
(ii) Additionally, budget tracking is necessary to avoid reverse-engineering the outputs; the only library that does not include budget tracking is diffpriv. 
(iii) Lastly, libraries should include features to help practitioners find the clipping bounds of a dataset's value range without looking at the data for the sensitivity calculation. However, Google DP is the only library that includes private sensitivity calculation for unbounded datasets.

\input{quantitative_overview}

\paragraph{RQ2.} 
Based on the utility benchmark, we provide the library recommendations of Table~\ref{tab:quantitative_comparison} and highlight the following points:
\newline 
\indent \textbf{Count.}
Based on the experiments of fig.~\ref{fig:Synthetic_count}, (i) for small and medium-sized datasets (1000 to 10000 records), diffprivlib performs notably better in terms of accuracy. 
(ii) The Geometric mechanism (SmartNoise) performs similar to the Laplace mechanism in accuracy, while the Geometric Truncated (diffprivlib) brings a substantial improvement.
(iii) However, diffprivlib's implementation of the count is a conditional sum of non-zero values. 
The drawbacks of this implementation is noticeable in the experiments of Fig.~\ref{fig:real_count}, where $156$ records have zero values, which are omitted and make the true count from diffprivlib's perspective equal to $493$ instead of $649$, generating significant MRE values.
(iv) Lastly, diffpriv and Chorus do not round the query’s output.
\newline 
\indent \textbf{Mean.}
Looking at Fig.~\ref{fig:Synthetic_mean} as a whole, (i) Chorus performs slightly worse regarding precision and significantly worse regarding accuracy. 
(ii) There exists similarities among the libraries' mean algorithms, for instance, the Laplace Truncated mechanism used by diffprivlib is similar to the one used by Google DP (Algorithm 2.4 of Li~et~al.~\cite{DP_in_practice}) in that they both clamp-down outlier outputs. 
The similarity is reflected in the uniformity of the mean outputs across libraries and, therefore, shows no indication to choose one library over another (excluding Chorus). 
(iii) The use of diffpriv's sensitivity sampler\footnote{For the sensitivity sampler: $\gamma = 0.1$; as per the example provided by the diffpriv team~\cite{diffpriv_repo}} improves utility substantially.
The sensitivity sampler constitutes part of diffpriv's unique selling value proposition, as it can calculate the local sensitivity of any query, including machine learning algorithms.
However, a query using the sensitivity sampler took roughly $2208$ times more to execute than without the sampler for a dataset sizes of $100000$~\cite{diffpriv_time_waste}. 
\newline 
\indent \textbf{Sum.}
Fig.~\ref{fig:Synthetic_sum} depicts a similar behavior to the mean query. 
Nonetheless, for higher dataset spreads, Google DP outputs higher and Chorus lower accuracy than in the mean query.  
This difference in accuracy among libraries disappears as the datasets' spread become smaller.
\newline 
\indent \textbf{Var.}
According to Fig.~\ref{fig:Synthetic_var}, (i) diffprivlib performs significantly worse in terms of accuracy; however, it is the only library whose outputs are consistent with the var, i.e., its outputs are $> 0$.
diffprivlib achieves output consistency with the Laplace Bounded Domain for the var query at the expense of more noise, i.e., bounding the output domain consumes privacy budget.
One may detect diffprivlib's under-performance in the results of MRE, which is lower than Google DP's and SmartNoise's, even when these libraries double the noise executing the mean query twice to calculate the var ($Var(X) = \mathbb{E}[X^2] - \mathbb{E}[X]^2$). 
(ii) Furthermore, the difference in SASE (and in MRE) between the two seemingly matching algorithms of SmartNoise and Google DP comes from the different low-level implementations of the Snapping Laplace mechanism and the use of the Pure Geometric mechanism by SmartNoise. 
The results show that the combination of their algorithm choice for the var query makes Google DP more accurate for datasets of large spreads and for $\varepsilon \leq 1$, while with lower spreads, SmartNoise provides more utility.
(iii) Additionally, Google DP and SmartNoise at least ensure var outputs not to be lower than $0$; however, their use of clamping algorithms makes $0$ too frequent.
(iv) Moreover, SmartNoise and Google DP perform slightly worse regarding precision than the rest of the libraries. 
\newline 
\indent \textbf{Across all queries.}
The libraries show similar performance in utility. However, while this is mostly consistent regarding precision, there are notable differences in accuracy in scenarios with datasets of $10000$ data points or less and with $\varepsilon \leq 1$ (range recommended by the inventors of DP Dwork~et~al.~\cite{dwork_exposed_2017}). 
Table~\ref{tab:quantitative_comparison} shows our library recommendations per query based on their utility performance and other factors such as output consistency.
Lastly, except for diffprivlib's count query in the presence of zero-value input data, the MRE results (accuracy) do not show the presence of algorithm bias across libraries.
\smallskip 

\noindent \textbf{RQ3.} 
Chorus outperforms the other libraries in memory consumption and diffpriv in execution time, but all libraries are only ready for small-scale deployment.

\section{Guidance for Library Designers}
\label{sec:recommendations_designers}

We extracted the following actionable advice for library designers, who may also use Tables~\ref{tab:Comparison_Table},~\ref{tab:Default_mechanisms},~\ref{tab:quantitative_comparison}, and~\ref{tab:DP_Mechs_Table} for guidance. 

\paragraph{Differences in utility.}
 No single library excels at every task (see Table~\ref{tab:quantitative_comparison}), suggesting that library designers can learn from one another. For settings with datasets of $10000$ data points or less and with $\varepsilon \leq 1$, we recommend: (i) diffprivlib’s Geometric Truncated mechanism for counts (as long as the practitioner does not expect zero values in the dataset). (ii) Google DP’s Snapping Laplace mechanism for sums, as it fares better with datasets of a large dataset spread. (iii) Avoid Chorus for the mean, and (iv) employ diffprivlib for the var as it avoids values $\leq 0$.

\paragraph{Maturity.}
As differential privacy becomes more popular, available tools are beginning to demonstrate increasing maturity. diffprivlib, Google DP, and SmartNoise are more advanced tools than Chorus and diffpriv regarding functionality and onboarding practitioners with a basic DP understanding. Moreover, the teams behind diffprivlib (IBM), SmartNoise (Microsoft), and Google DP are responsive to reported bugs and answer questions swiftly in our experience.

\begin{figure} [htpb]
    \centering
\includegraphics[scale=0.26]{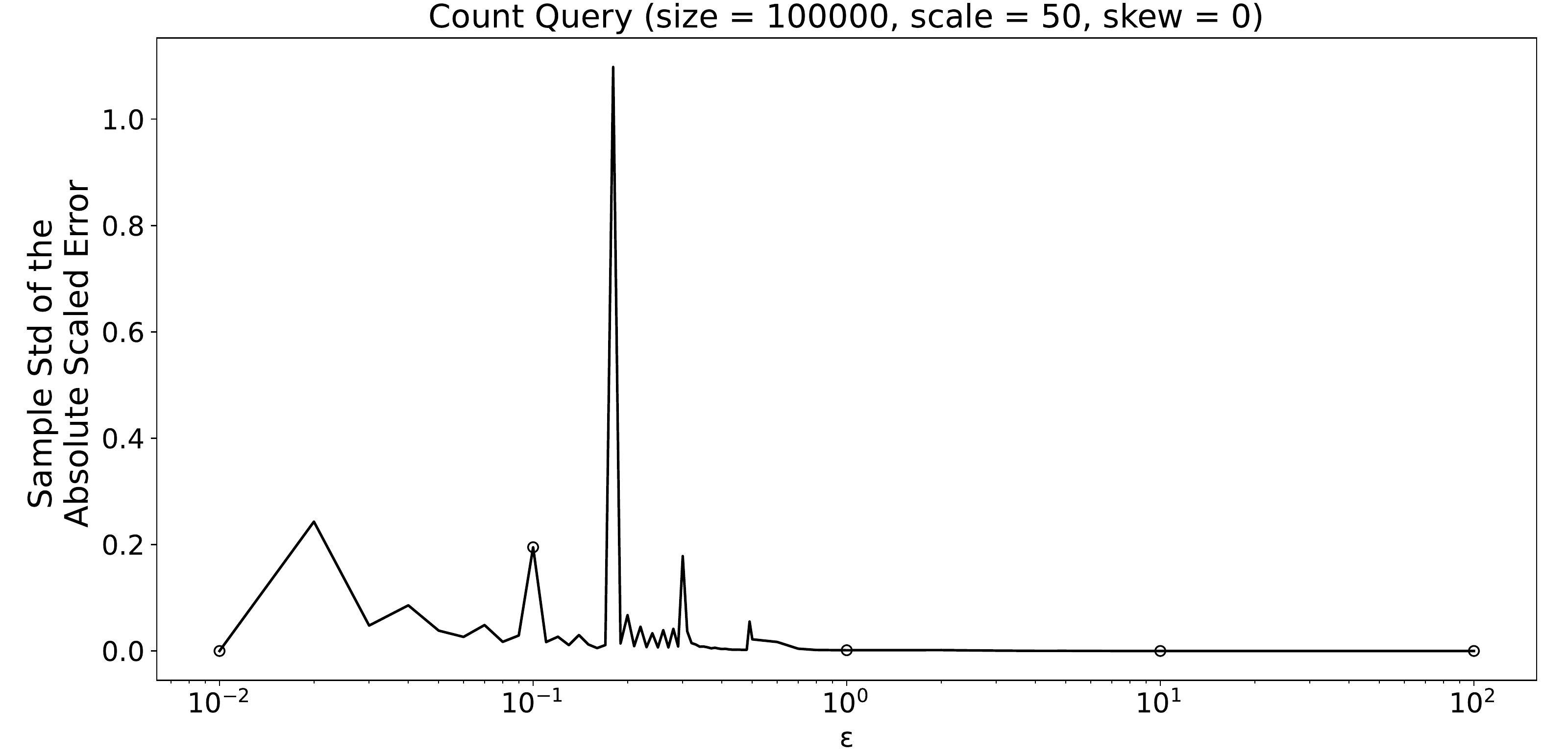}
\caption{Anomaly caused by a bug encountered in WhiteNoise version 0.1.3 in 2020.}
\label{fig:MS_bug}
\end{figure}

\paragraph{Implementation bugs.}
We call for caution in using any implementation of differential privacy, as existing libraries are relatively young tools. There might still be bugs, such as the one we found in SmartNoise when it was still called WhiteNoise (see Fig.~\ref{fig:MS_bug}), and the one we fixed in Chorus~\cite{bug_fix}. Additionally, using diffpriv’s sensitivity sampler on a count query yields an error~\cite{count_error}.

\paragraph{Floating point and side channels.}
diffprivlib, diffpriv, and Chorus and other future library designers should consider implementing the Snapping mechanism. Library designers should also consider addressing other side channels such as privacy accounting and private sensitivity calculation.

\paragraph{Practical value of the Geometric mechanism.}
Geometric mechanisms are not often used in the differential privacy literature; however, for integer-valued queries, they show considerably better utility (when truncated) than the rest of mechanisms (notably in the count), and it does not have a floating-point vulnerability because its domain is the set of integer numbers.

\paragraph{Output consistency.}
Except for output rounding in the count query, diffprivlib is the only library that introduces mechanisms such as Laplace Truncated or Bounded Domain for the sum, mean, and var that prevent output inconsistency (e.g., var $\leq 0$). 
All libraries would benefit from the development of improved mechanisms concerning output consistency similar to diffprivlib's.

\paragraph{Operational performance.}
All libraries offer sufficient performance for small-scale analysis.
On the other hand, none of the available libraries appears to be ready for immediate large-scale deployment.

\section{Guidance for Researchers}
\label{key_findings}

\paragraph{Understanding and communicating theoretical privacy bounds for mechanisms.}
A practitioner might see that most libraries use a variation of the Laplace mechanism and, consequently, may expect the libraries to agree on how much noise to add; however, our experiments conclude otherwise in specific scenarios (see experiments with $\epsilon < 1$).
This is because a privacy guarantee is an \emph{upper} bound on $\epsilon$, not always an exact bound, e.g., a library that adds enough noise for $\varepsilon = 1$ also satisfies DP for $\varepsilon = 2$; practitioners should have in mind that the algorithms' input is an upper bound on $\varepsilon$.
In our benchmark, some implementations have proven better than others at achieving \emph{tighter} upper bounds for $\varepsilon$, resulting in more utility.
Our results suggest small but important differences between libraries in terms of the utility obtained for a given $\varepsilon$.
Furthermore, features like floating-point protection may yield lower utility for the same input $\varepsilon$, leading to additional confusion for practitioners.
Likewise, ensuring var values $> 0$ may generate more noise than others for the same input $\varepsilon$, yet the added utility is not tangible in the error curves.
While these mechanism variations are detailed in their respective literature, the effects on utility may be surprising for a practitioner who is only familiar with the basics of DP.
We suggest that additional research is needed on the \emph{concrete} application of DP mechanisms in practice, to understand the gap between theoretical bounds and actual utility and develop ways to communicate that information to practitioners.

\paragraph{Understanding the utility gain and implications of post-processing.}
The mechanisms benchmarked here are simple, but deployments of differential privacy increasingly make use of more complex mechanisms~\cite{range_queries_DP}.
These mechanisms increase utility, but this gain is sometimes hard to bound analytically.
Moreover, these mechanisms may introduce bias or other artifacts in unpredictable ways, as demonstrated by the results in the 2020 US Census~\cite{DP_census_problems}.
As more advanced mechanisms make their way to deployments, it is important to understand both the utility gain they provide \emph{and} their other implications on analysis results.


\paragraph{Understanding how much utility to expect.}
Based on our benchmark, depending on the dataset and the query, the utility may vary, sometimes significantly.
More specifically, this dependency is observed for datasets containing at least $10000$ data points, and for values of $\varepsilon$ lower than $1$; Figures ~\ref{fig:Synthetic_count} and ~\ref{fig:real_var} are examples of these conditions, among others in the Appendix.
Clearly, utility depends not just on $\varepsilon$, but also on the \emph{scale of the data}.
This situation stands in contrast to the way we usually communicate about differential privacy---we typically focus heavily on the setting of $\varepsilon$.
This focus is appropriate when communicating with data subjects, whose privacy is of primary concern.
However, for analysts attempting to learn from the data, additional guidance is needed about \emph{how to obtain better utility}.
In many cases, the answer may be simple---collect more data---but other strategies (like considering alternative mechanisms) may be helpful too.
We suggest that additional research is needed to understand how to maximize utility in practice, and to develop effective communication strategies to inform practitioners.


%% file: quantitative_overview.tex
\begin{table}
\caption{Library recommendations based on the quantitative benchmark of the selected five open-source libraries.}
\label{tab:quantitative_comparison}

    \centering
    \begin{tabular}{ll}
    
        \toprule
        \textbf{Utility benchmark} &  \\ 
        \toprule 
        
        Count & diffprivlib \\ \hline
        Sum & Google-DP \\ \hline
        Mean & Not Chorus  \\ \hline
        Var & diffprivlib  \\ 

        \toprule 
        \textbf{Scalability benchmark} &  \\ 
        \toprule 
        
        Execution time & diffpriv \\ \hline
        Memory consumption & Chorus \\
        \bottomrule
        
    \end{tabular}
\end{table}

%% file: 09_Conclusion.tex
\section{Conclusion}
\label{conclusion}

To help guide practitioners, library designers, and researchers in navigating the adoption of tools for DP, we performed a qualitative comparison and a benchmark to validate the utility and the scalability of the five libraries towards which we believe a large body of practitioners will gravitate.
Based on our results, we recommend practitioners facing a choice among libraries to prioritize protection against side-channels and analyst support; however, there exist enough similarities to feel comfortable choosing any.
No single library excels in all aspects, indicating that library designers can learn from one another. 
In particular, we suggest increasing efforts in output consistency and towards large-scale deployment.
We conclude that, as long as practitioners account for potential side-channels, these libraries provide the privacy \emph{they need}.

\paragraph{Limitations.} Our benchmark is a snapshot, and our conclusions may become out of date in the future---in fact, we hope they soon will be, due to continued rapid development of the tools! 
We hope that our findings help library designers (with whom we were in contact throughout our study) improve their products and drive tool support for differential privacy.
We also hope that this work highlights potential pitfalls for future library designers.
Finally, by releasing the code for our benchmark, we hope this work will form the basis for future evaluations.

\paragraph{Future work.}
Researchers should consider studying the gap between theoretical bounds and actual utility, researching the practical implications of increasing utility with more complex mechanisms, and discovering other ways to attain more utility.
Regarding our benchmark, in particular, researchers may compare SmartNoise and Google DP in their native libraries (Rust and C\Plus\Plus, respectively) with our studies' datasets.
Moreover, practitioners may benchmark the DP machine learning algorithms proposed by diffprivlib and diffpriv, and others offered by PyTorch and TensorFlow.
Furthermore, benchmarking other queries like quantile and mechanisms outside the default set, like the Gaussian and the exponential mechanism, may reveal other obscure differences among libraries.
Additionally, other libraries and platforms exist, which, while they did not comply with our inclusion criteria or were recently released, propose DP functionalities that are worth clustering and benchmarking.
We have found the following tools: Google's Rappor~\cite{rappor_google}, DJoin~\cite{narayan_djoin_nodate}, ARX~\cite{arx_dp}, PSI \cite{kacsmar_differentially_2020}, Arivat~\cite{roy_airavat_nodate}, a DP violation detector~\cite{ding_detecting_nodate}, and Google's ZetaSQL~\cite{ZetaSQL}.
We do not recommend benchmarking PINQ~\cite{mcsherry_privacy_nodate} or GUPT~\cite{gupt} as they are deprecated.
Finally, we encourage researchers to select one of the five benchmarked libraries (or a combination) based on this publication and test them in real-world use cases.

%% file: DP_Mechs_Table.tex
\begin{table*}
\centering
\small
\begin{tabular}{ |M{1.5cm}|M{2cm}|M{1.5cm}|M{11cm}| } 

\hline
\textbf{Mechanism}  & \textbf{Implementation} & \textbf{Libraries} & \textbf{Description} \\
\hline\hline

Laplace& Pure & All & Noise is sampled from a Laplace distribution of domain $(-\inf, \inf)$. This noise is added to the true value without post-processing. \cite{dwork_algorithmic_2013} \\ \cline{2-4}
    &Truncated  & diffprivlib & If after adding noise to the true value this output falls outside a pre-defined range, then the output is mapped to the closest bound of the output range, e.g. a count output of less than zero could be mapped to the lower bound 0. If the domain bounds coincide with e.g. changes in behavior because the probability of returning values at the domain bounds is non-zero, it is recommended to use the Bounded Domain mechanism instead \cite{bounded_truncated_IBM}.\\ \cline{2-4}
    & Bounded Domain & diffprivlib & Samples outputs until one falls within the pre-defined range. While this mechanism is suitable to prevent not desired values, e.g. a negative var value, or implemented in classifiers, it requires more tailoring so that it satisfies DP \cite{bounded_truncated_IBM}. \\ \cline{2-4}
    &Bounded Noise & diffprivlib & Samples from a truncated domain of a Laplacian distribution \cite{bounded_noise_IBM}. \\ \cline{2-4}
    &Folded & diffprivlib & Similar to Truncated, but instead of outputting the closest bound, the output outside a pre-defined range is folded around this range until the output falls within, e.g. with a pre-described range of $[l, u]$, a count output of $c < l$ could be folded by recursively computing $n*l - (c)$ until the count falls within $[l, u]$, where $n \in \mathbb{N}$. \\ \cline{2-4}
\hline
Gaussian & Pure & All including Google-DP, but except PyDP &  Noise is sampled from a Gaussian distribution of domain $(-\inf, \inf)$. This noise is added to the true value without post-processing. \cite{dwork_algorithmic_2013}\\ \cline{2-4}
 & Analytic Gaussian & diffprivlib & It removes at least a third of the variance of the Pure Gaussian mechanism by a novel noise calibration strategy and a post-processing technique \cite{analytical_Gaussian}. \\  \cline{2-4}
  & Discrete Gaussian & diffprivlib & Modifies the Gaussian mechanism so that discrete noise may be sampled without losing privacy or accuracy guarantees \cite{discrete_Gaussian}. \\
\hline 

Exponential & Pure & All including Google-DP, but except PyDP & Achieves differential privacy for categorical query outputs by randomly choosing a category proportionally to its utility value \cite{mcsherry_mechanism_nodate}, i.e. a category with a higher utility score than another is as much more likely to be picked. the utility values decay exponentially, making the true results more likely to be picked while maintaining DP. \\ \cline{2-4}
 & Permute and Flip & diffprivlib & It randomly selects a value, then, the algorithm flips a biased coin based on the utility of the selected value \cite{permute_and_flip}, and releases the output if the results it heads. \\ \cline{2-4}
 & Hierarchical & diffprivlib & Adapts the Pure exponential mechanism to hierarchical data, so that the utility calculation is less complex due to the intrinsic hierarchy of the data. \\ \cline{2-4}
  & With Base-$2$ DP & SmartNoise \break (In progress) & By switching the DP definition from base $e$ to base $2$, one is able to perform precise base $2$ arithmetic, and, thus, avoid floating-point vulnerability. \cite{floating_point_exponential}. \\ 
\hline

Geometric& Pure & diffprivlib, SmartNoise & Employs a discrete variant of the Laplace mechanism by satisfying DP with equality, and thus, producing tighter guarantees for integer-value outputs, and, in turn, higher accuracy \cite{geometric_mech_IBM}. \\ \cline{2-4}
& Truncated & diffprivlib &   Uses the same technique as Laplace Truncated but the underlying mechanism is the Geometric. 
\\ \cline{2-4} 
 & Folded & diffprivlib & Uses the same technique as Laplace Folded but the underlying mechanism is the Geometric. \\ 
\hline
Staircase & Pure & diffprivlib & Optimizes the Laplace Mechanism, obtaining more accuracy for moderate-low $\varepsilon$ values. 
The shape of the noise distribution is a \textit{staircase}, technically considered as a \textit{geometric mixture of uniform probability distributions.} \cite{staircase_mech}\\
\hline
Binary & Pure & diffprivlib & Specifically designed for binary inputs, this mechanism adapts randomized response to ($\varepsilon$, $\delta$)-differential privacy. In effect, the logic flips a biased coin to output the true input or its complementary \cite{binary_mech_IBM}. \\
\hline
Bingham & Pure & diffprivlib & A differentially-private mechanism built upon the Bingham distribution, exclusively used for estimating the first eigenvector of a covariance matrix.  \cite{Bingham_mech_IBM} \\
\hline
Vector & Pure & diffprivlib & Employed for perturbing convex objective functions of a machine learning classifier before optimization \cite{vector_mech_IBM}. \\
\hline
Uniform & Pure & diffprivlib & Derived from the edge case where the Laplace Bounded Noise has an $\varepsilon$ = 0 \cite{bounded_noise_IBM}. \\
\hline
Snapping & Laplace & SmartNoise, Google-DP & A modification of the Laplace mechanism to protect against the floating-point vulnerability of the theoretical Laplace continuous distribution.
Among other steps, the key of the mechanism is rounding to the closest multiple of a power of $2$. 
Furthermore, note that the Snapping mechanism implementation from SmartNoise and Google-DP differ, and in turn, both differ from the original work from \cite{mironov_significance_2012}, as they achieve floating-point safety while adding less noise.\\ \cline{2-4}
 & Gaussian & Google-DP & Similar to the Snapping mechanism for the Laplace distribution, this variation modifies the Gaussian mechanism to protect against the floating-point vulnerability of the theoretical Gaussian continuous distribution.\cite{mironov_significance_2012}\\
\hline

\end{tabular}

\caption{ Overview of all the mechanisms implemented by the benchmarked libraries, among others.} 
\label{tab:DP_Mechs_Table}

\end{table*}

%% file: plots_table.tex








\begin{table*}
    \centering
\begin{tabular}{m{0.5\linewidth}m{0.5\linewidth}}

\includegraphics[ width=\linewidth, height=\linewidth, keepaspectratio]{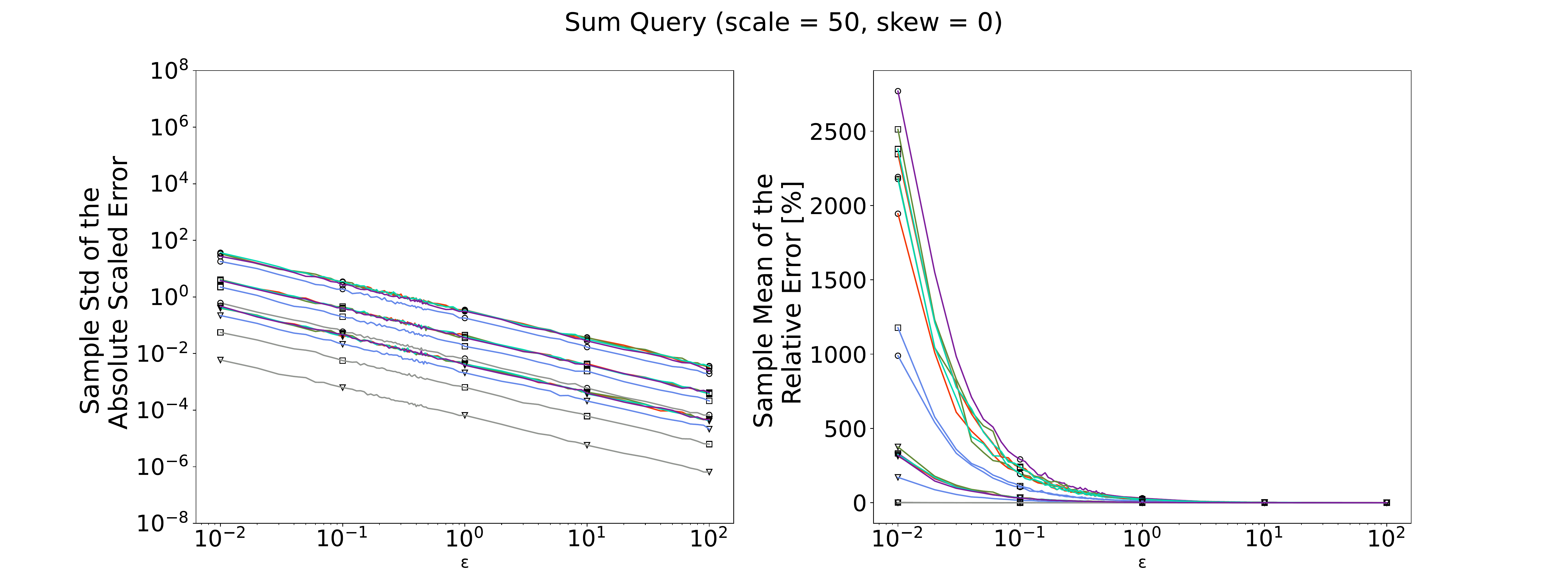} 
&
\includegraphics[ width=\linewidth, height=\linewidth, keepaspectratio]{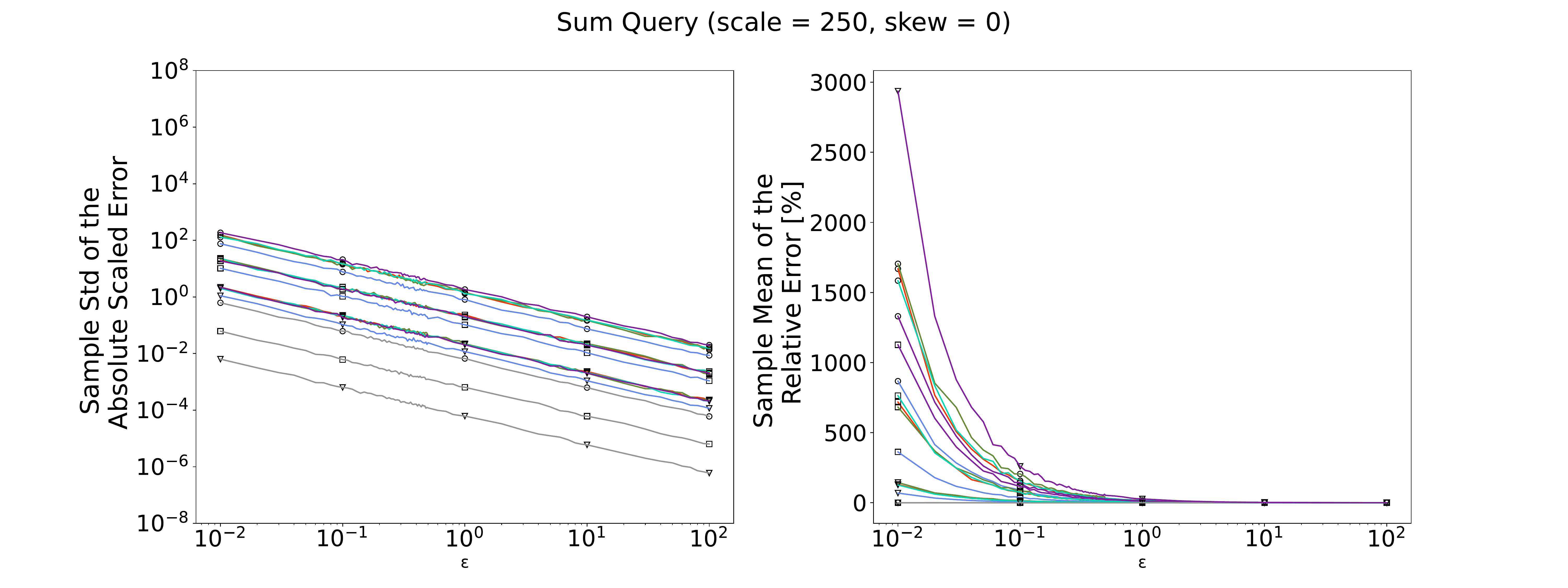} \\

\includegraphics[ width=\linewidth, height=\linewidth, keepaspectratio]{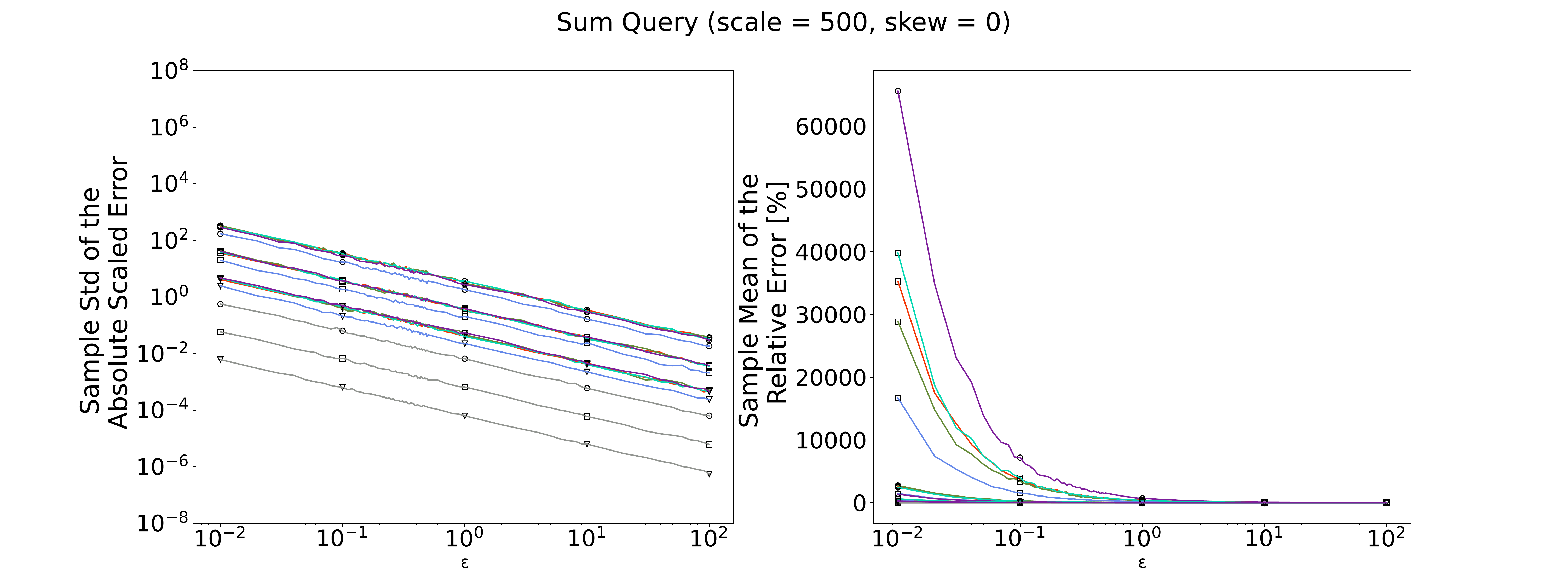} 
&
\includegraphics[ width=\linewidth, height=\linewidth, keepaspectratio]{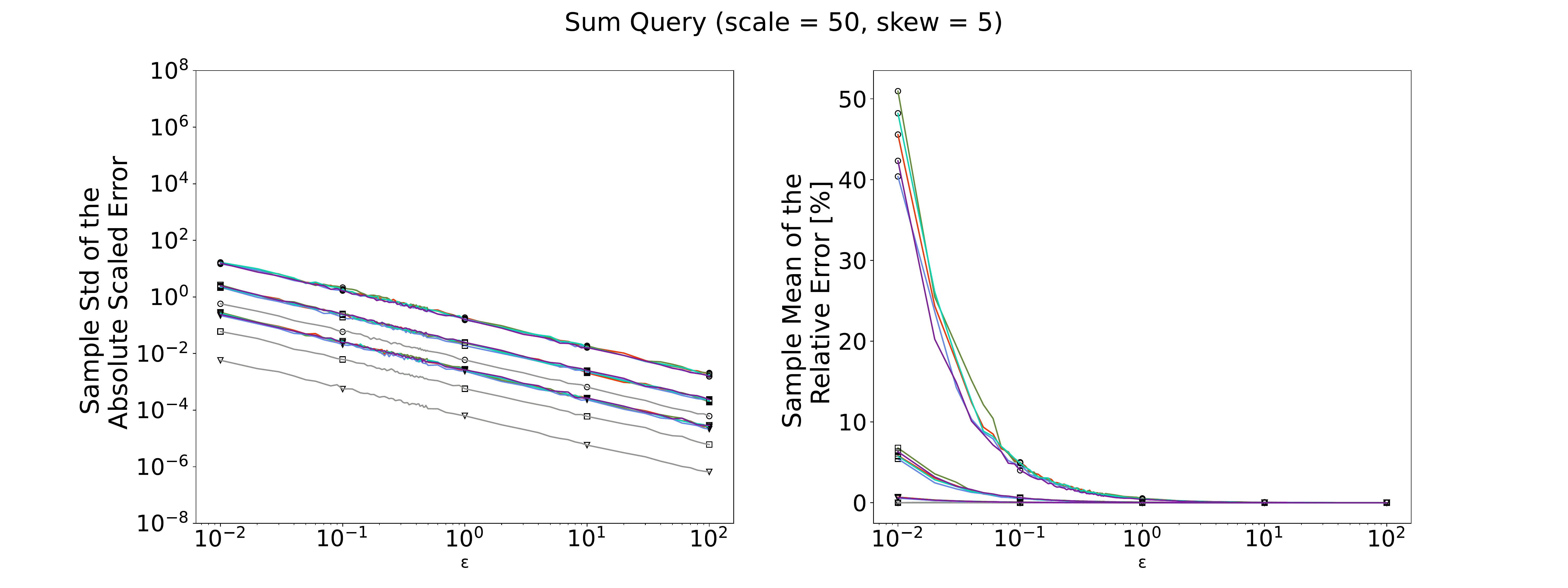} \\

\includegraphics[ width=\linewidth, height=\linewidth, keepaspectratio]{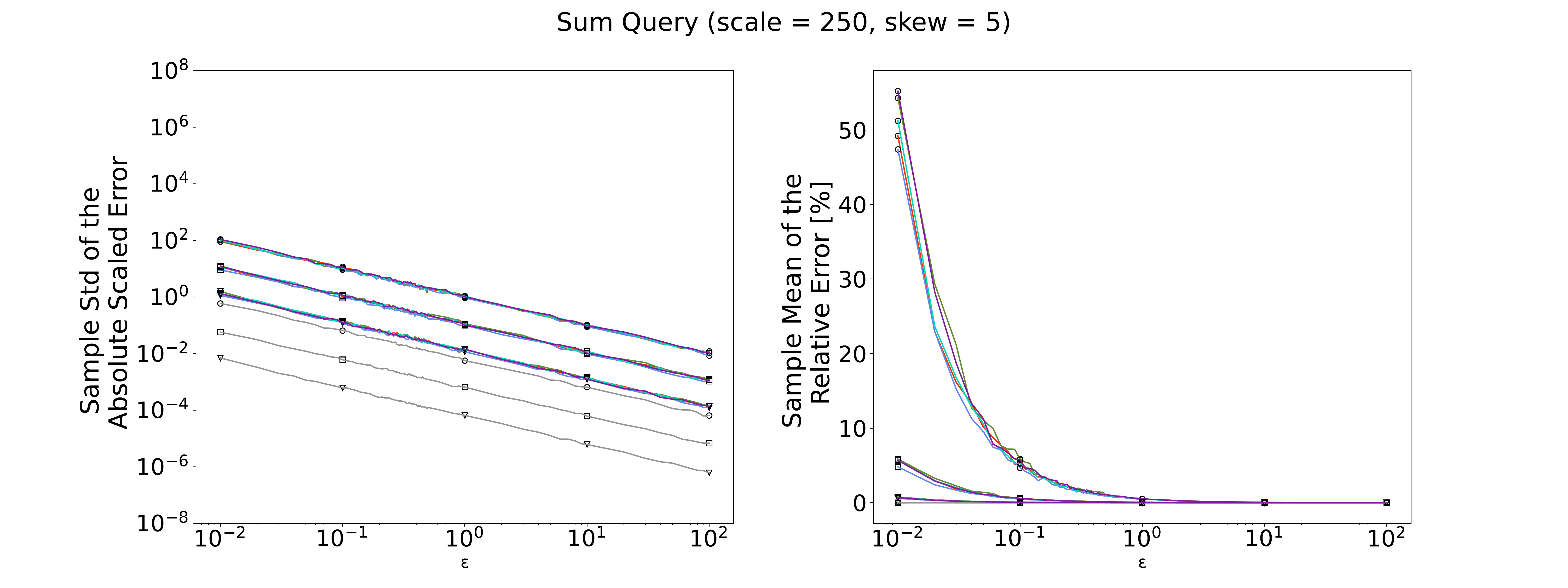} 
&
\includegraphics[ width=\linewidth, height=\linewidth, keepaspectratio]{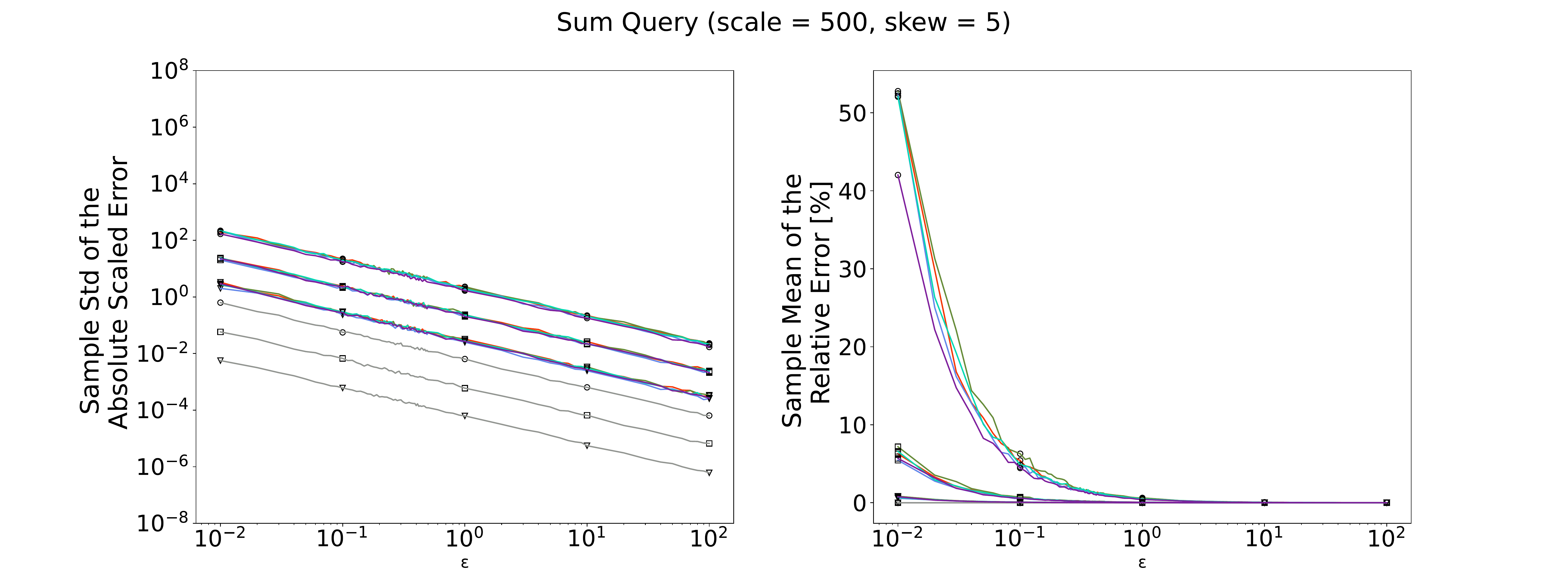} \\

\includegraphics[ width=\linewidth, height=\linewidth, keepaspectratio]{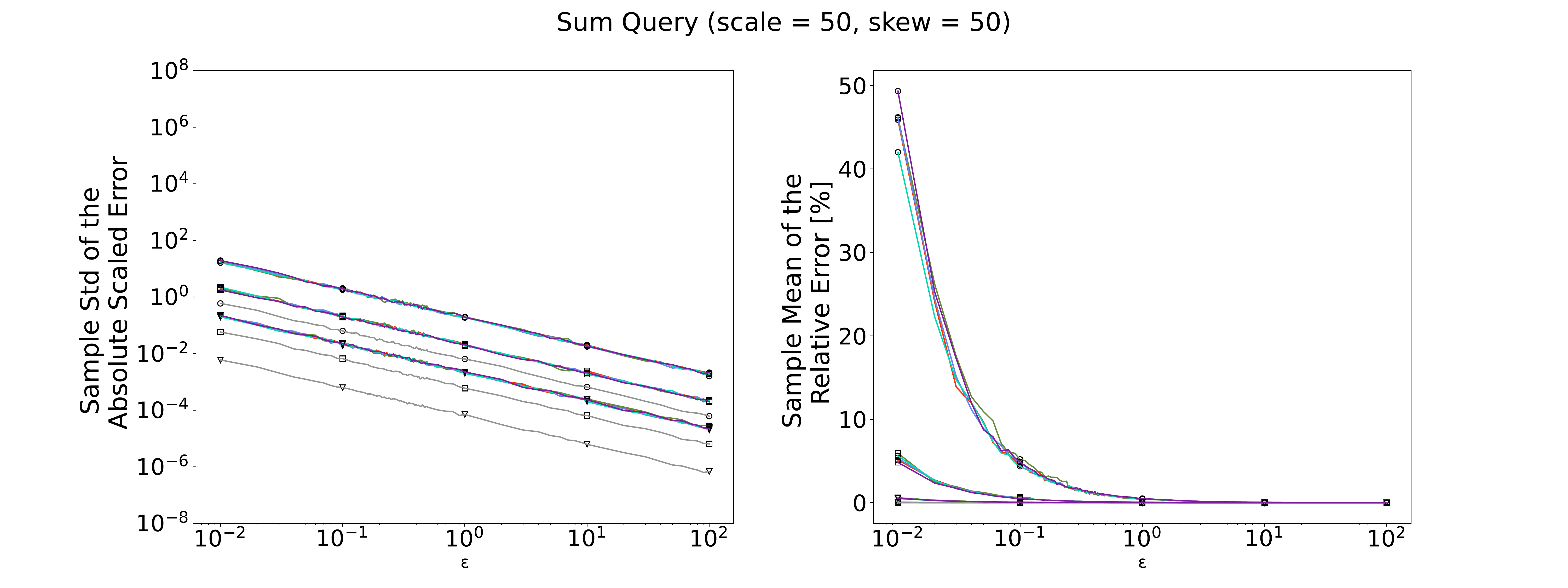} 
&
\includegraphics[ width=\linewidth, height=\linewidth, keepaspectratio]{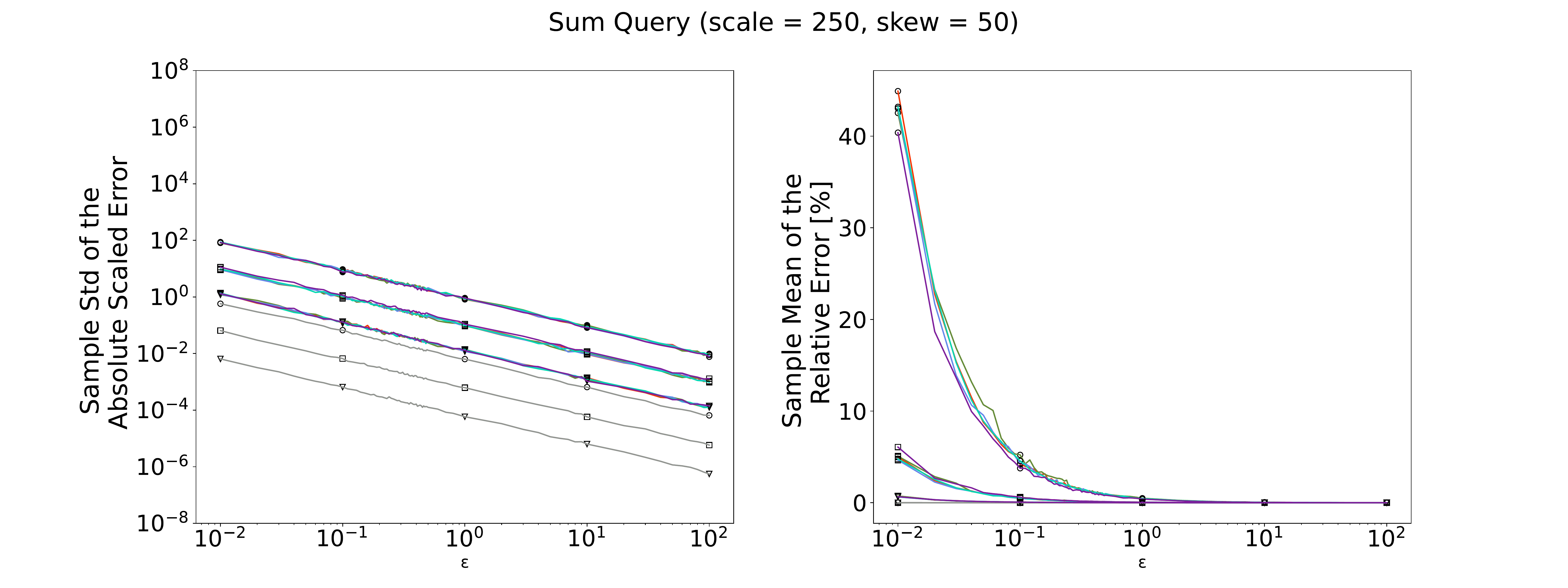} \\

\includegraphics[ width=\linewidth, height=\linewidth, keepaspectratio]{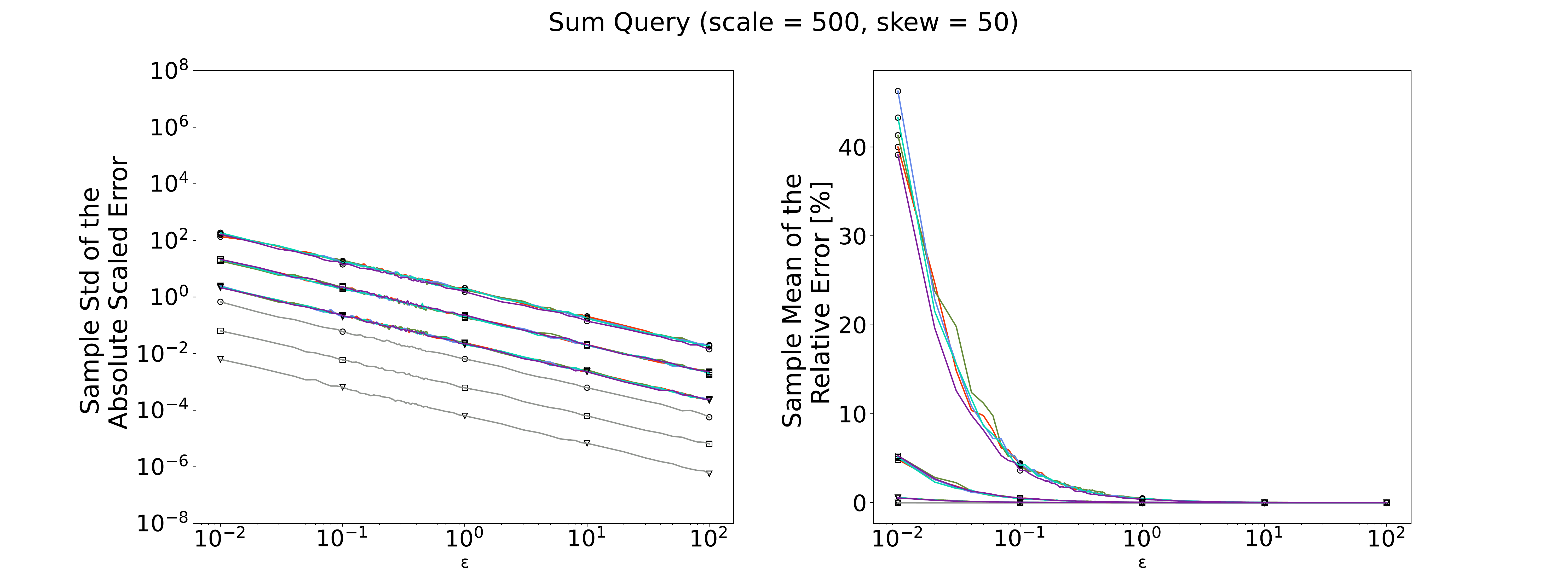} 
&
\includegraphics[width=\linewidth, height=\linewidth, keepaspectratio, scale=0.28]{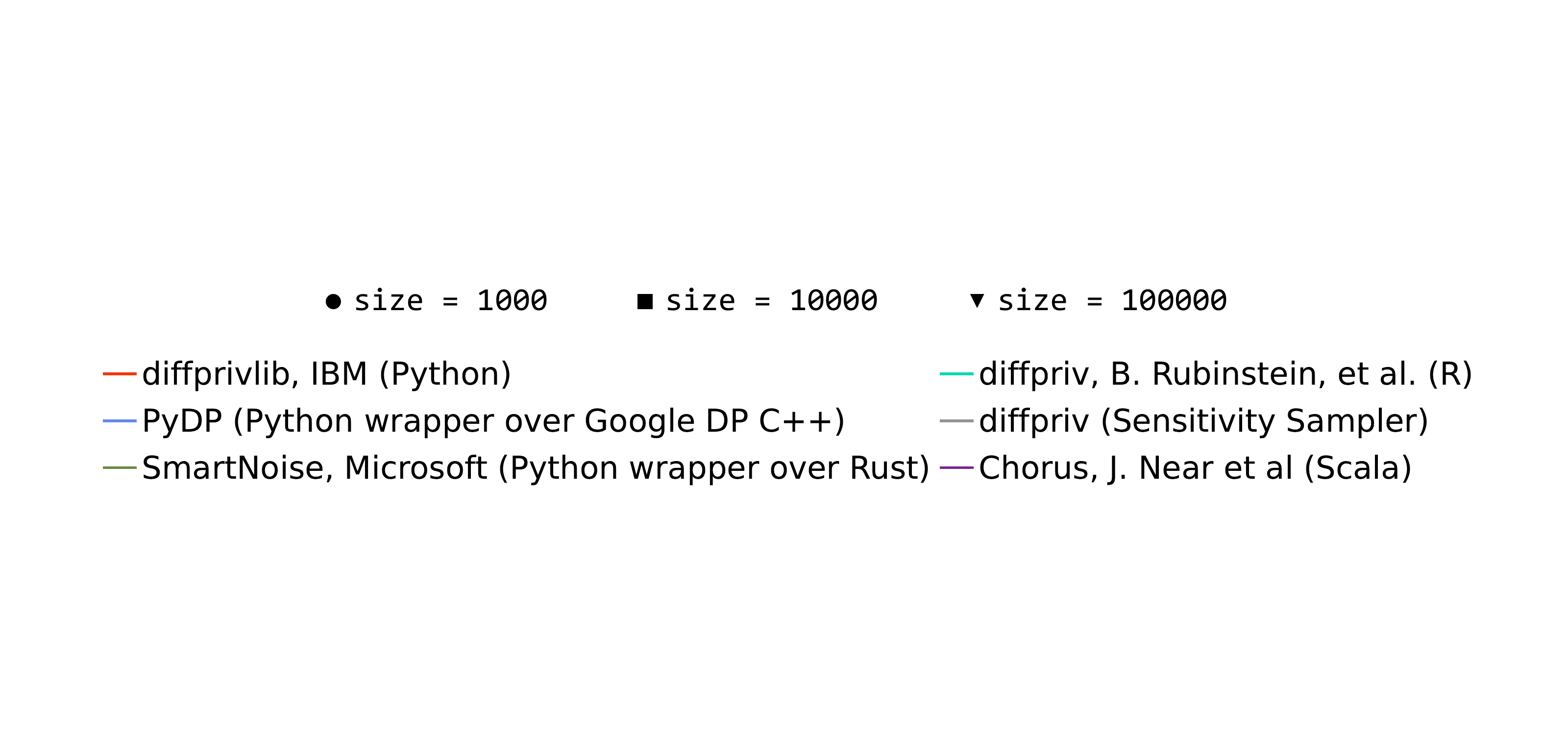}
\\

\end{tabular}
    \caption{ Set of detailed experiments' outputs for all three independent variables (dataset size, scale, and skewness) of the sum query (500 experiments per \textbf{$\varepsilon$}).}
    \label{tab:all_sum_exps}
\end{table*}


\begin{table*}
    \centering
\begin{tabular}{m{0.5\linewidth}m{0.5\linewidth}}

\includegraphics[ width=\linewidth, height=\linewidth, keepaspectratio]{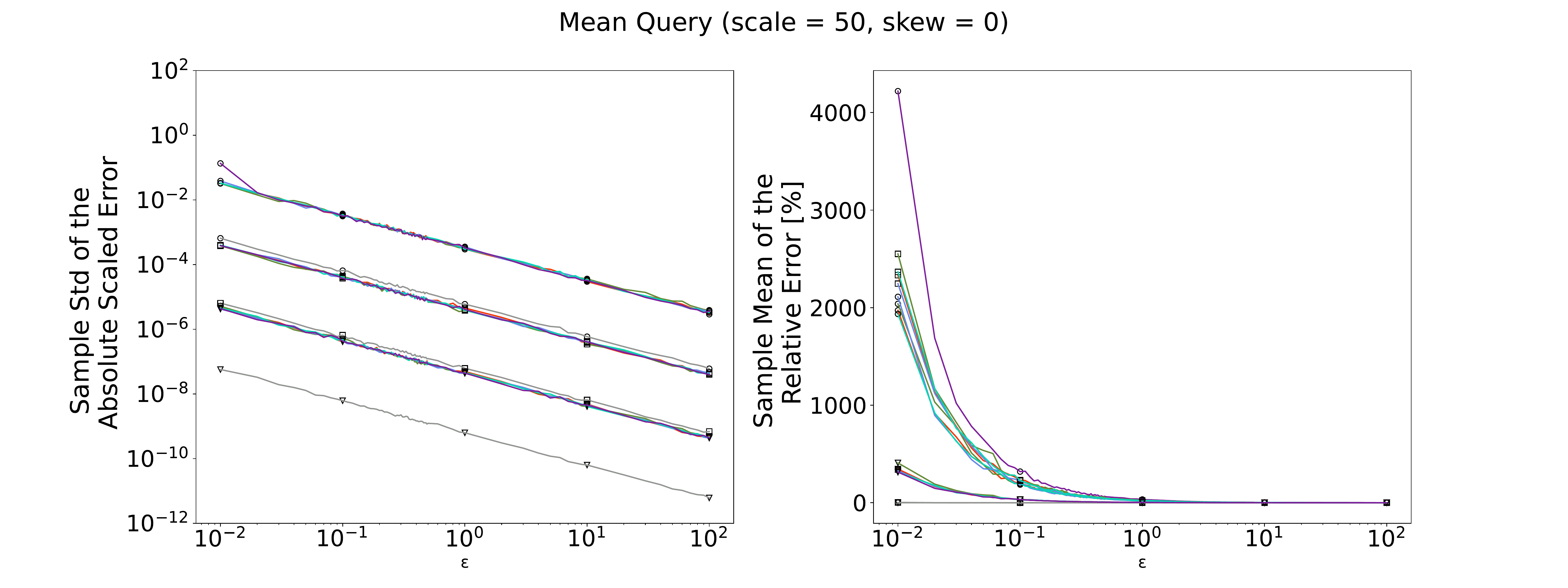} 
&
\includegraphics[ width=\linewidth, height=\linewidth, keepaspectratio]{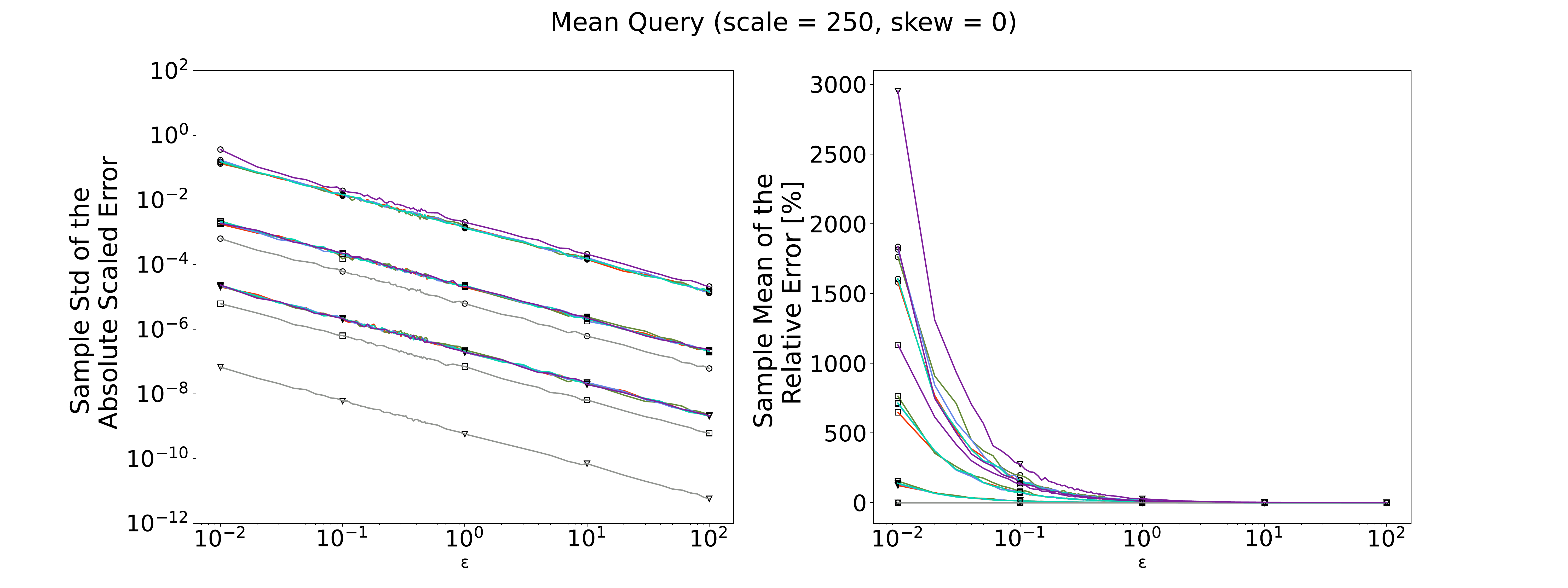} \\

\includegraphics[ width=\linewidth, height=\linewidth, keepaspectratio]{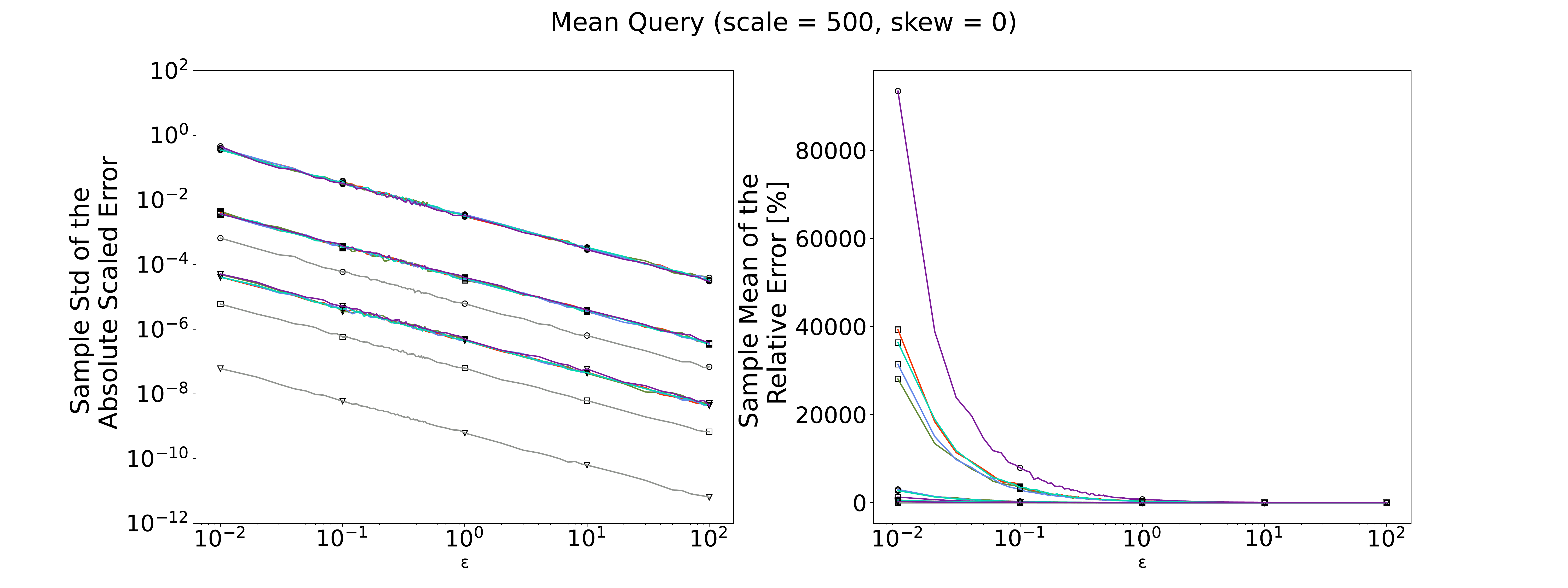} 
&
\includegraphics[ width=\linewidth, height=\linewidth, keepaspectratio]{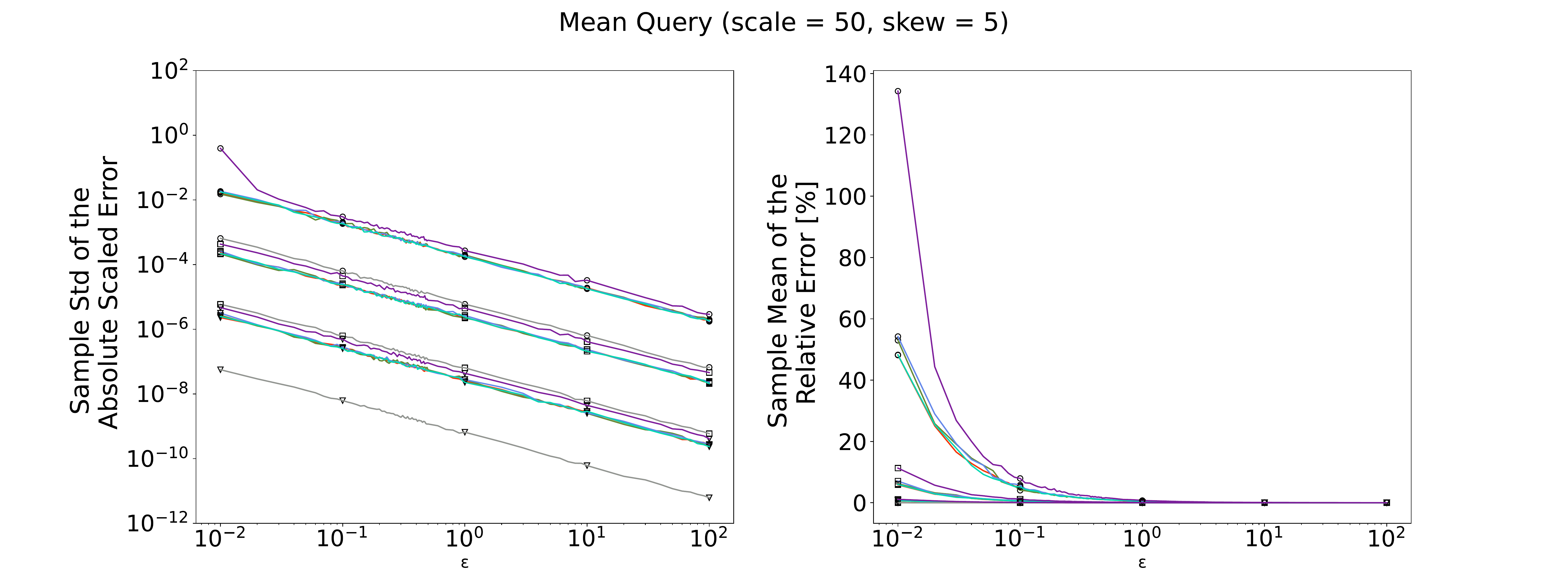} \\

\includegraphics[ width=\linewidth, height=\linewidth, keepaspectratio]{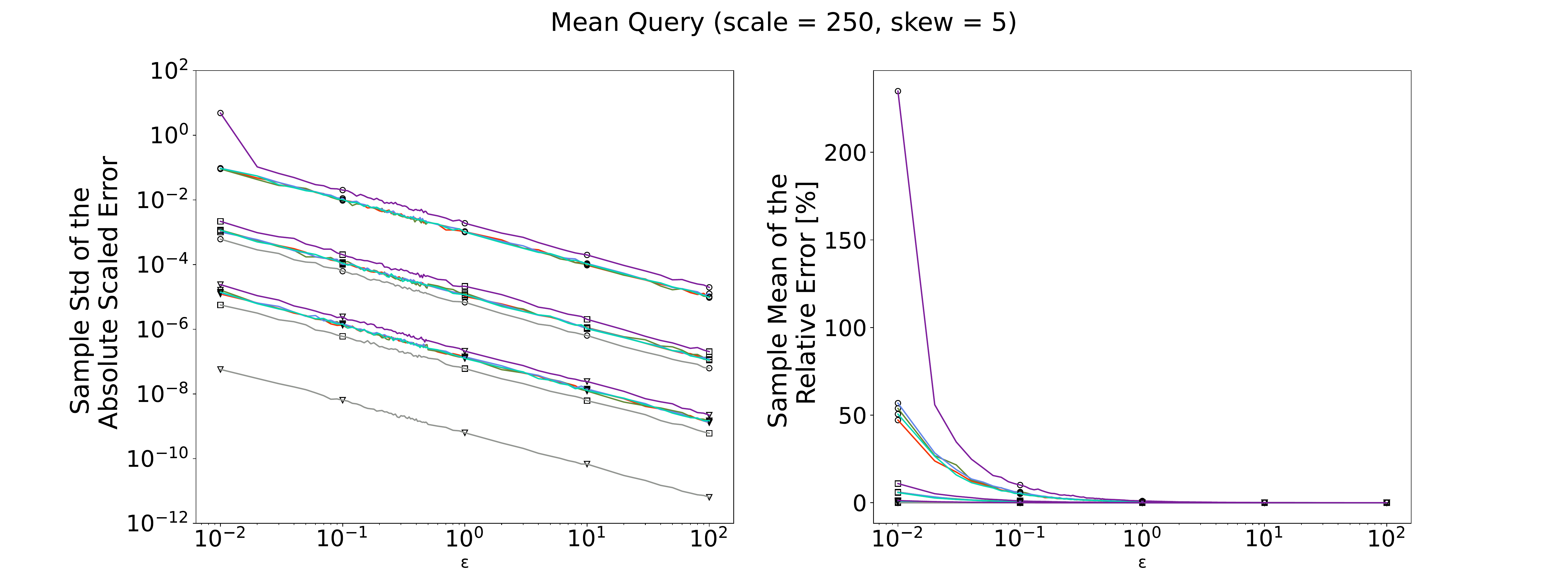} 
&
\includegraphics[ width=\linewidth, height=\linewidth, keepaspectratio]{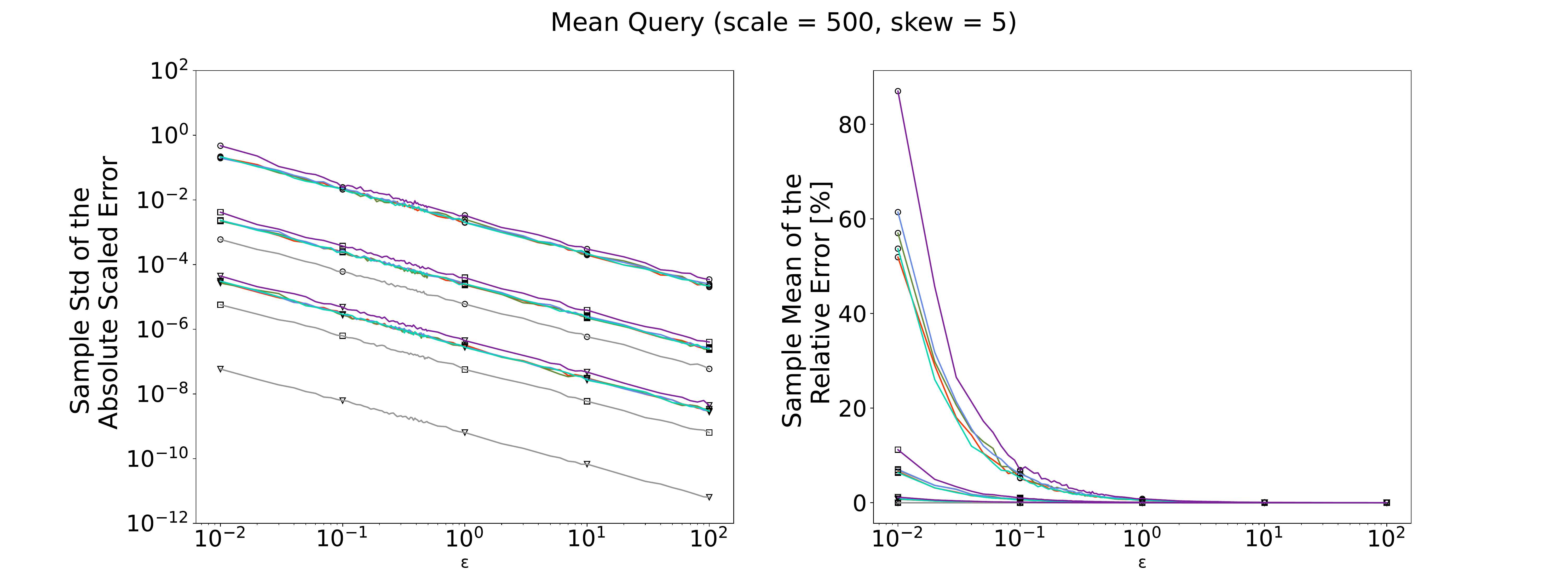} \\

\includegraphics[ width=\linewidth, height=\linewidth, keepaspectratio]{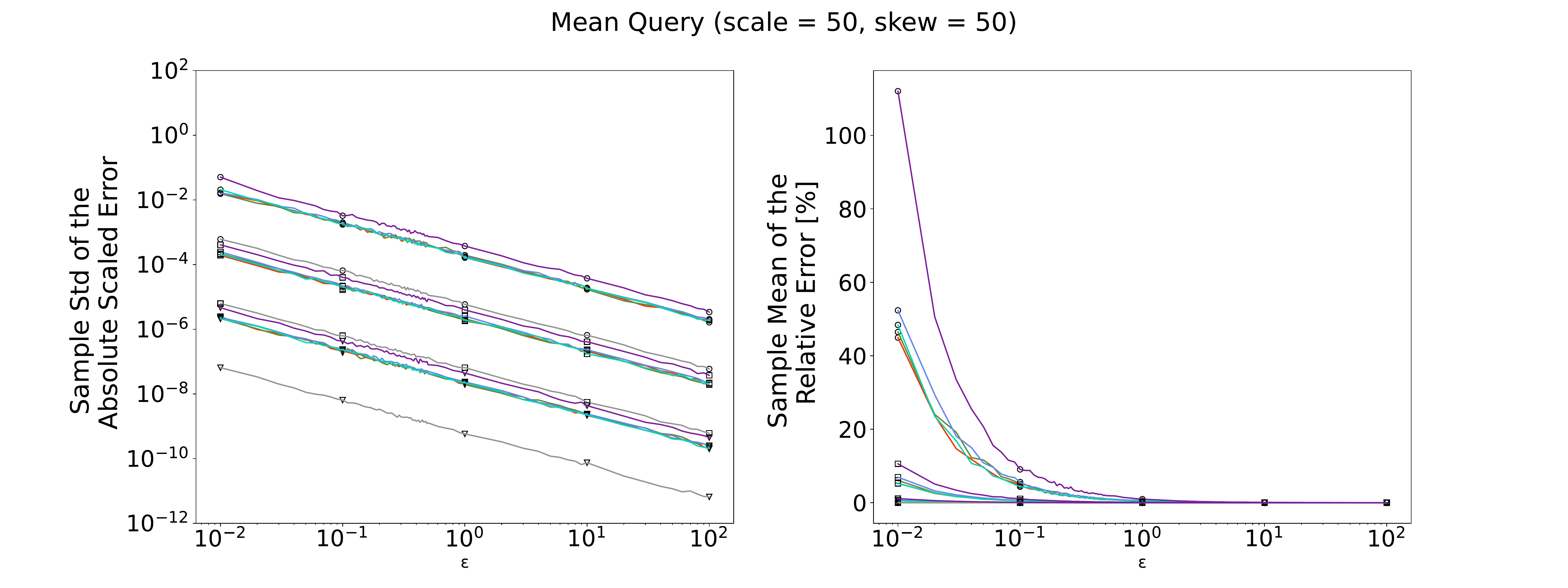} 
&
\includegraphics[ width=\linewidth, height=\linewidth, keepaspectratio]{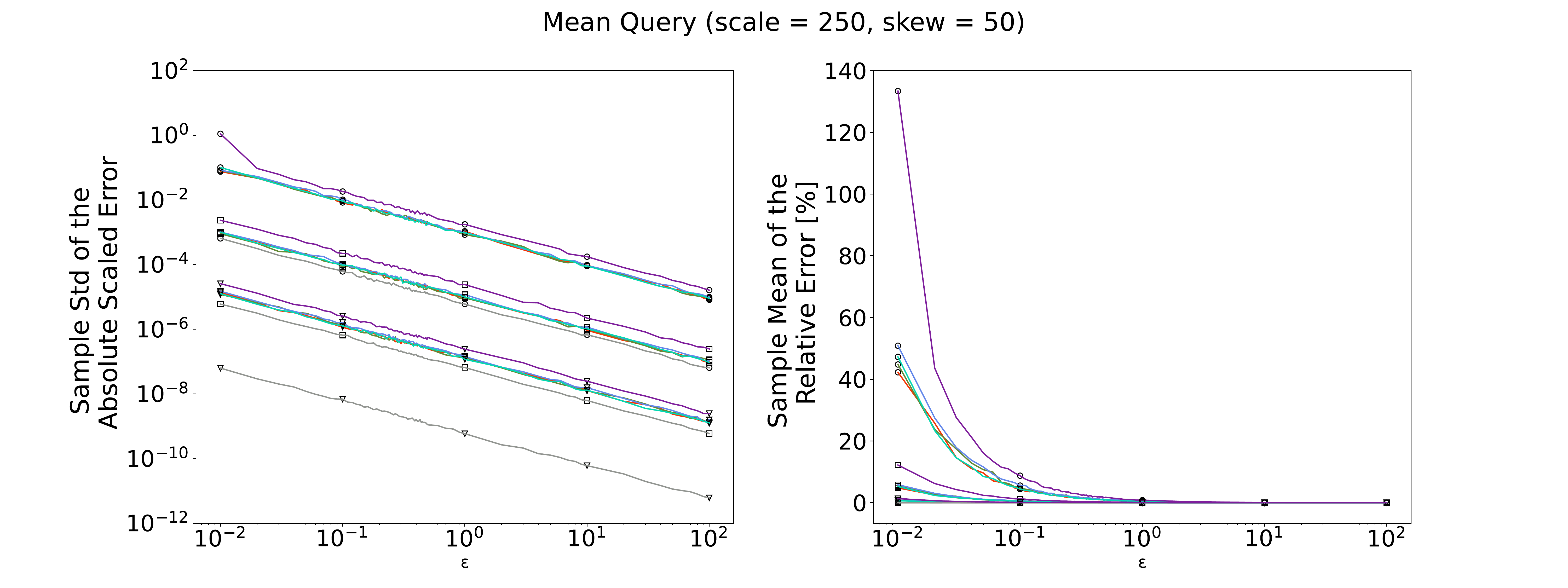} \\

\includegraphics[ width=\linewidth, height=\linewidth, keepaspectratio]{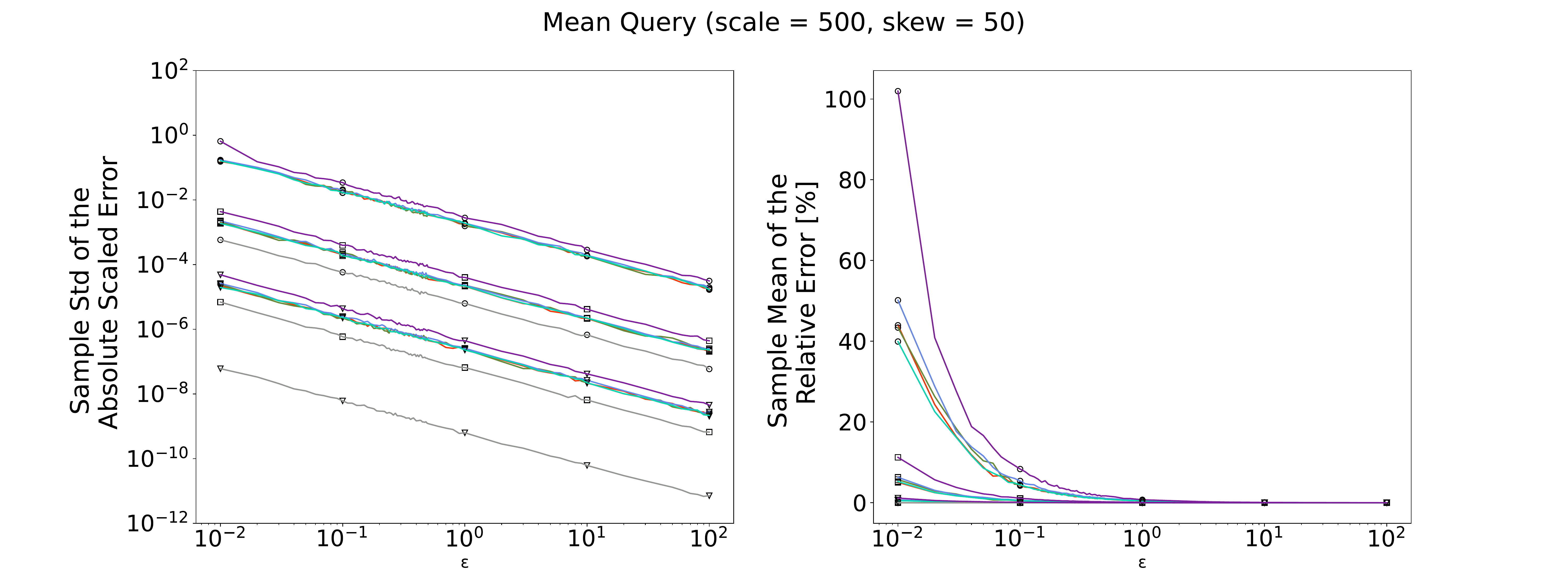} 
&
\includegraphics[width=\linewidth, height=\linewidth, keepaspectratio, scale=0.28]{legends.pdf}
\\

\end{tabular}
    \caption{ Set of detailed experiments' outputs for all three independent variables (dataset size, scale, and skewness) of the mean query (500 experiments per \textbf{$\varepsilon$}).}
    \label{tab:all_mean_exps}
\end{table*}


\begin{table*}
    \centering
\begin{tabular}{m{0.5\linewidth}m{0.5\linewidth}}

\includegraphics[ width=\linewidth, height=\linewidth, keepaspectratio]{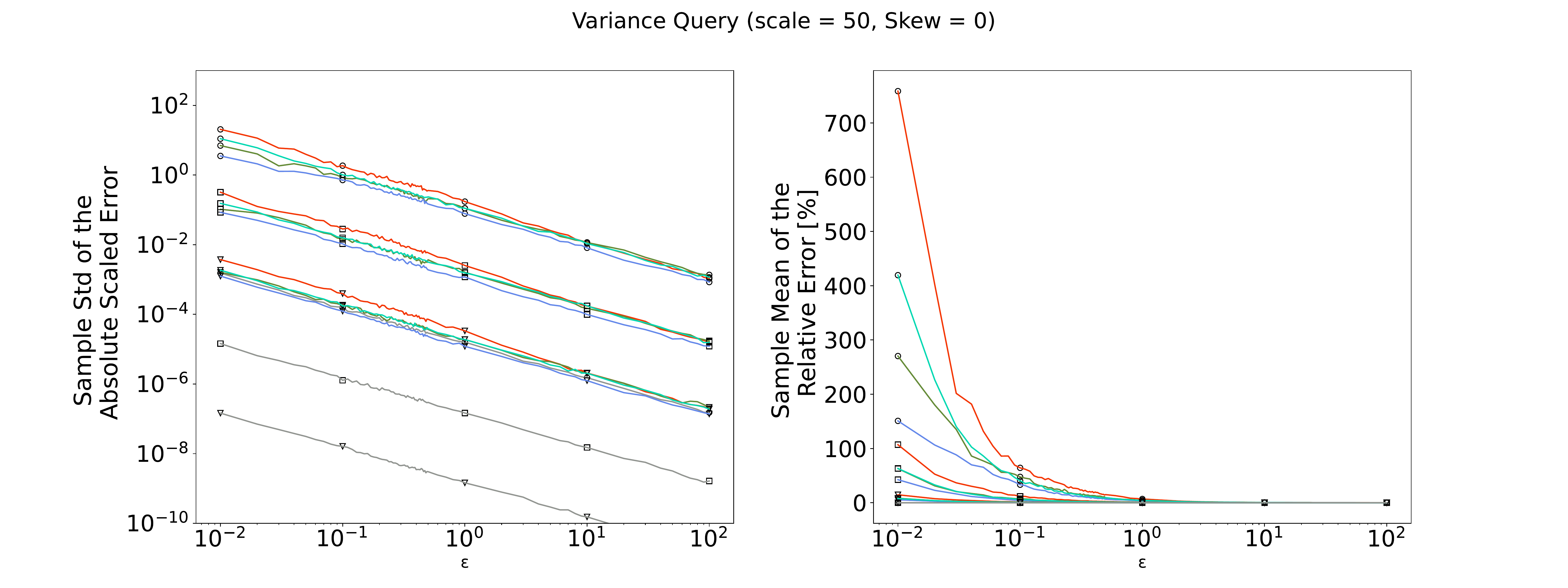} 
&
\includegraphics[ width=\linewidth, height=\linewidth, keepaspectratio]{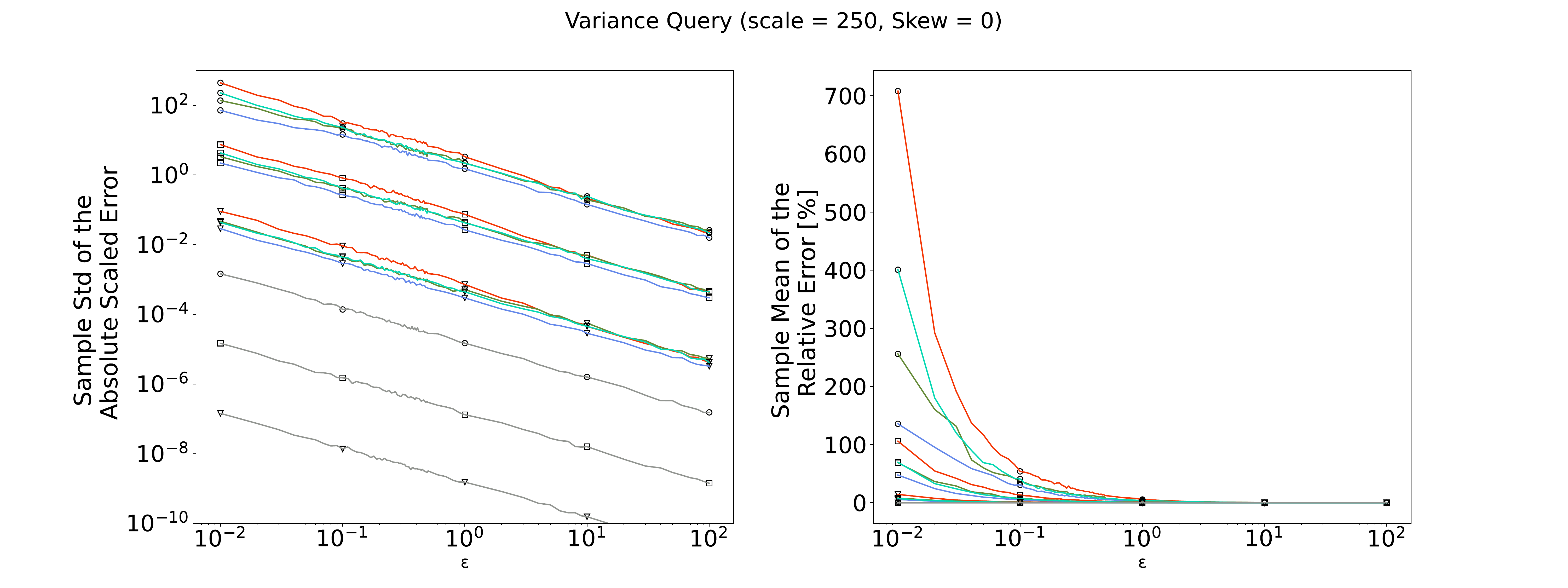} \\

\includegraphics[ width=\linewidth, height=\linewidth, keepaspectratio]{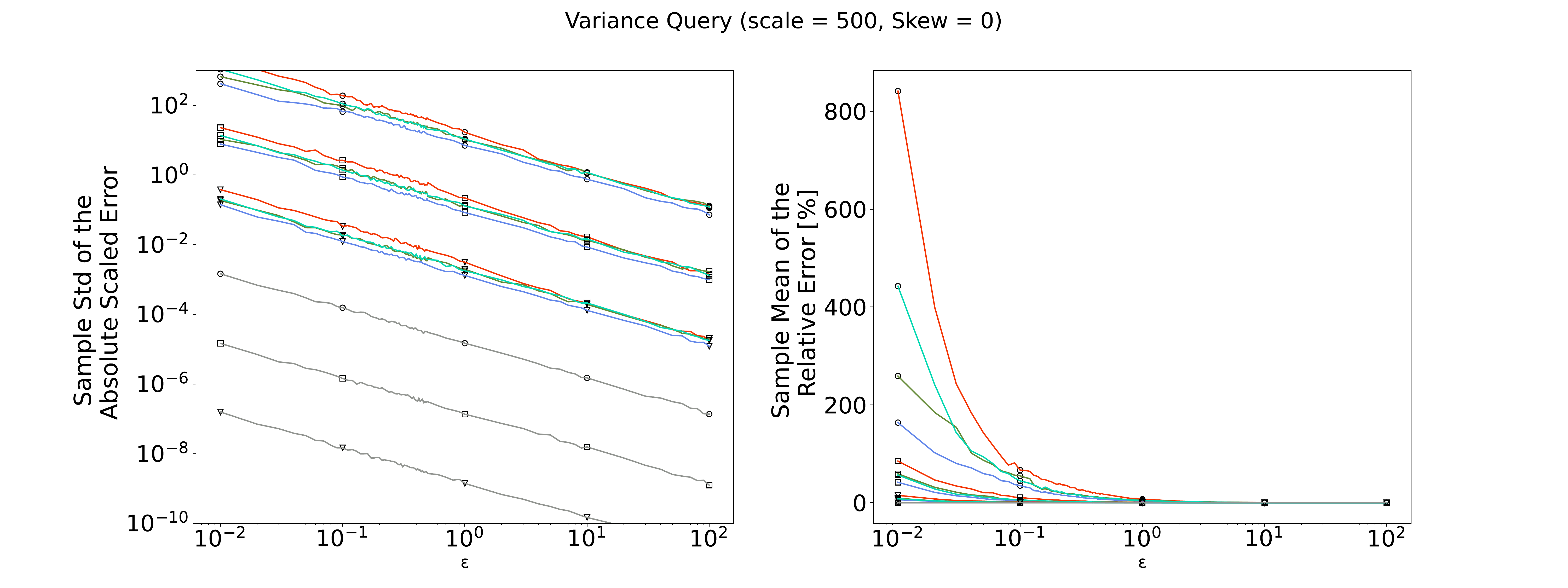} 
&
\includegraphics[ width=\linewidth, height=\linewidth, keepaspectratio]{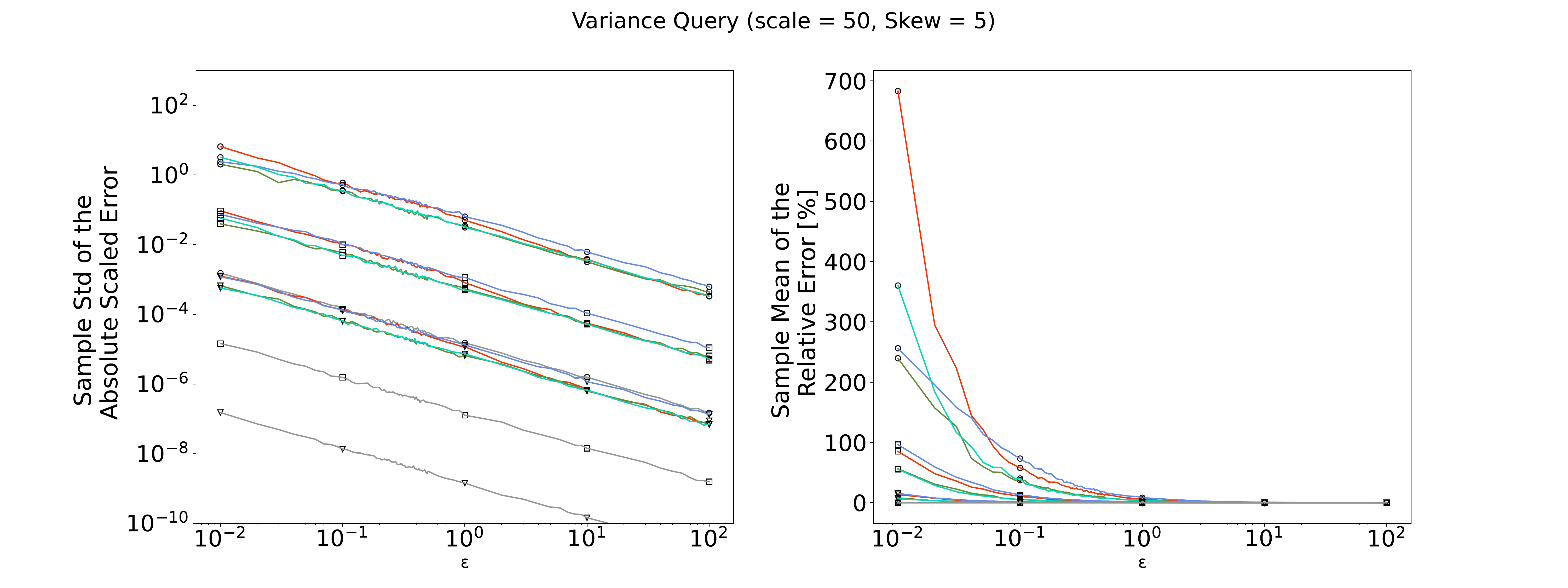} \\

\includegraphics[ width=\linewidth, height=\linewidth, keepaspectratio]{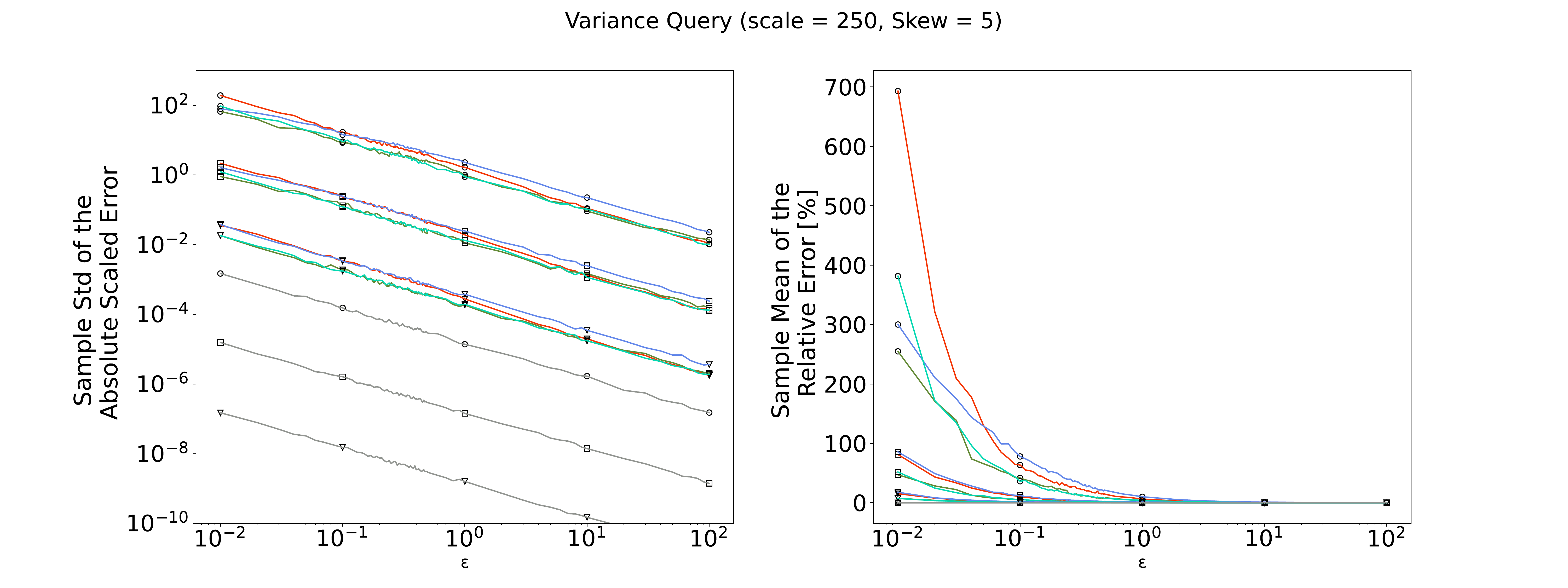} 
&
\includegraphics[ width=\linewidth, height=\linewidth, keepaspectratio]{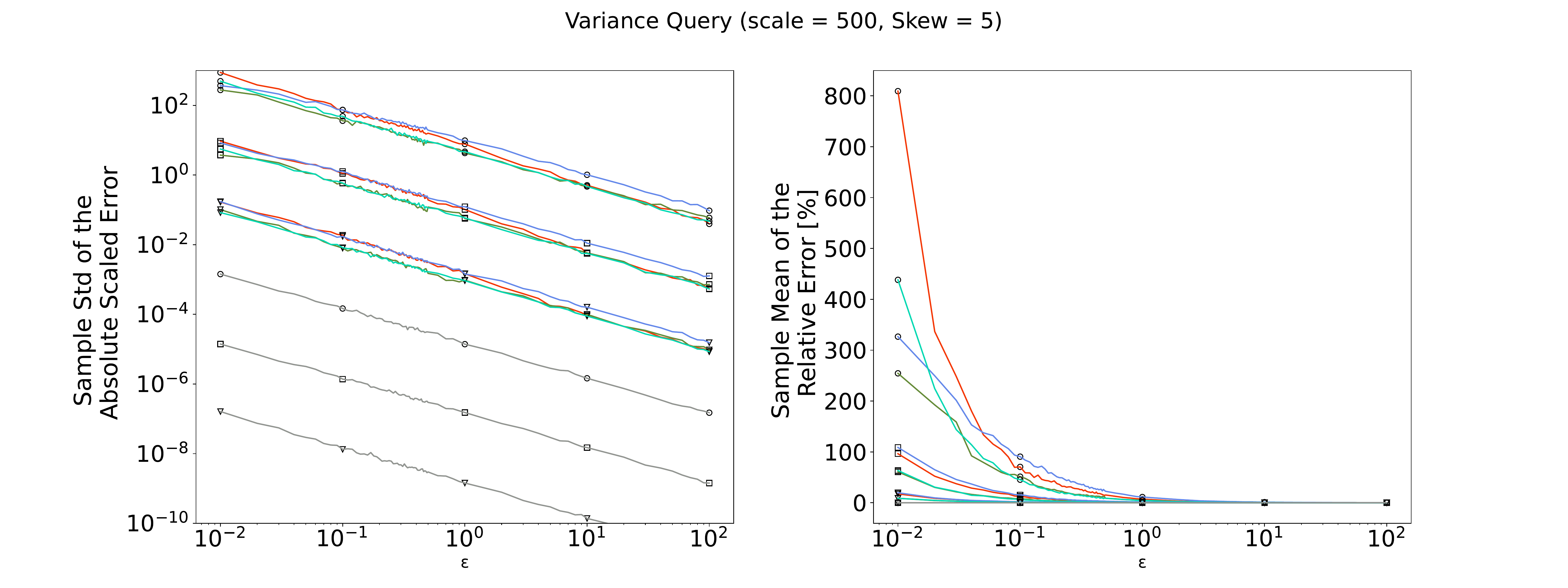} \\

\includegraphics[ width=\linewidth, height=\linewidth, keepaspectratio]{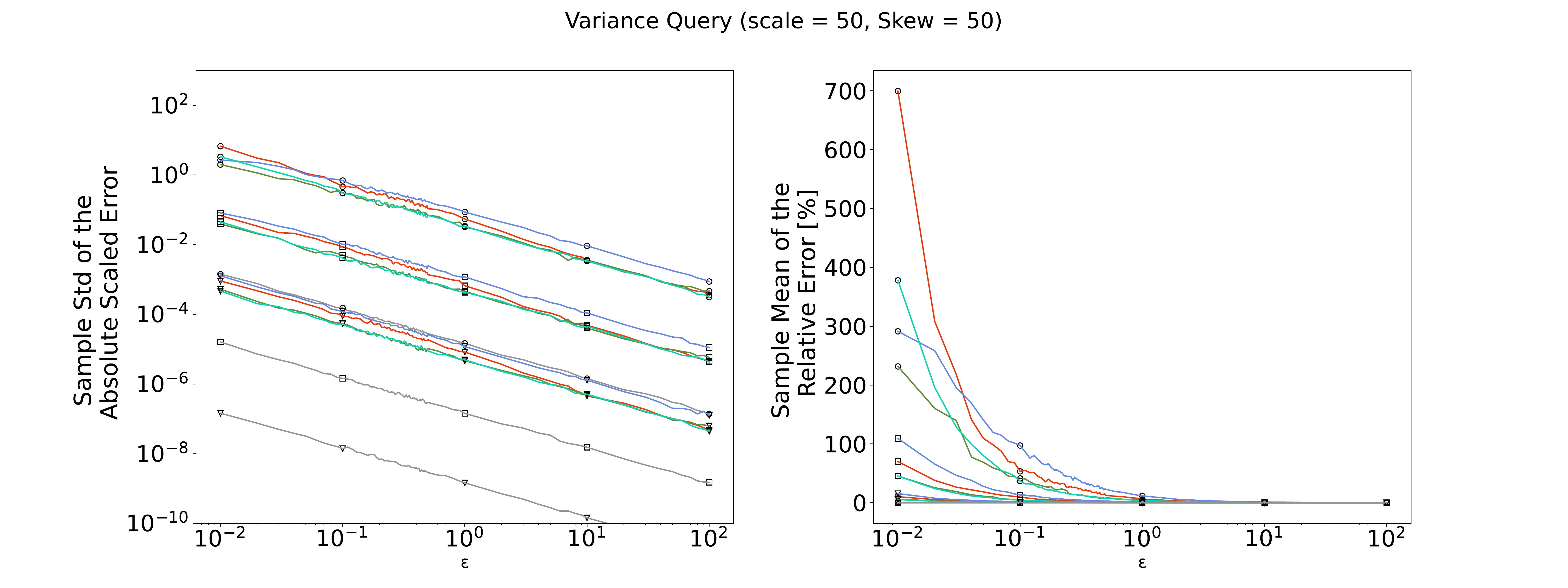} 
&
\includegraphics[ width=\linewidth, height=\linewidth, keepaspectratio]{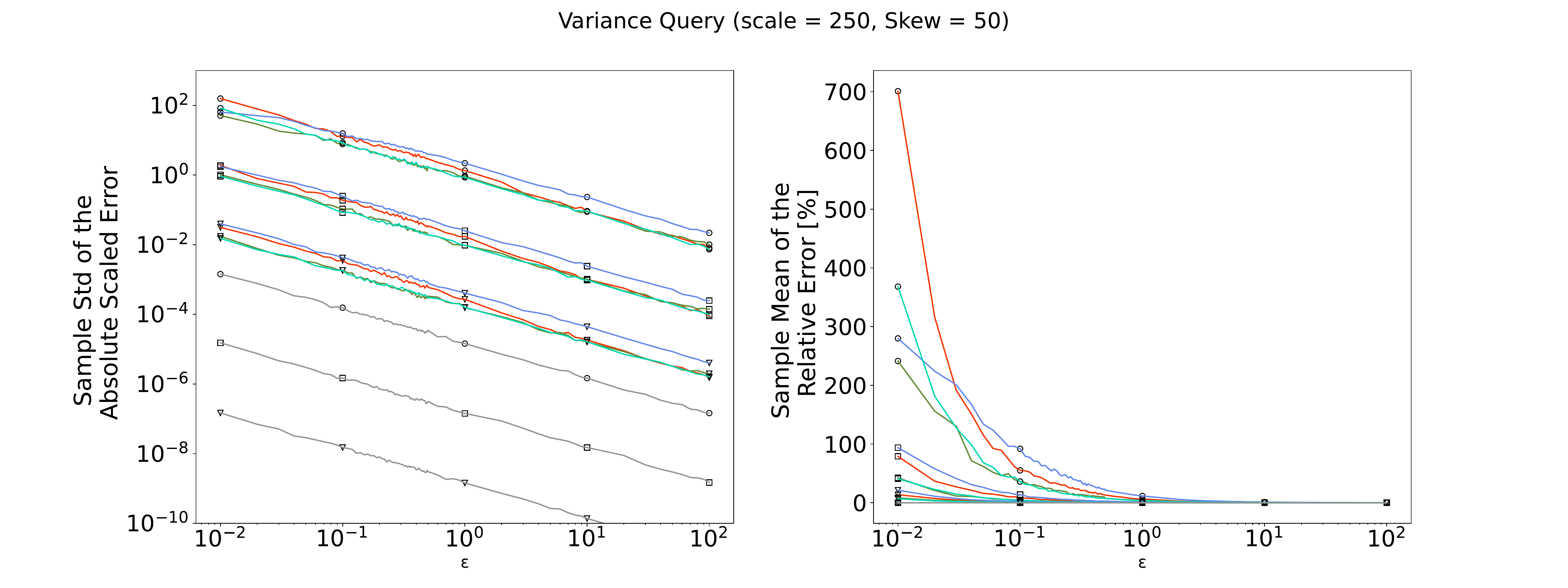} \\

\includegraphics[ width=\linewidth, height=\linewidth, keepaspectratio]{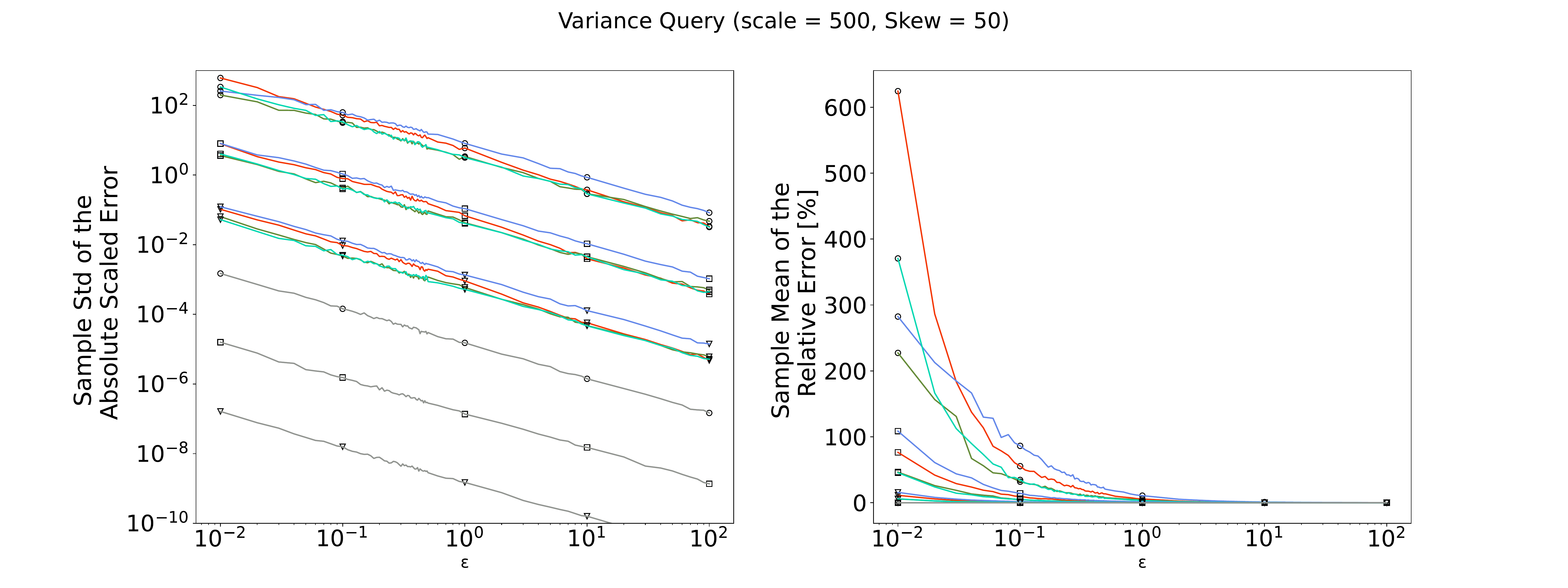} 
&
\includegraphics[width=\linewidth, height=\linewidth, keepaspectratio, scale=0.28]{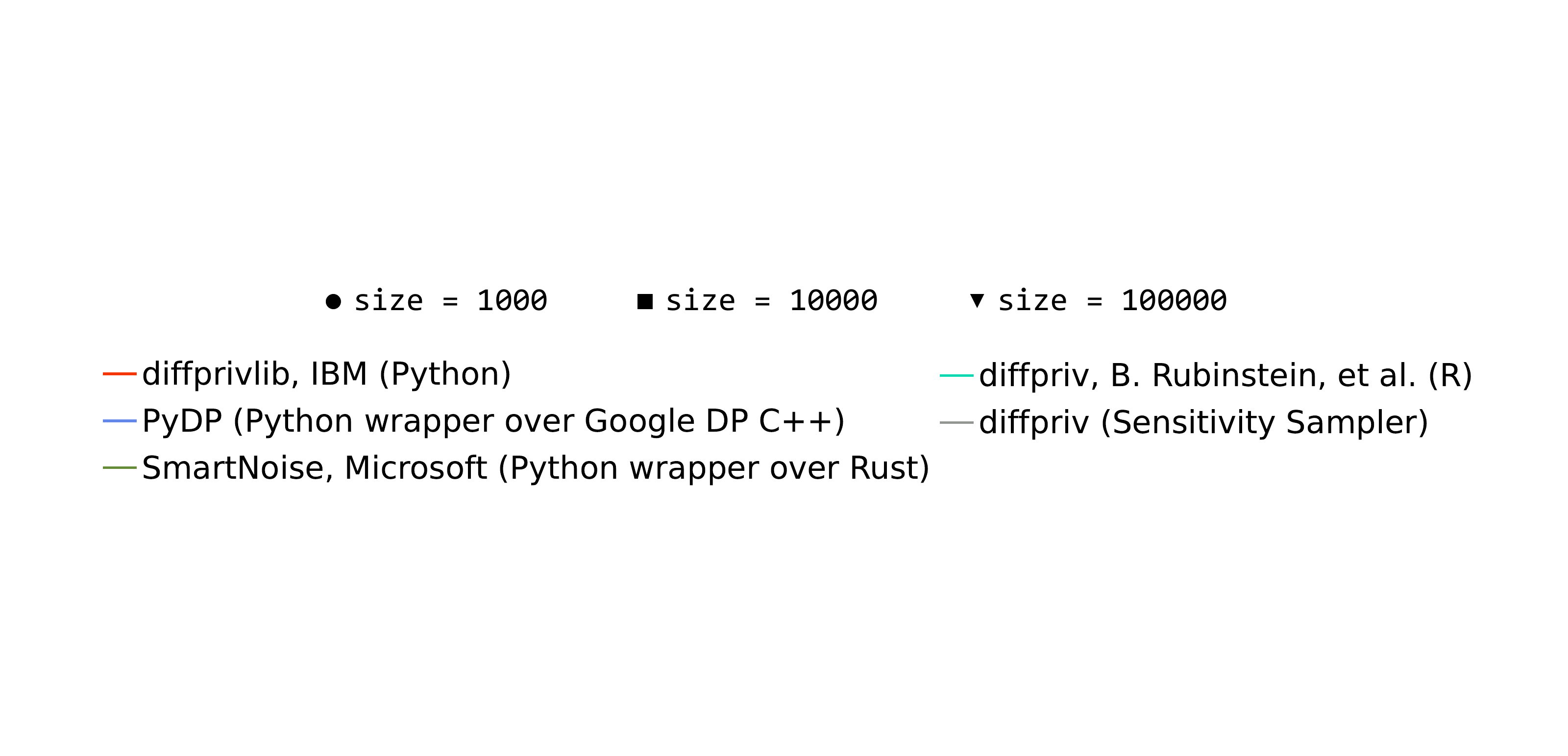}
\\

\end{tabular}
    \caption{ Set of detailed experiments' outputs for all three independent variables (dataset size, scale, and skewness) of the var query (500 experiments per \textbf{$\varepsilon$}).}
    \label{tab:all_var_exps}
\end{table*}

\begin{table*}
    \centering
\begin{tabular}{p{0.48\linewidth} p{0.48\linewidth}}

\includegraphics[ width=\linewidth, height=\linewidth, keepaspectratio]{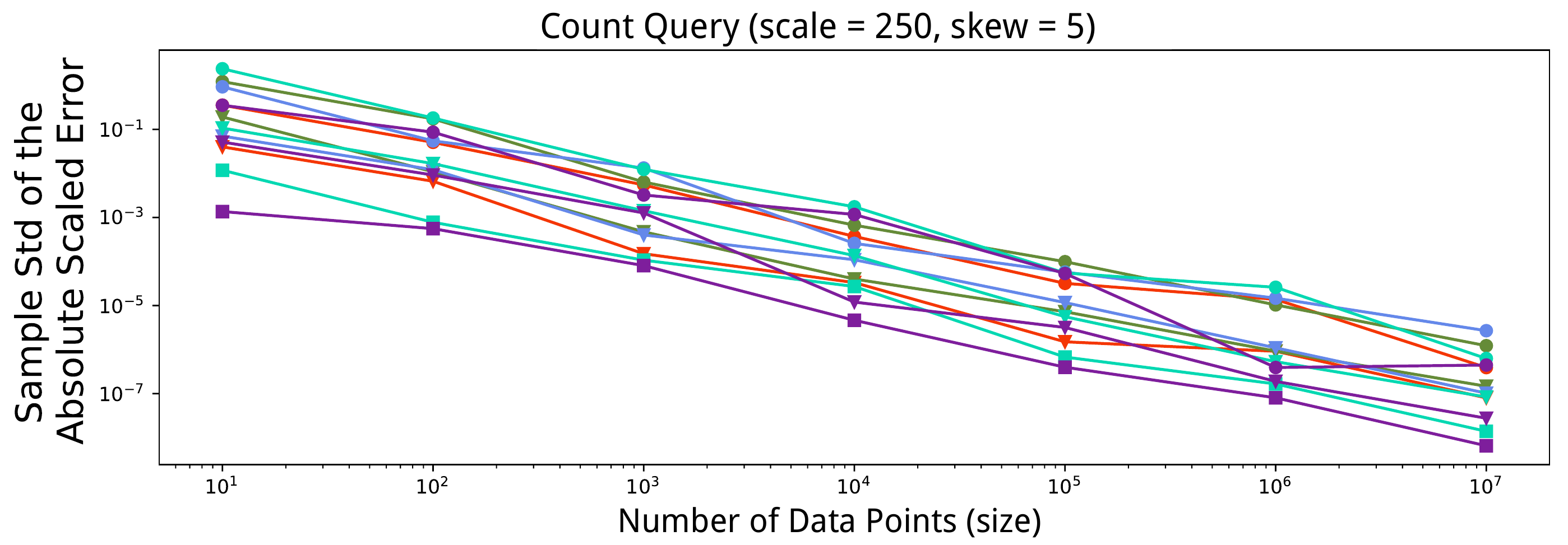}
&
\includegraphics[ width=\linewidth, height=\linewidth, keepaspectratio]{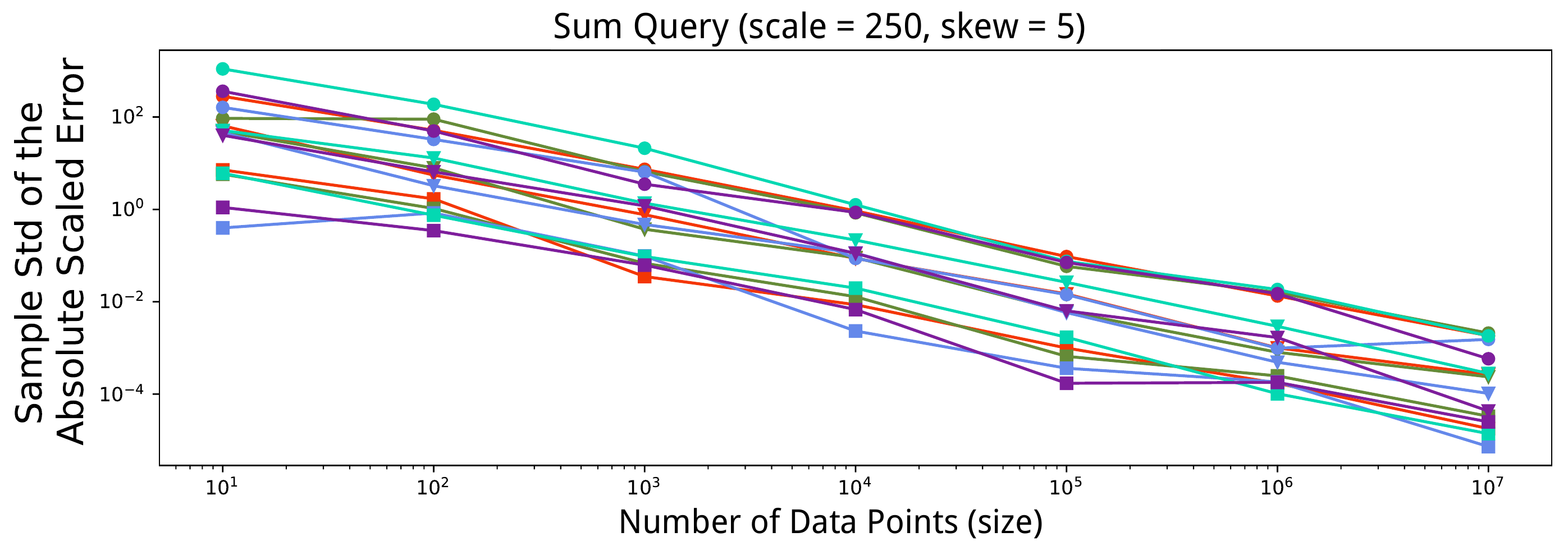} \\

\includegraphics[ width=\linewidth, height=\linewidth, keepaspectratio]{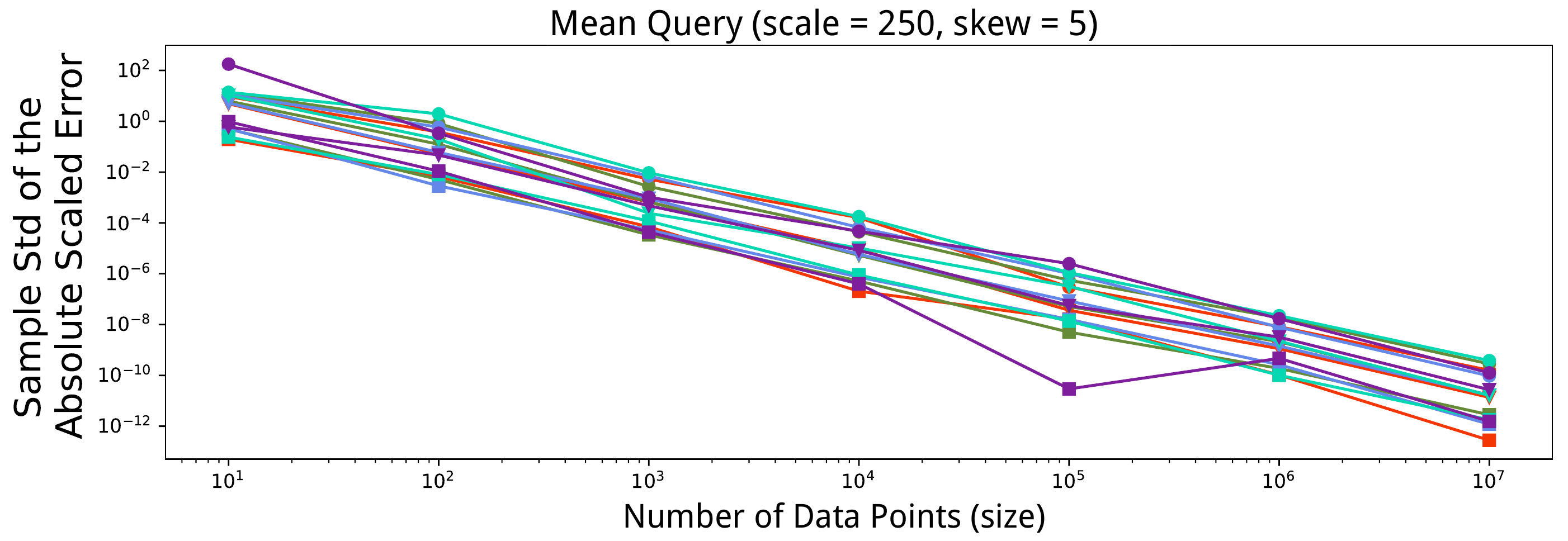}
&
\includegraphics[ width=\linewidth, height=\linewidth, keepaspectratio]{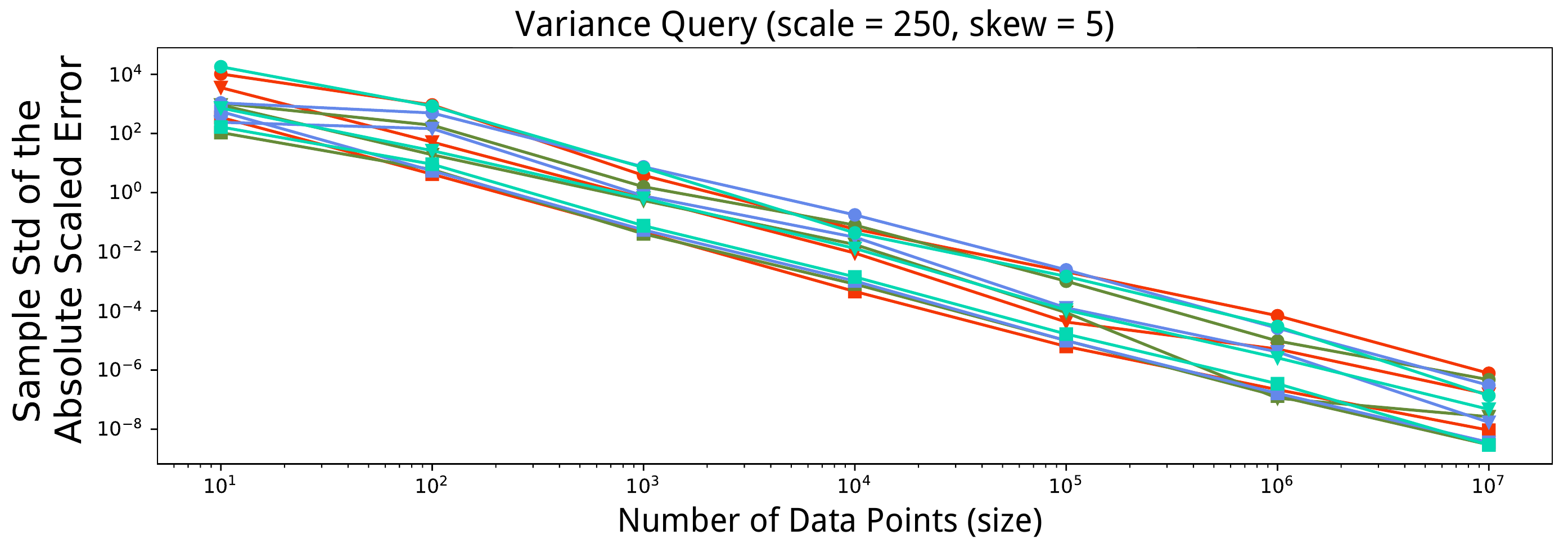} \\

\multicolumn{2}{c}{\includegraphics[scale=0.3]{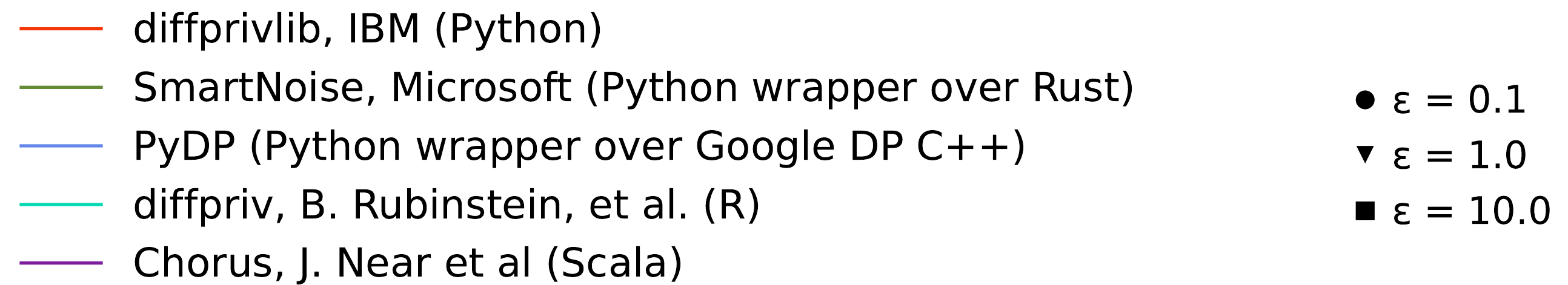}}

\end{tabular}
    \caption{Sample std of the absolute scaled error across varying dataset scales (5 experiments per \textbf{$\varepsilon$} and dataset scale).}
    \label{tab:all_execution_time}
\end{table*}


\begin{table*}
    \centering
\begin{tabular}{p{0.48\linewidth} p{0.48\linewidth}}

\includegraphics[ width=\linewidth, height=\linewidth, keepaspectratio]{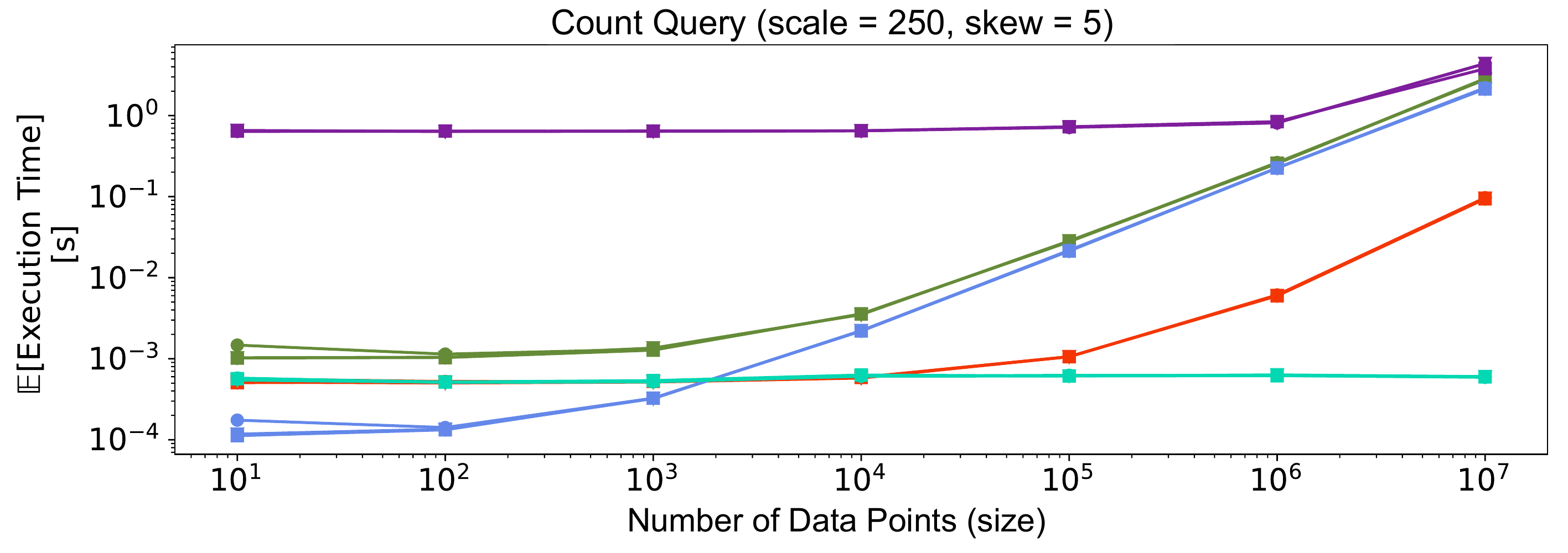}
&
\includegraphics[ width=\linewidth, height=\linewidth, keepaspectratio]{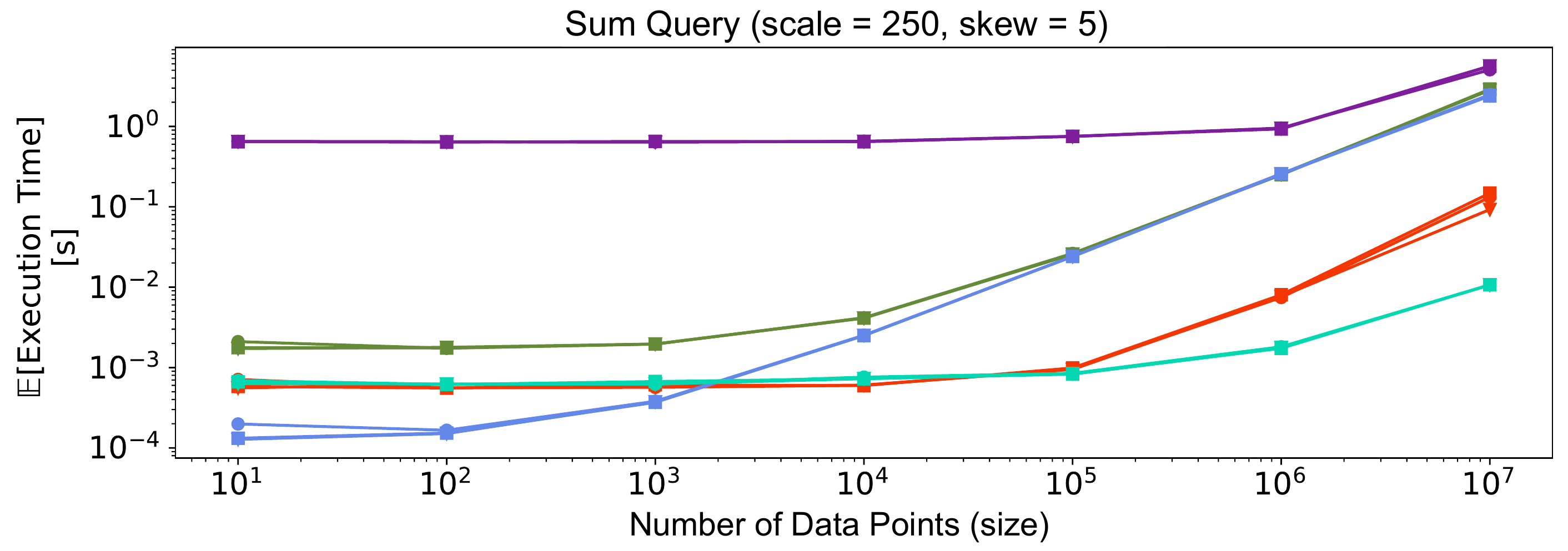} \\

\includegraphics[ width=\linewidth, height=\linewidth, keepaspectratio]{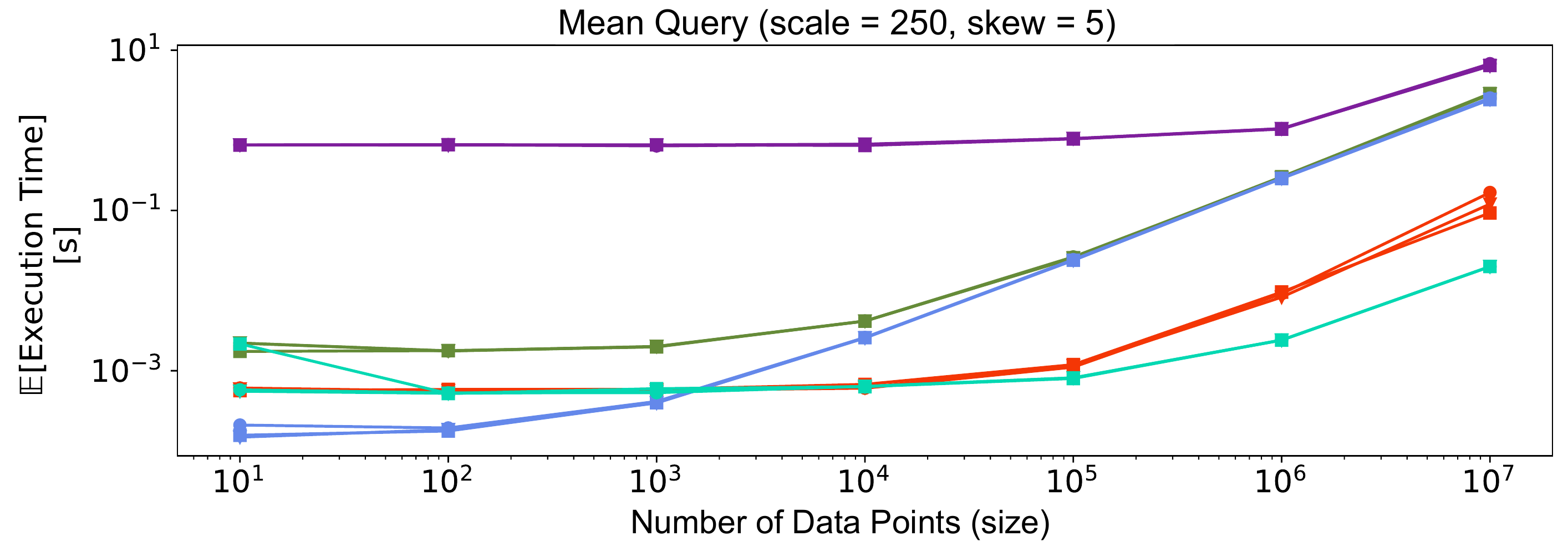}
&
\includegraphics[ width=\linewidth, height=\linewidth, keepaspectratio]{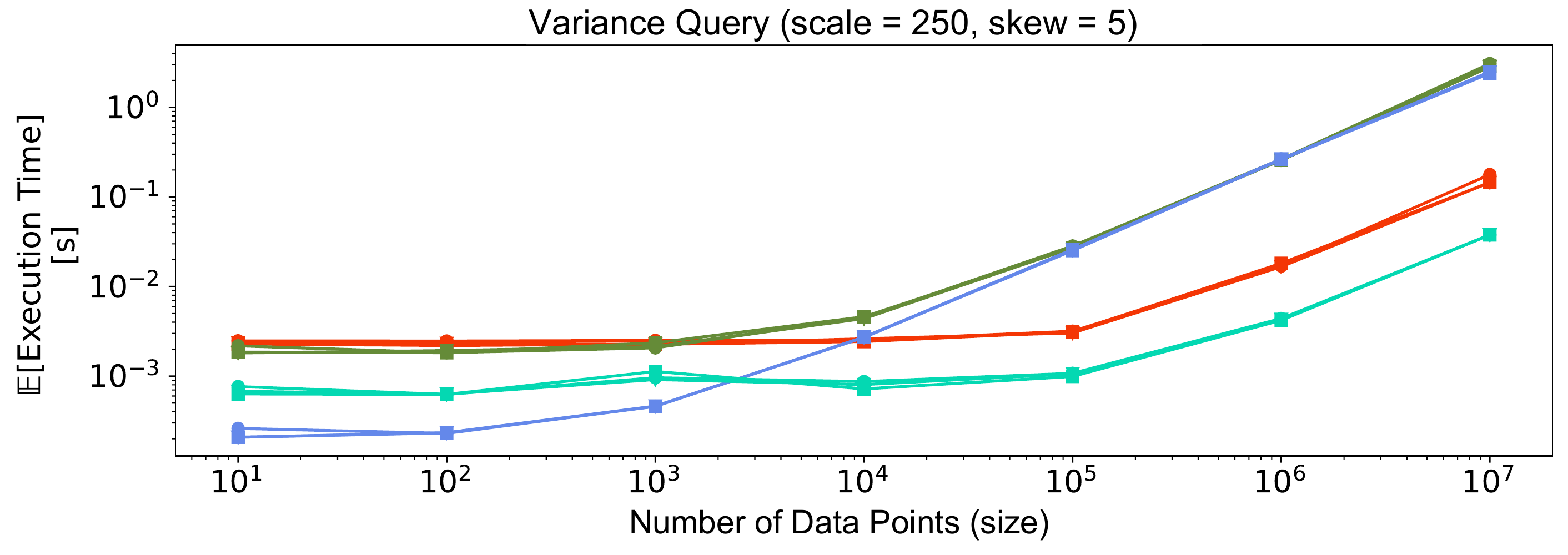} \\

\multicolumn{2}{c}{\includegraphics[scale=0.3]{scale_legends.pdf}}

\end{tabular}
    \caption{ Execution time for a range of an increasing number of data points for all queries: count, sum, mean, and var (5 experiments per \textbf{$\varepsilon$} and dataset size).}
    \label{tab:all_execution_time}
\end{table*}

\begin{table*}
    \centering
\begin{tabular}{p{0.48\linewidth} p{0.48\linewidth}}

\includegraphics[ width=\linewidth, height=\linewidth, keepaspectratio]{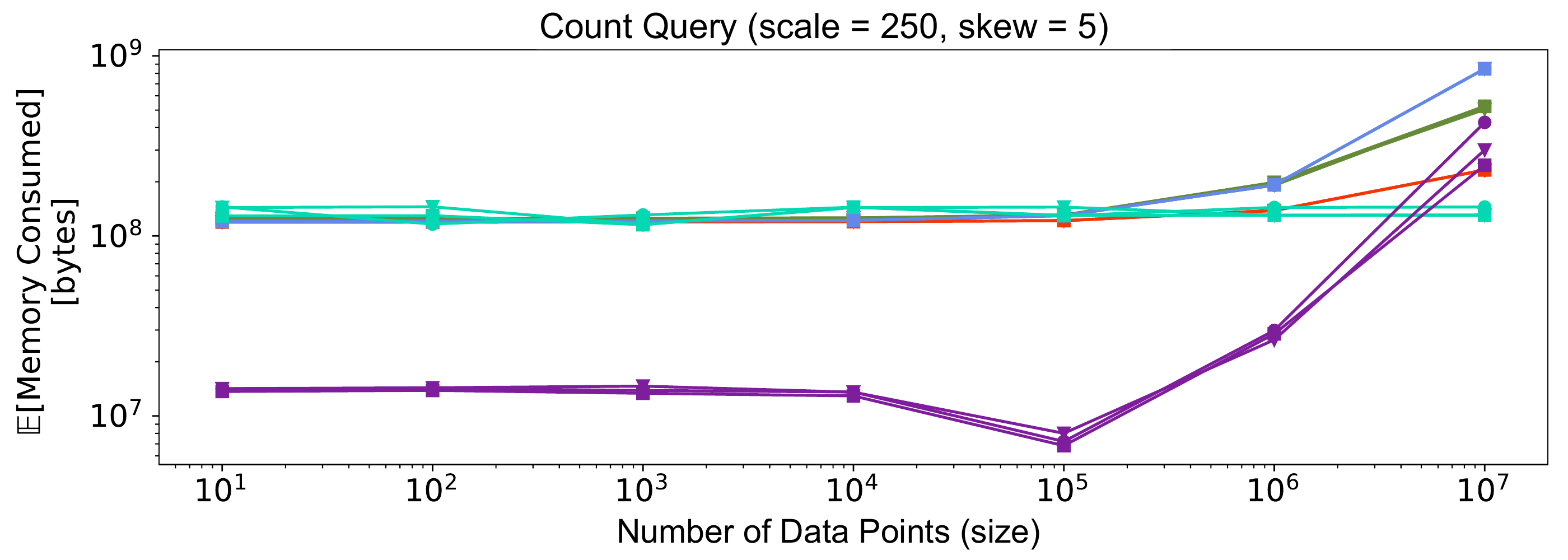}
&
\includegraphics[ width=\linewidth, height=\linewidth, keepaspectratio]{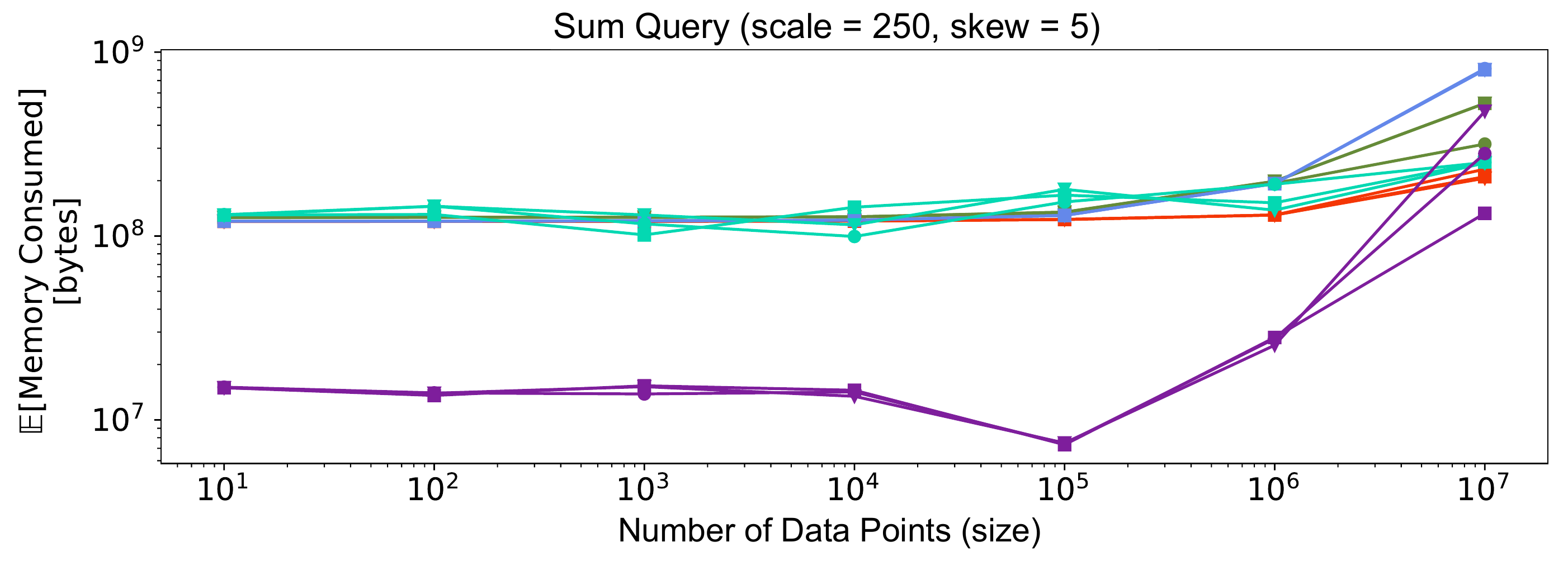} \\

\includegraphics[ width=\linewidth, height=\linewidth, keepaspectratio]{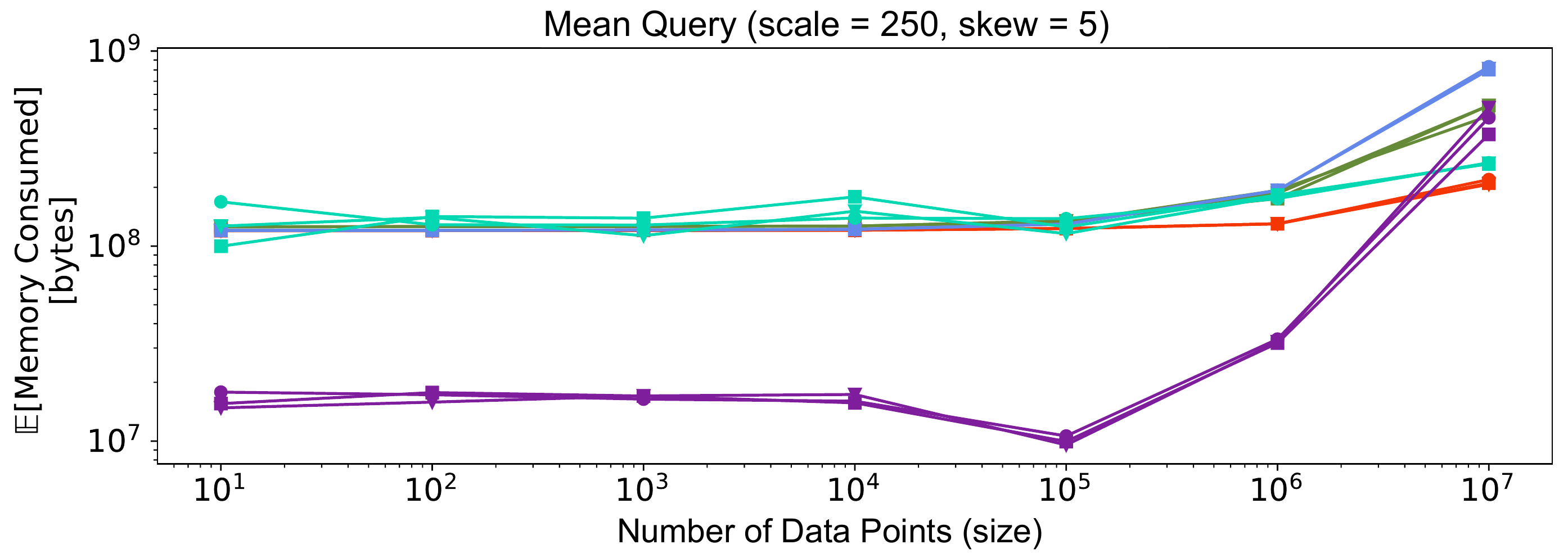}
&
\includegraphics[ width=\linewidth, height=\linewidth, keepaspectratio]{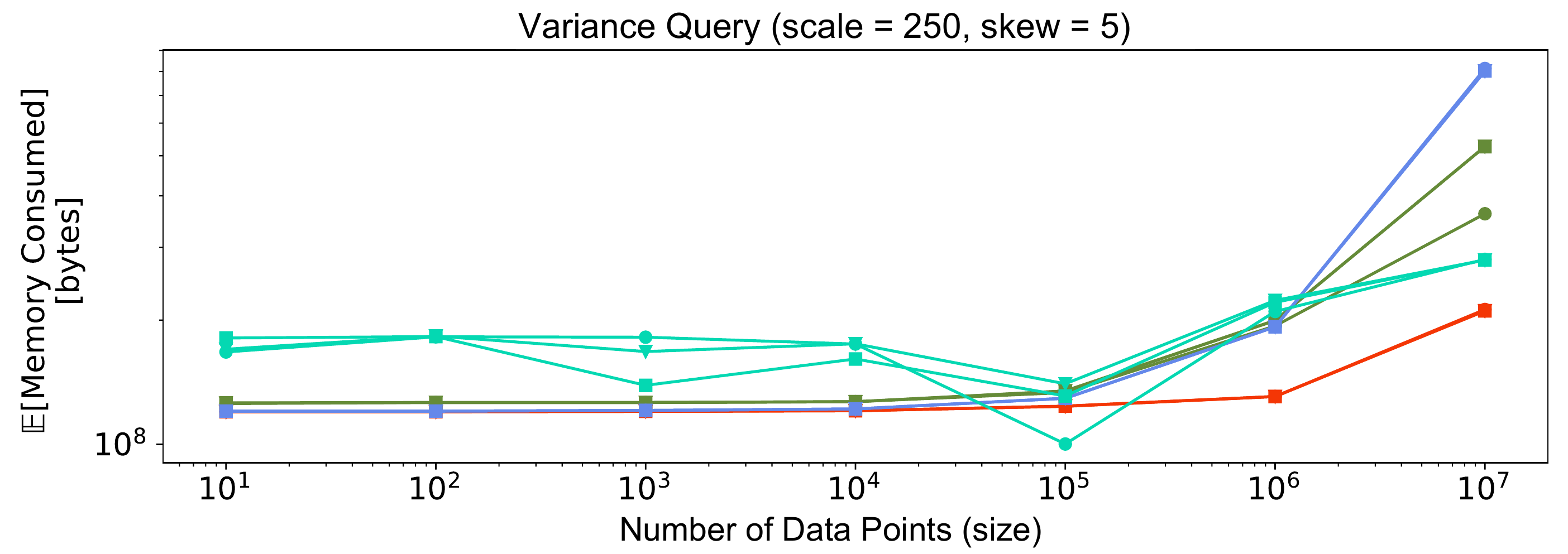} \\

\multicolumn{2}{c}{\includegraphics[scale=0.3]{scale_legends.pdf}}

\end{tabular}
    \caption{ Memory consumption for a range of an increasing number of data points for all queries: count, sum, mean, and var (5 experiments per \textbf{$\varepsilon$} and dataset size).}
    \label{tab:all_memory_comsumption}
\end{table*}


\begin{table*}
    \centering
\begin{tabular}{p{0.48\linewidth} p{0.48\linewidth}}

\includegraphics[ width=\linewidth, height=\linewidth, keepaspectratio]{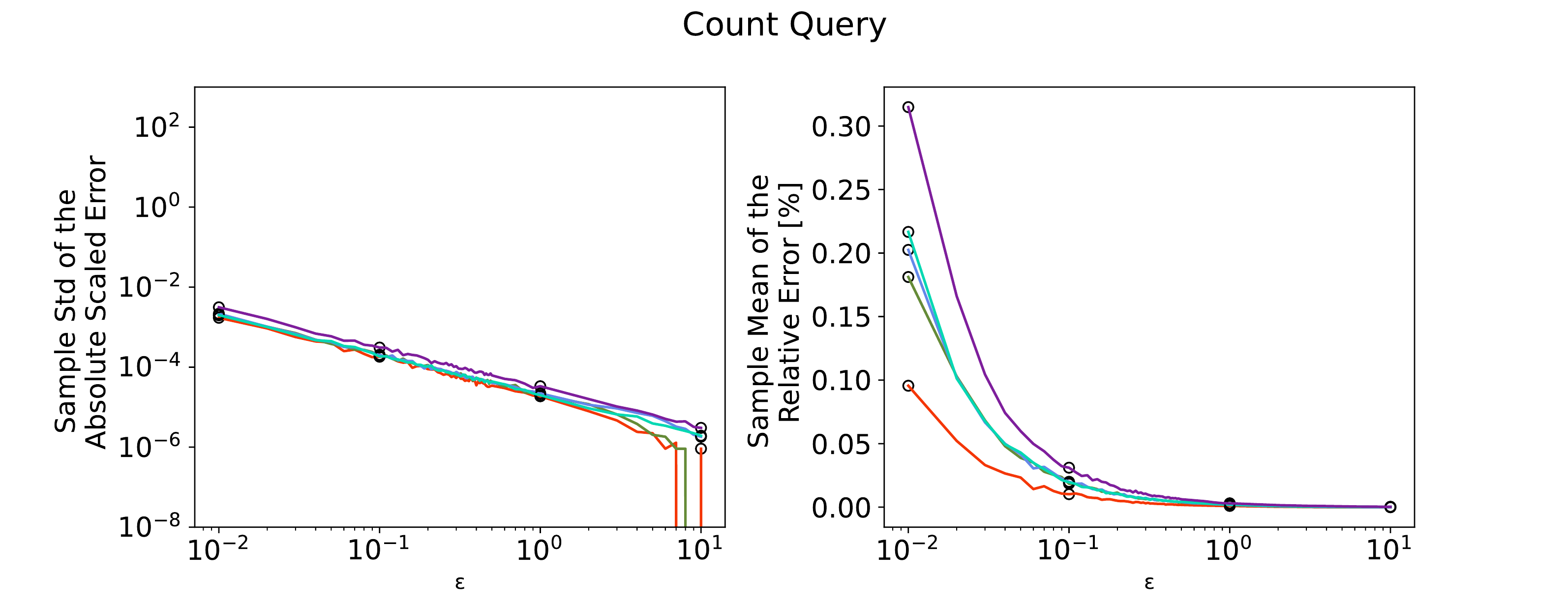}
&
\includegraphics[ width=\linewidth, height=\linewidth, keepaspectratio]{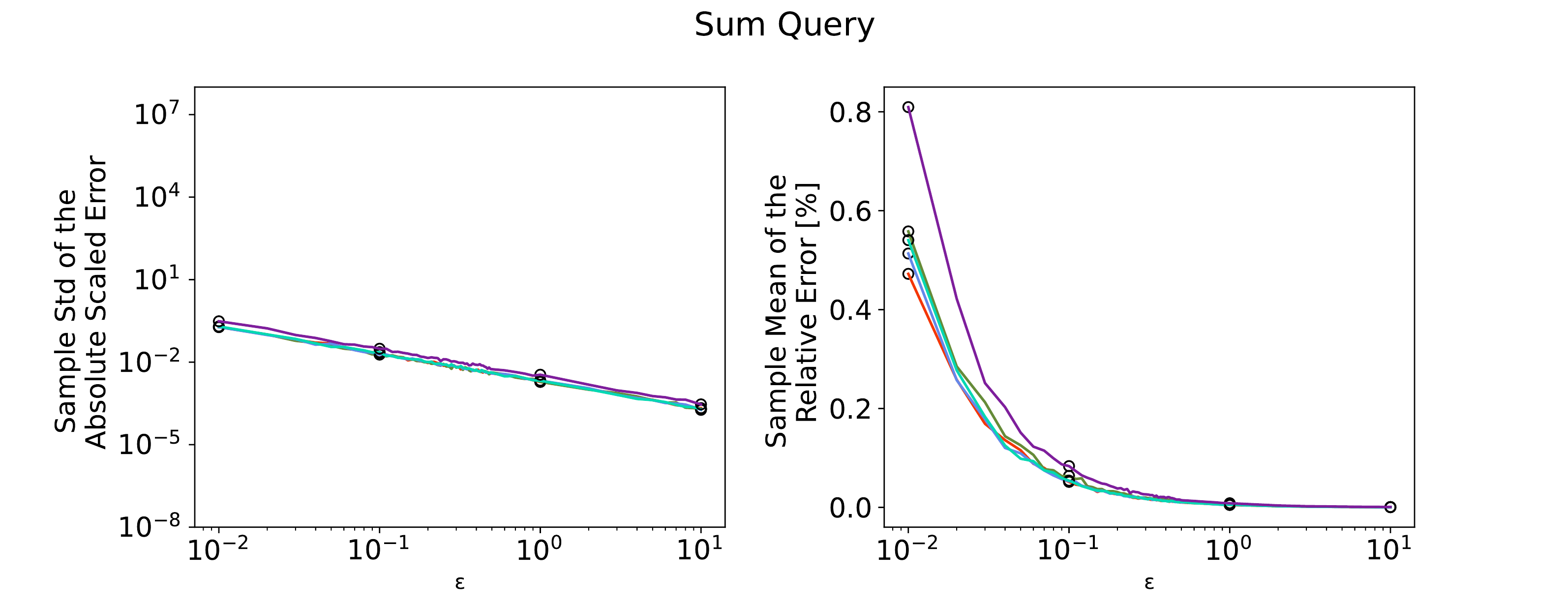} \\

\includegraphics[ width=\linewidth, height=\linewidth, keepaspectratio]{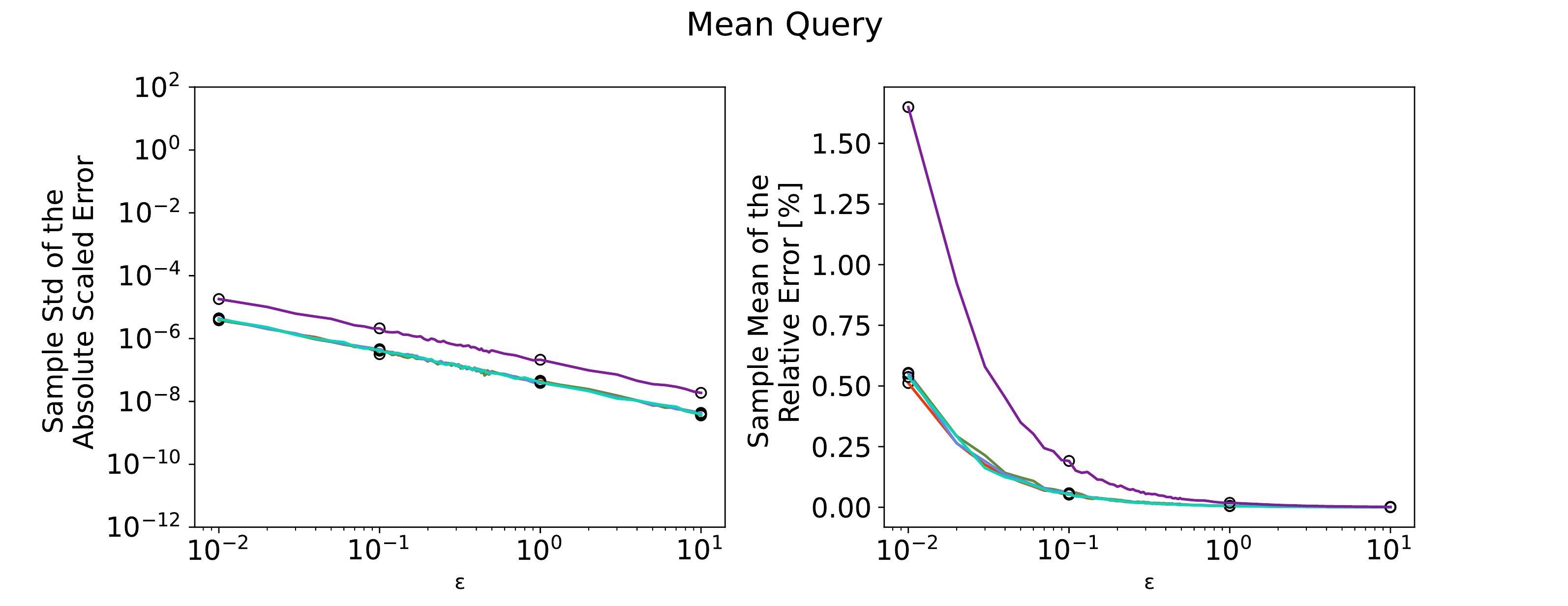}
&
\includegraphics[ width=\linewidth, height=\linewidth, keepaspectratio]{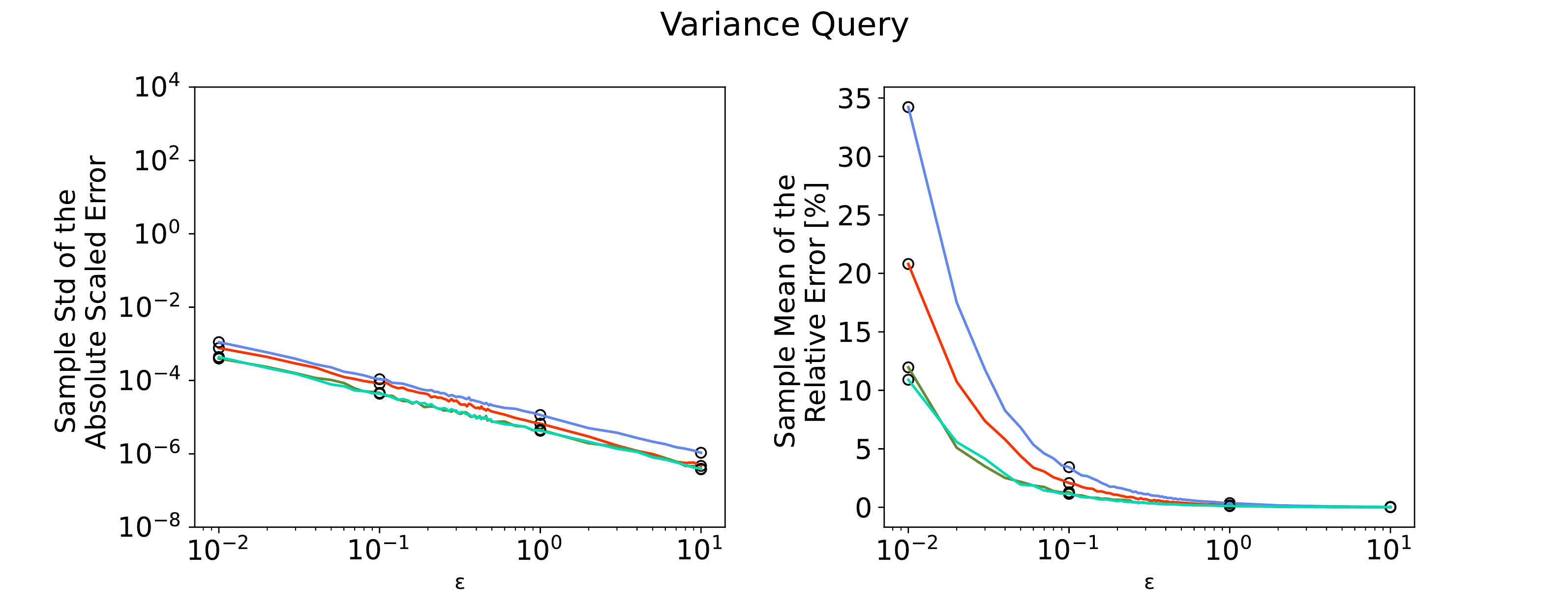} \\

\multicolumn{2}{c}{\includegraphics[ scale=0.3]{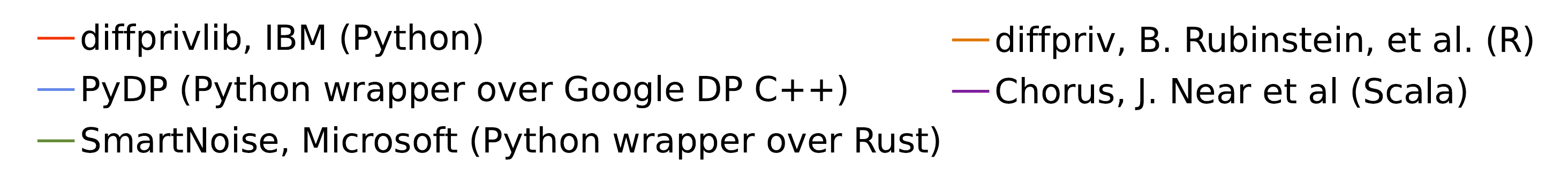}}

\end{tabular}
    \caption{ Experiments of the queries count, sum, mean, and var on the attribute Age of the U.S.A census dataset containing 48842 individuals (500 experiments per \textbf{$\varepsilon$}).}
    \label{tab:real_Age}
\end{table*}

\begin{table*}
    \centering
\begin{tabular}{p{0.48\linewidth} p{0.48\linewidth}}

\includegraphics[ width=\linewidth, height=\linewidth, keepaspectratio]{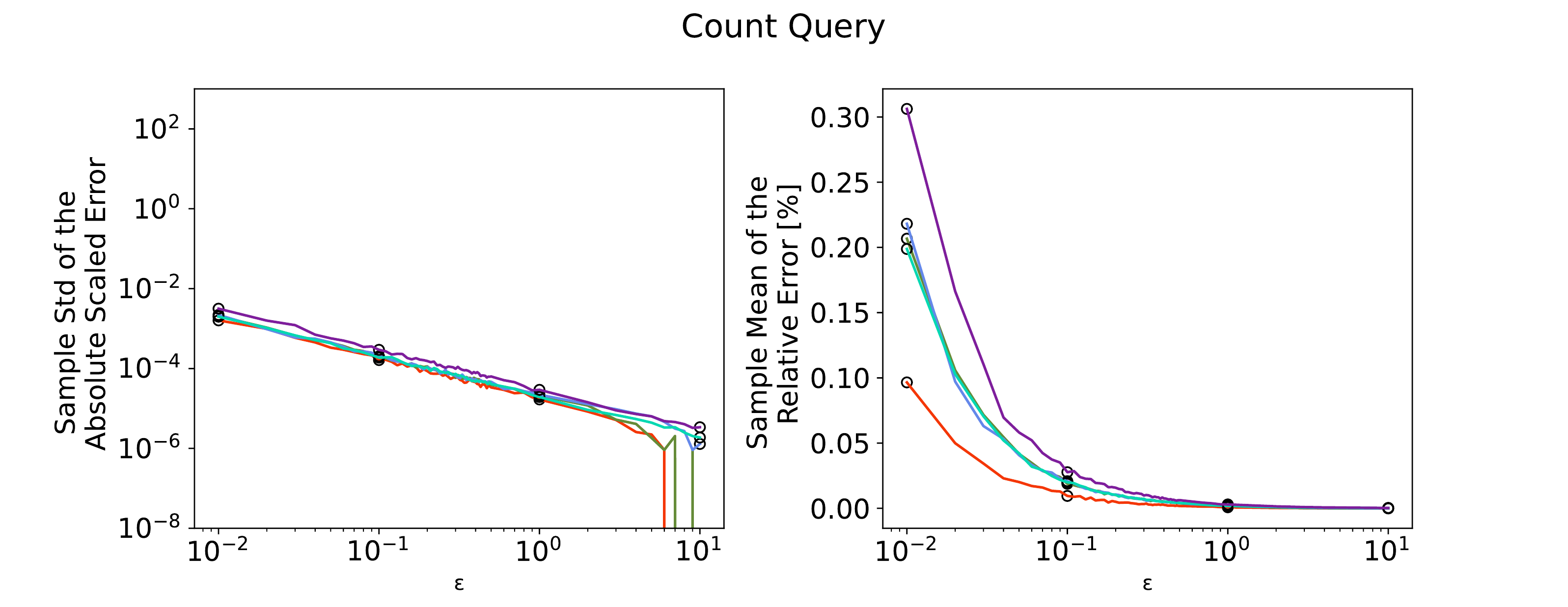}
&
\includegraphics[ width=\linewidth, height=\linewidth, keepaspectratio]{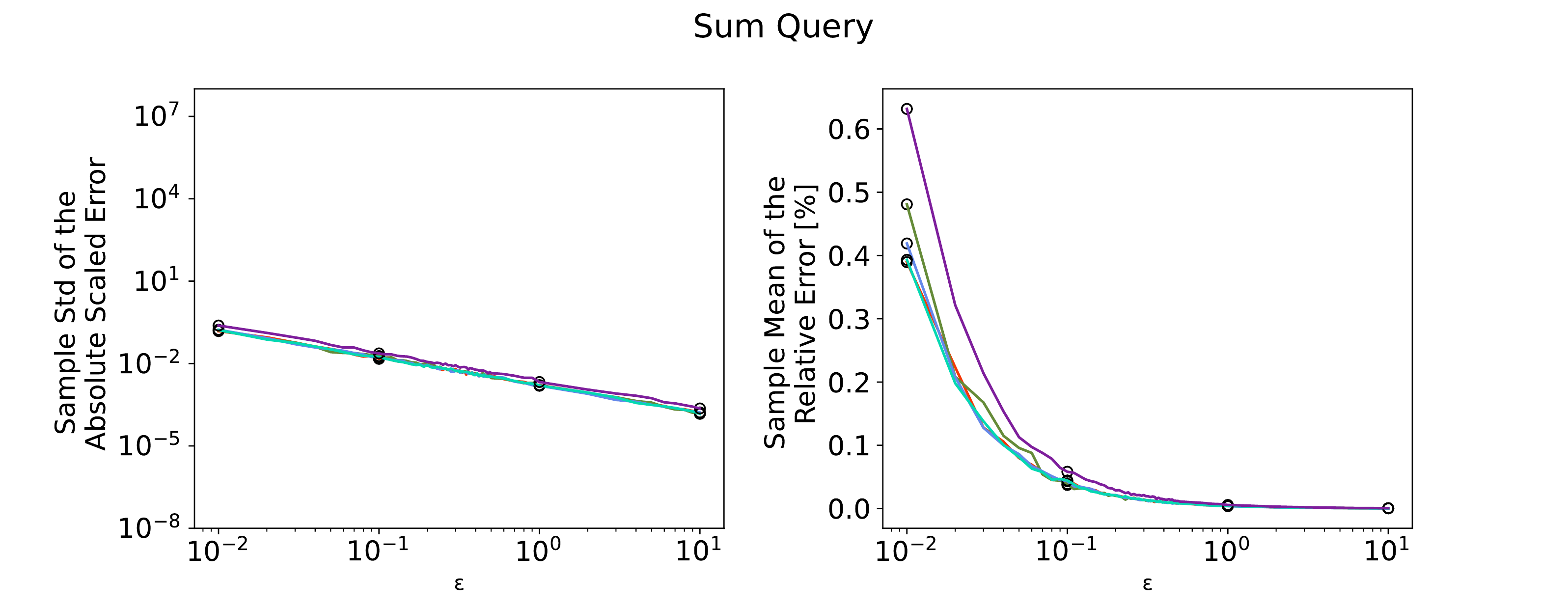} \\

\includegraphics[ width=\linewidth, height=\linewidth, keepaspectratio]{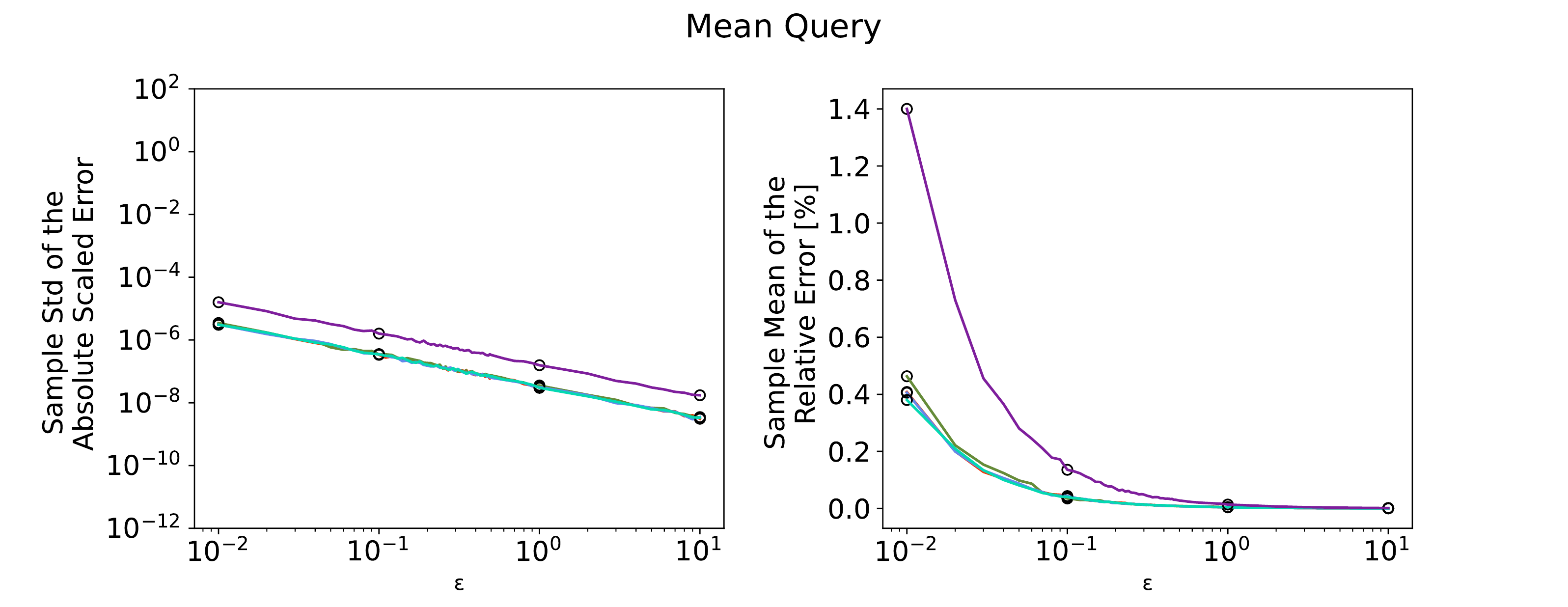}
&
\includegraphics[ width=\linewidth, height=\linewidth, keepaspectratio]{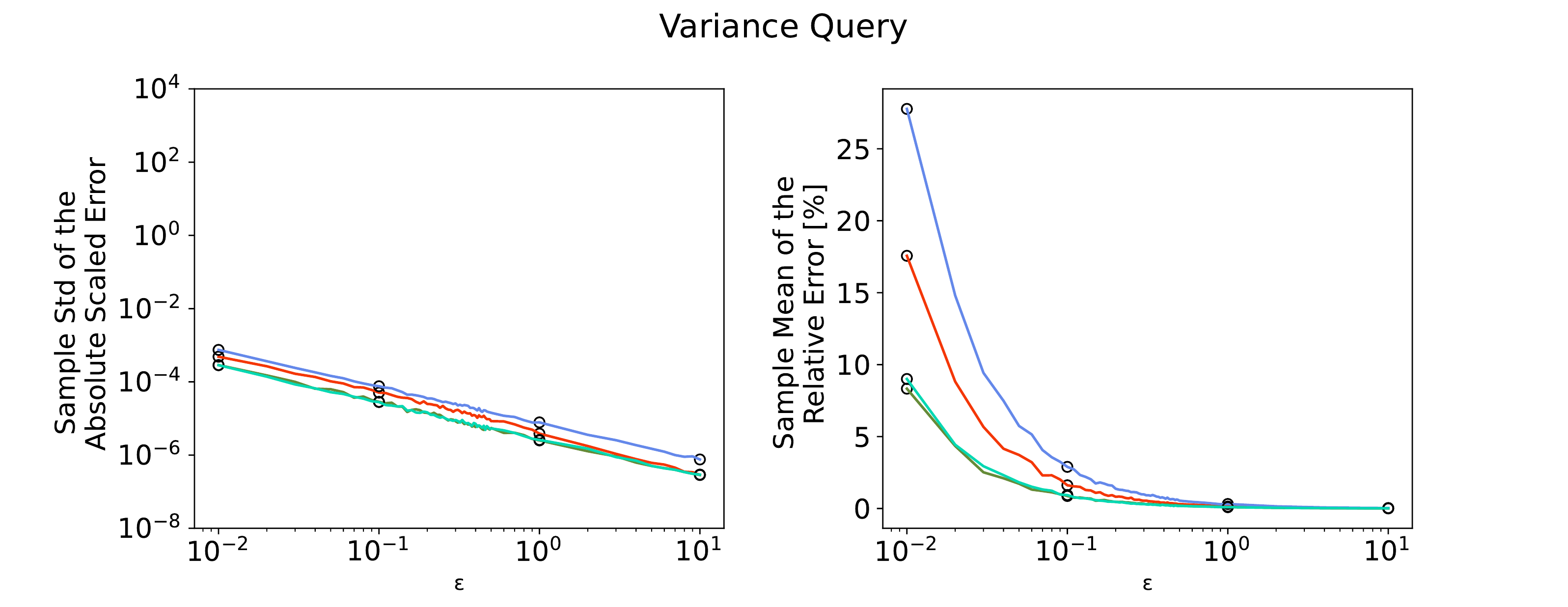} \\

\multicolumn{2}{c}{\includegraphics[ scale=0.3]{legend_read_dataset.pdf}}

\end{tabular}
    \caption{ Experiments of the queries count, sum, mean, and var on the attribute Hours of the U.S.A census dataset containing 48842 individuals (500 experiments per \textbf{$\varepsilon$}).}
    \label{tab:real_hours}
\end{table*}

\begin{table*}
    \centering
\begin{tabular}{p{0.48\linewidth} p{0.48\linewidth}}

\includegraphics[ width=\linewidth, height=\linewidth, keepaspectratio]{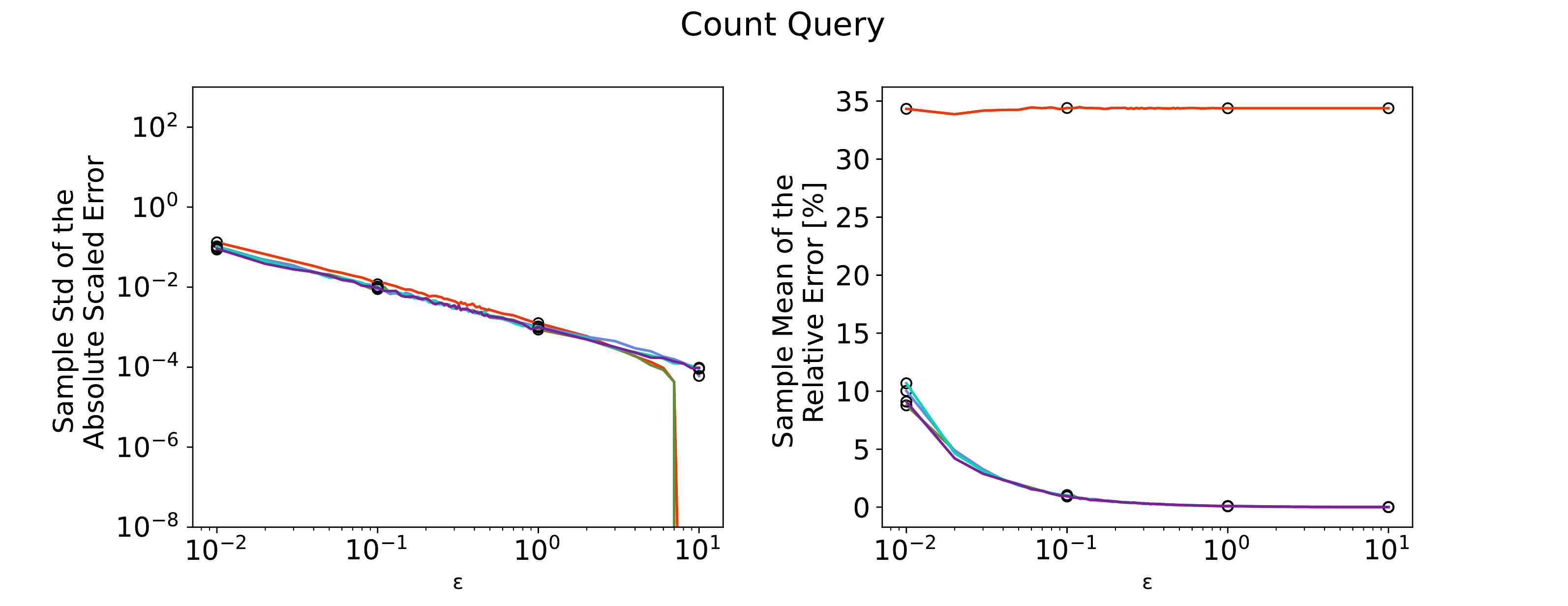}
&
\includegraphics[ width=\linewidth, height=\linewidth, keepaspectratio]{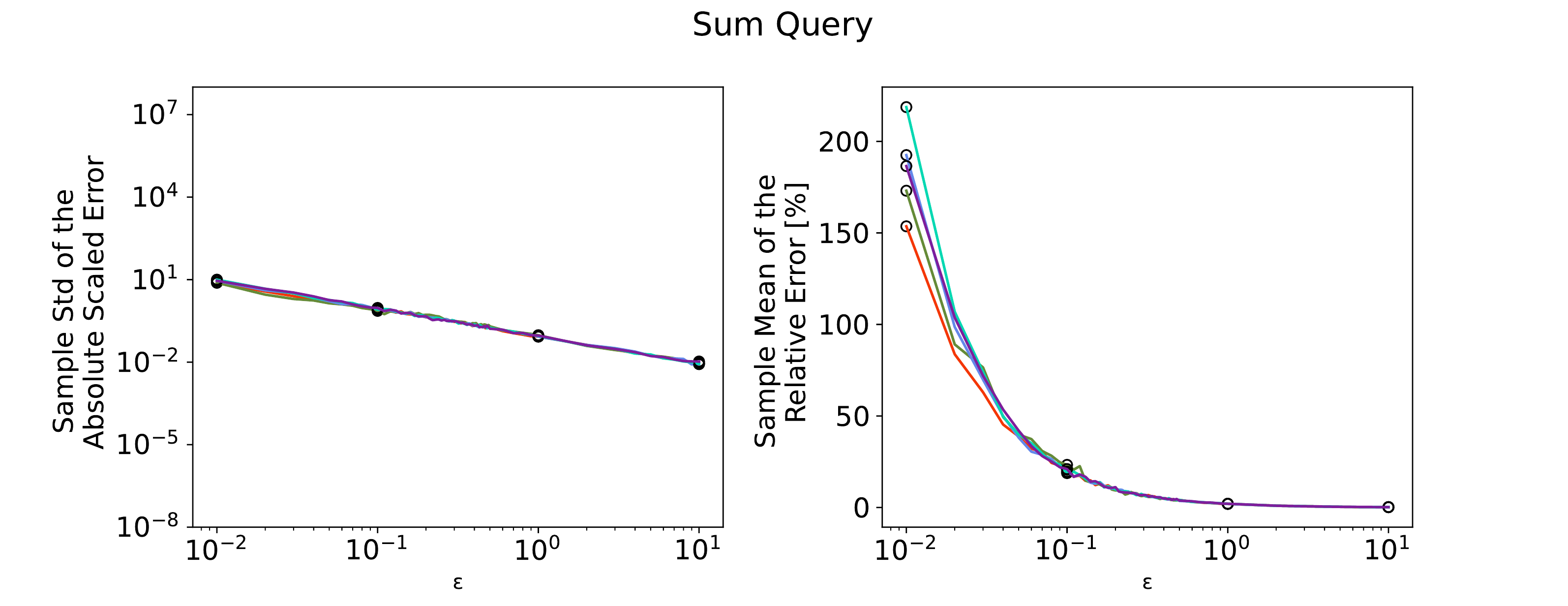} \\

\includegraphics[ width=\linewidth, height=\linewidth, keepaspectratio]{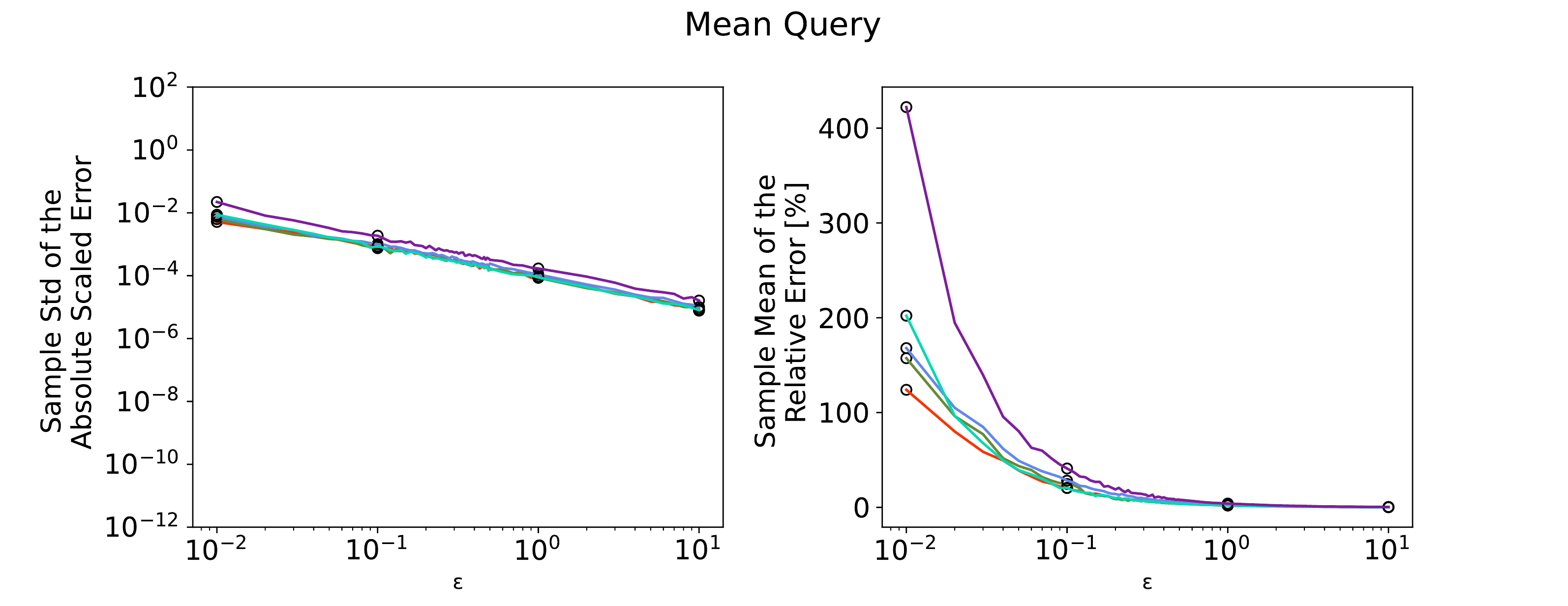}
&
\includegraphics[ width=\linewidth, height=\linewidth, keepaspectratio]{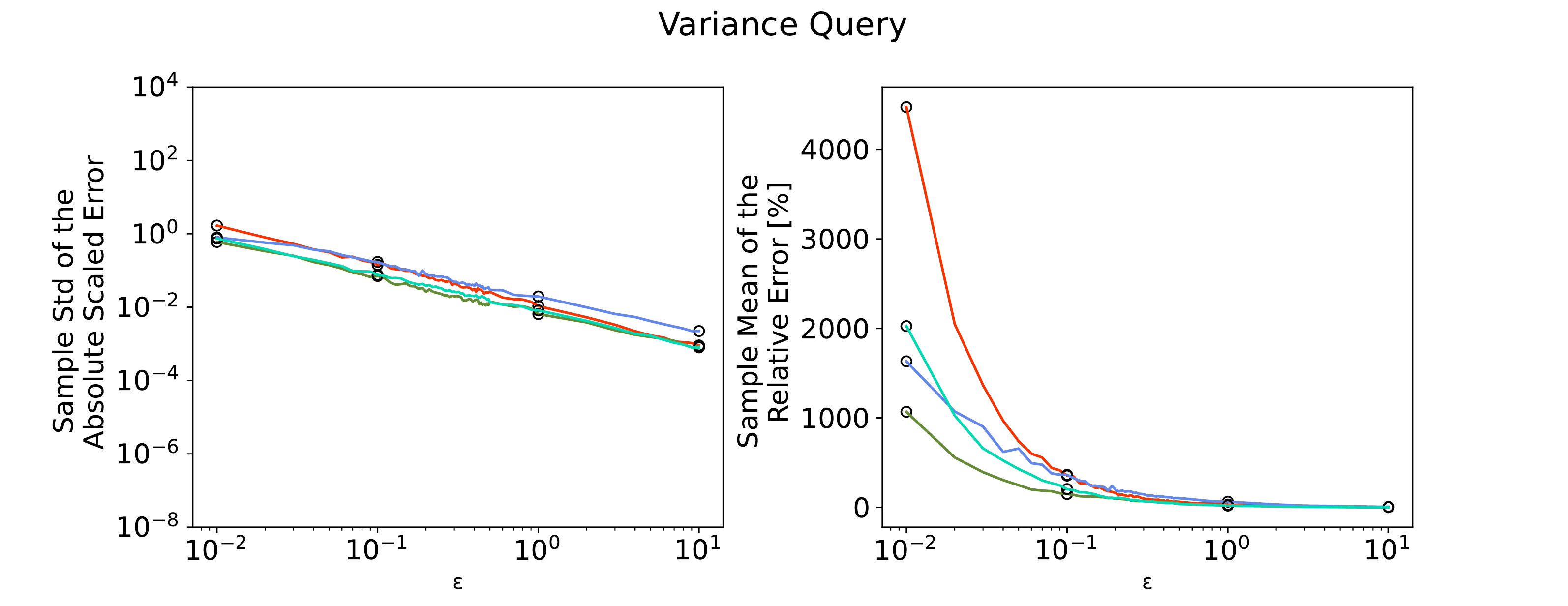} \\

\multicolumn{2}{c}{\includegraphics[scale=0.3]{legend_read_dataset.pdf}}

\end{tabular}
    \caption{ Experiments of the queries count, sum, mean, and var on the attribute Absences of the Portuguese education dataset containing 649 individuals (500 experiments per \textbf{$\varepsilon$}).}
    \label{tab:real_absences}
\end{table*}

\begin{table*}
    \centering
\begin{tabular}{p{0.48\linewidth} p{0.48\linewidth}}

\includegraphics[ width=\linewidth, height=\linewidth, keepaspectratio]{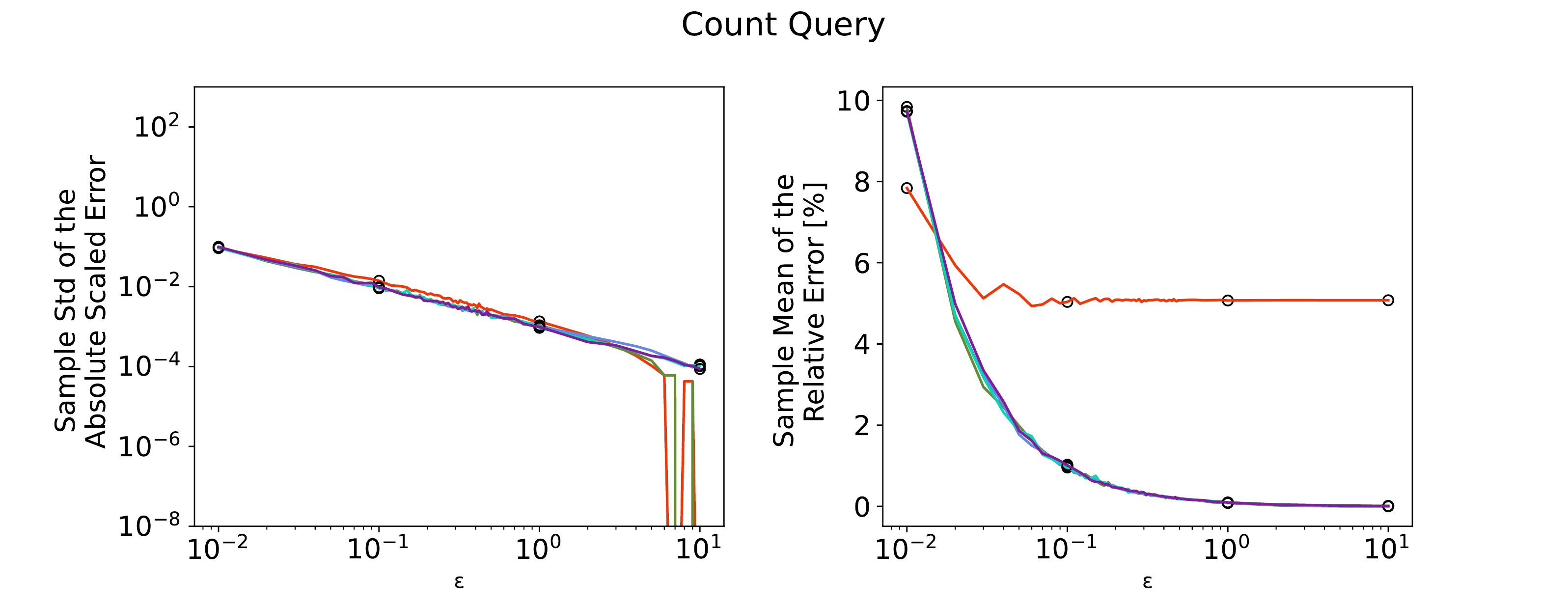}
&
\includegraphics[ width=\linewidth, height=\linewidth, keepaspectratio]{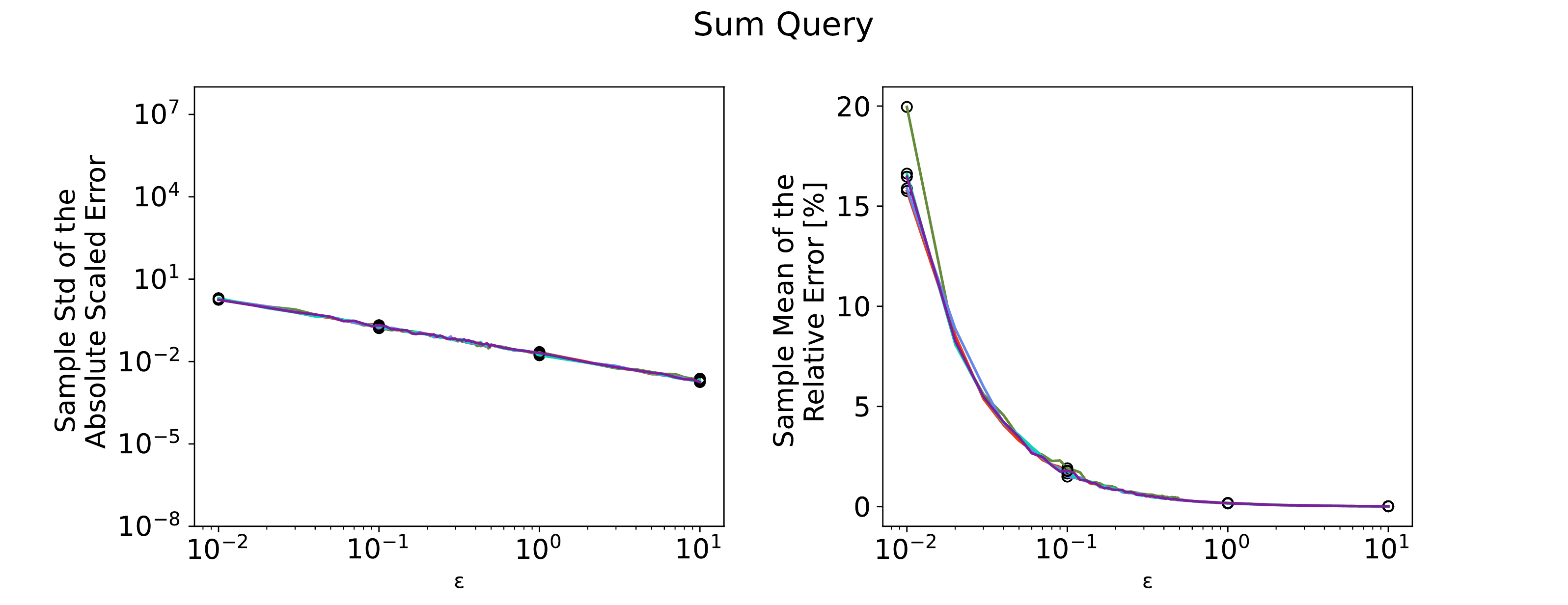} \\

\includegraphics[ width=\linewidth, height=\linewidth, keepaspectratio]{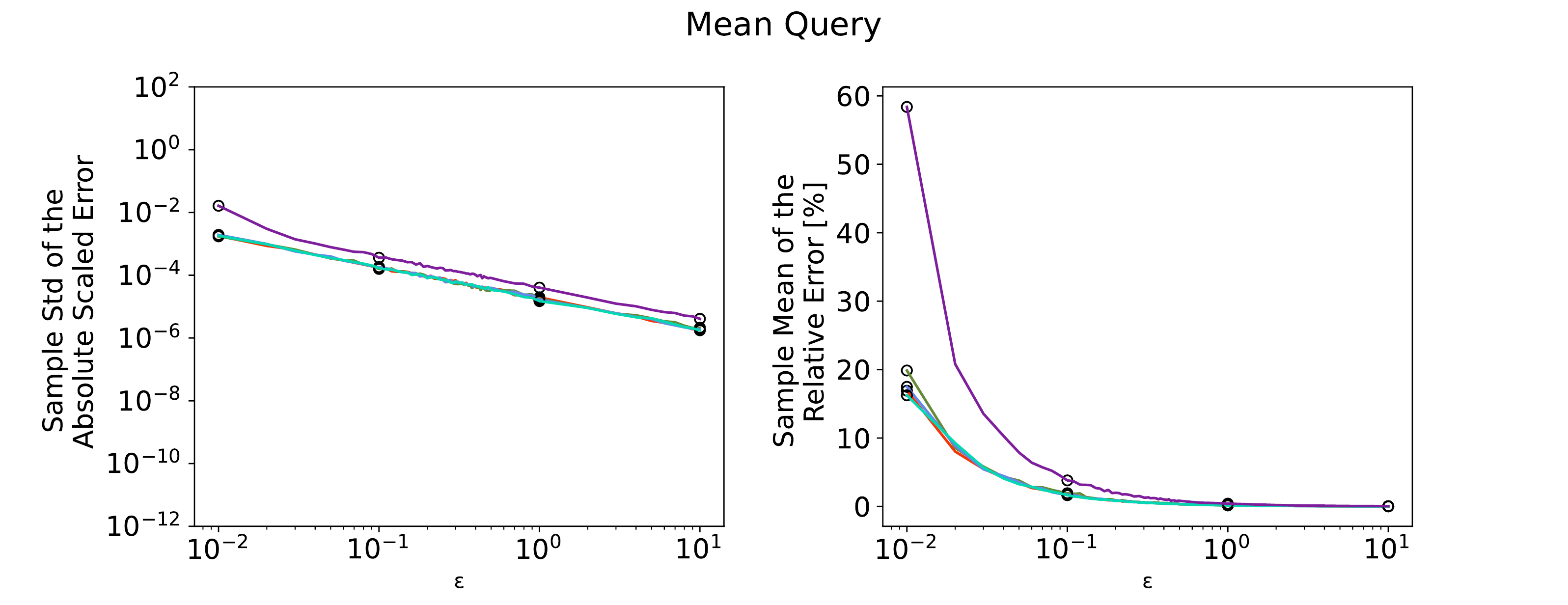}
&
\includegraphics[ width=\linewidth, height=\linewidth, keepaspectratio]{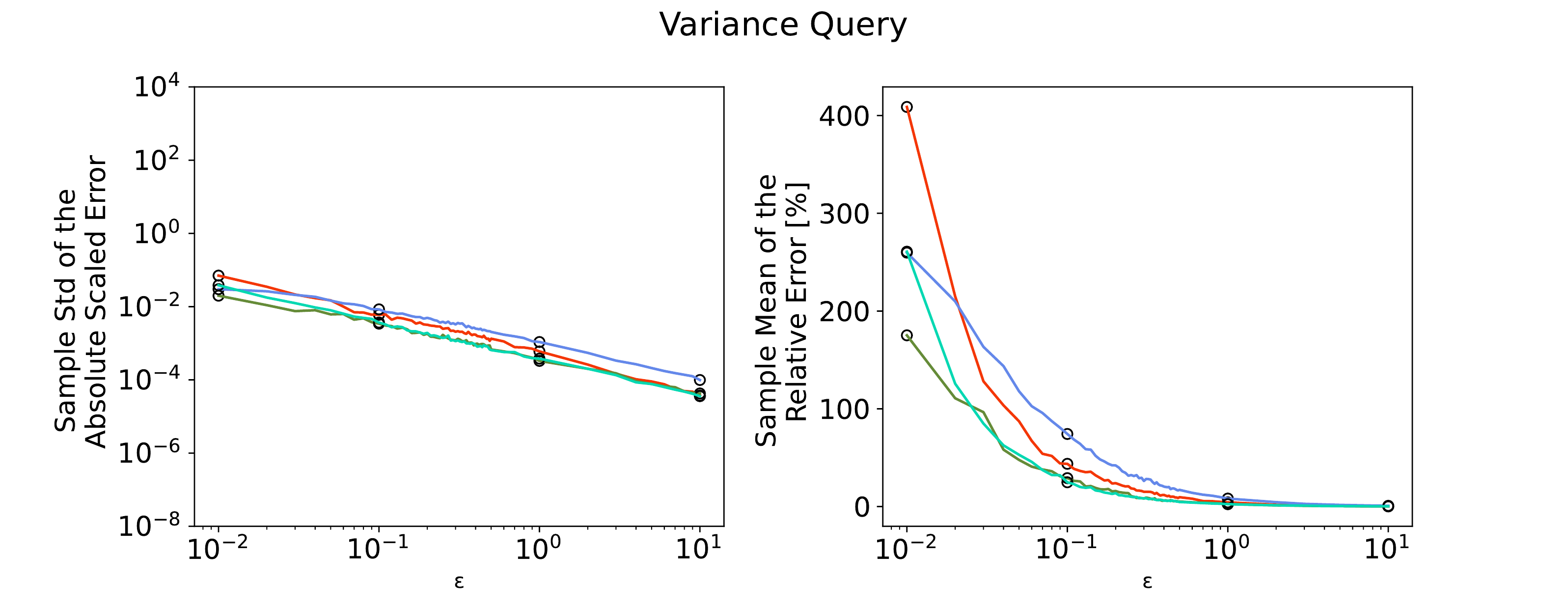} \\

\multicolumn{2}{c}{\includegraphics[scale=0.3]{legend_read_dataset.pdf}}

\end{tabular}
    \caption{ Experiments of the queries count, sum, mean, and var on the attribute Grades of the Portuguese education dataset containing 649 individuals (500 experiments per \textbf{$\varepsilon$}).}
    \label{tab:real_grades}
\end{table*}